\documentclass[nofootinbib,amsmath,prd,aps,superscriptaddress,tightenlines,12pt]{revtex4}
\usepackage{graphicx}
\usepackage{epstopdf}
\usepackage{bm}
\usepackage{epsfig}
\usepackage{graphics}
\usepackage{xspace}
\usepackage{subfigure}

\def\bnslash{\bar n\!\!\!\slash}

\def\OMIT#1{}

\newcommand{\nn}{\nonumber}

\newcommand{\bn}{{\bar n}}
\newcommand{\bea}{\begin{eqnarray}}
\newcommand{\eea}{\end{eqnarray}}

\newcommand{\gsim}{\mathrel{\rlap{\lower4pt\hbox{\hskip1pt$\sim$}}\raise1pt\hbox{$>$}}}

\newcommand{\be}{\begin{equation}}
\newcommand{\ee}{\end{equation}}

\begin{document}

\setlength\baselineskip{17pt}


\title{\bf Probing nuclear dynamics in  jet production \\ with a global event shape}

\author{Zhong-Bo Kang}
\affiliation{Los Alamos National Laboratory,
                   Theoretical Division,
                   Los Alamos, NM 87545}
                   
\author{Xiaohui Liu}
\affiliation{High Energy Division, 
                  Argonne National Laboratory, 
                  Argonne, IL 60439}
\affiliation{Department of Physics and Astronomy, 
                   Northwestern University,
                   Evanston, IL 60208}
                   
\author{ Sonny Mantry}
\affiliation{High Energy Division, 
                  Argonne National Laboratory, 
                  Argonne, IL 60439}
\affiliation{Department of Physics and Astronomy, 
                   Northwestern University,
                   Evanston, IL 60208}

\author{Jian-Wei Qiu}
\affiliation{Physics Department, 
                   Brookhaven National Laboratory, 
                   Upton, NY 11973}
\affiliation{C.N. Yang Institute for Theoretical Physics, 
                   Stony Brook University, 
                   Stony Brook, NY 11794}



\newpage
\makebox[6.5in][r]{\hfill ANL-HEP-PR-13-18}


\begin{abstract}
  \vspace*{0.3cm}

We study single jet production in electron-nucleus collisions $e^- + N_A \to J + X$, using the 1-jettiness ($\tau_1$) global event shape. It inclusively quantifies  the pattern of radiation in the final state,  gives enhanced sensitivity to soft radiation at wide angles from the nuclear beam and final-state jet, and facilitates the resummation of large Sudakov logarithms associated with the veto on additional jets. Through their  effect on the observed pattern of radiation, 1-jettiness can be a useful probe of  nuclear PDFs and power corrections from dynamical effects in the nuclear medium. This formalism allows for the standard jet shape analysis while simultaneously providing sensitivity to soft radiation at wide angles from the jet.    We use a factorization framework  for cross-sections differential in $\tau_1$ and the transverse momentum ($P_{J_T}$) and rapidity ($y$) of the jet, in the region $\tau_1\ll P_{J_T}$. The restriction $\tau_1\ll P_{J_T}$ allows only soft radiation between the nuclear beam and jet directions, thereby acting as a veto on  additional jets. This region is also insensitive to the details of the jet algorithm, allowing for better theoretical control over resummation, while providing enhanced sensitivity to nuclear medium effects.  We give numerical results at leading twist, with resummation at the next-to-next-to-leading logarithmic (NNLL) level of accuracy, for a variety of nuclear targets. Such studies would be ideal for the EIC and the LHeC proposals for a future electron-ion collider, where a range of nuclear targets are planned.

\end{abstract}

\maketitle

\newpage
\tableofcontents

\section{Introduction}
\label{introduction}

The discovery of the quark-gluon plasma (QGP) in heavy-ion collisions at RHIC and the LHC, has made possible, for the first time,  laboratory studies of quark-gluon matter at the high densities and temperatures that existed only a few microseconds after the Big Bang. One of the key pieces of evidence in the discovery of the QGP was the observed  \cite{Arsene:2004fa,Back:2004je,Adams:2005dq,Adcox:2004mh,Muller:2012zq,Aamodt:2010jd,CMS:2012aa,Milov:2011jk} suppression of high transverse momentum hadrons or jets in heavy-ion collisions compared to that in proton-proton collisions. This suppression can be understood in terms of the energy loss \cite{Gyulassy:1993hr,Baier:1996sk,Zakharov:1997uu,Wiedemann:2000za,Gyulassy:2000er,Wang:2001ifa,Arnold:2002ja,Ovanesyan:2011xy}  experienced by  fast-moving partons propagating through the QGP plasma, formed during the heavy-ion collision, before emerging as final-state hadrons or jets. Such nuclear medium effects also induce additional radiation, associated with the energy-loss mechanisms,  that can  alter the characteristics, such as the overall jet shape,  of the observed  radiation in the final state. Such a medium modification of jet shape or jet quenching has been proposed in theory \cite{Vitev:2008rz,Vitev:2009rd,D'Eramo:2010xk} and  has been investigated at both RHIC and LHC \cite{Ploskon:2009zd,Kapitan:2011xy,:2012is}, where the nuclear medium effects are visualized by varying  jet-shape parameters such as the jet-cone size.

Studying the medium modification of jet shape and jet production in cold nuclear matter would provide independent tests of energy-loss mechanisms.  In addition, it provides new and complementary observables to study phenomena related to cold nuclear matter including shadowing, anti-shadowing, EMC, and fermi-motion effects that affect the properties of nuclear parton distribution functions (PDFs). This can provide complementary information to the analysis of jet quenching associated with the QGP, as well as independent tests of energy loss mechanisms. For example, one of the puzzling  results \cite{Tarafdar:2012ef}   observed at RHIC was that heavy meson production had the same level of suppression as light meson production, even though one expects heavy quarks to be less likely to lose energy due to medium induced effects in the QGP. Similar studies with cold nuclear matter could shed light on this puzzle.

The proposed electron-ion collider (EIC) \cite{Boer:2011fh, AbelleiraFernandez:2012cc,AbelleiraFernandez:2012ni}, aims to conduct detailed studies of electron-ion (e-A) collisions, at higher energies and luminosities than ever before, for a wide range of nuclear targets. Such a facility will be an ideal laboratory for nuclear studies including gathering detailed information on the momentum and spatial distributions of quarks and gluons in the nucleon, the correlations of these distributions with nucleon spin, low Bjorken-$x$ physics and the associated gluon saturation physics, and in particular the effects of the nuclear environment on these properties as well as  nuclear medium induced  effects on the distributions of hadrons and jets.

A powerful way to complement the nuclear studies mentioned above is through global event shape analyses that characterize the detailed properties of the radiation produced in e-A collisions. In particular, in the study of jet distributions, global event shapes which depend on the properties of radiation  throughout the event, can provide complementary information to results based on analyses that focus on the region in and near the boundary of the jet. For example, energy loss in the nuclear medium can produce soft radiation at wide angles from the nuclear beam and final-state jet directions. A global event shape will capture this wide-angle soft radiation in addition to the radiation inside and outside the boundary of the jet. Comparing the distributions for such global event shapes for different nuclei in the e-A collisions  can provide vital information on the relevant  nuclear dynamics.

The concept of event shapes for deep inelastic scattering (DIS) was first introduced and developed \cite{Antonelli:1999kx,Dasgupta:2001sh,Dasgupta:2001eq,Dasgupta:2002bw} more than a decade ago. Thrust \cite{Antonelli:1999kx} and Broadening \cite{Dasgupta:2001eq} distributions were studied at the next-to-leading-log (NLL) level of accuracy and matched at ${\cal O}(\alpha_s)$ to  fixed order results. A numerical comparison was also done against ${\cal O}(\alpha_s^2)$ results \cite{Catani:1996vz,Graudenz:1997gv}. Thrust distributions have also been measured at HERA by the H1\cite{Adloff:1997gq,Aktas:2005tz,Adloff:1999gn} and ZEUS\cite{Breitweg:1997ug,Chekanov:2002xk,Chekanov:2006hv} collaborations.

In this paper, we use a global event shape called 1-jettiness ($\tau_1$) \cite{Stewart:2010tn} to study single  jet production in  e-A collisions, 
\bea
\label{process}
e^- + N_A \to J + X,
\eea
where electron scatters off a nucleus $N_A$ with atomic weight  $A$, in the deep inelastic regime to produce one final state jet ($J$). In such processes, one usually detects the final state electron to determine the virtuality of the exchanged gauge boson. For sufficiently large virtuality of the exchanged gauge boson, the machinery of QCD factorization \cite{Collins:1989gx} can be used to separate short-distance physics from  non-perturbative effects which are absorbed into long distance parton correlation functions. Alternatively, one can consider jet production where the scattered electron is unobserved. In this case,  it is the large transverse momentum of the jet that plays the role of the hard scale in the process. Such a process has been studied in the past in the context of spin-dependent observables \cite{Kang:2011jw}.

\begin{figure}
\includegraphics[scale=0.35]{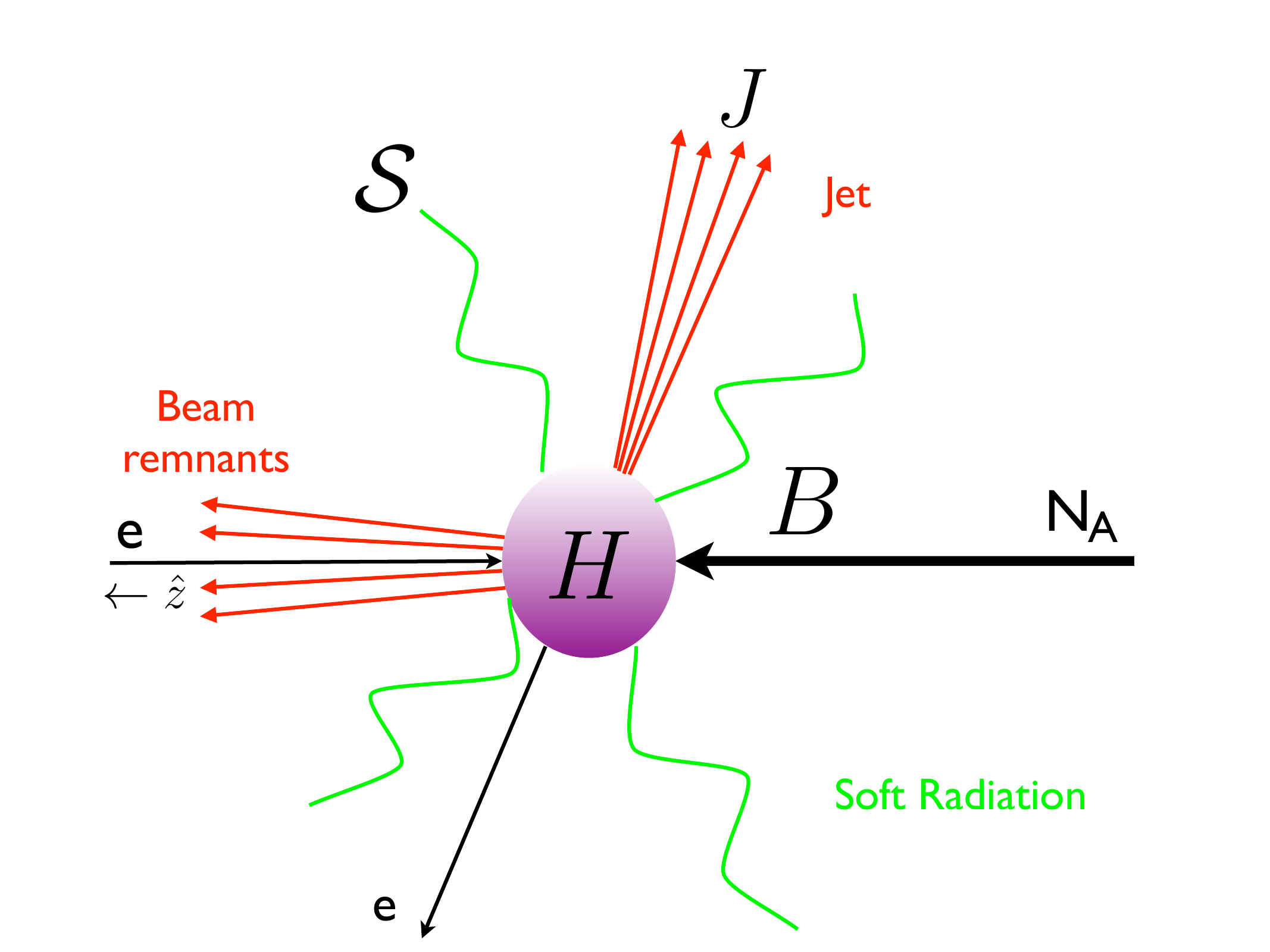}
\caption{Schematic figure of the process $e^- + N_A \to J + X$ in the limit $\tau_1\ll P_{J_T}$. The restriction $\tau_1\ll P_{J_T}$ allows only soft radiation between the beam and jet directions. The factorization framework for this process is schematically shown in Eqs.(\ref{schem-1}) and (\ref{schem-2}).}
\label{fig:process}
\end{figure}

In this work, we consider the process in Eq.(\ref{process}) 
with an additional constraint  imposed by  the $1$-jettiness event shape $\tau_1$. The use of 1-jettiness as a global DIS event shape was first proposed in Ref.~\cite{Kang:2012zr}. In particular, we are interested in the differential cross-section
\bea
\label{obs}
d\sigma_A \equiv \frac{d^3\sigma (e^- + N_A \to J + X)}{dy\> dP_{J_T}\>d\tau_1},
\eea
where $P_{J_T}$ and $y$ are the transverse momentum and rapidity of the jet $J$, respectively.
The event shape $\tau_1$ restricts the radiation between the final state jet and the nuclear beam directions. In the limit  $\tau_1\to 0$, the final state jet becomes infinitely narrow and only soft radiation (of energy $E\sim \tau_1$) is allowed between the nuclear beam and jet directions. Any energetic radiation must be closely  aligned with either the  beam or  jet directions. This is schematically illustrated in Fig. \ref{fig:process}. We restrict ourselves to such configurations  by imposing the phase space condition
\bea
\label{pscond}
\tau_1 \ll P_{J_T}.
\eea
A factorization and resummation framework for the 1-jettiness DIS event shape, in this region of phase space, was first derived in Ref.~\cite{Kang:2012zr}

The detailed properties of the radiation illustrated in Fig.~\ref{fig:process} will be affected by the nuclear target in the process. For example, for larger nuclei one typically expects enhanced hadronic activity between the jet and beam directions. The soft radiation between the beam and jet directions can be affected by jet quenching or energy loss as the jet emerges from the nuclear medium. This is because partons produced in the hard collisions could undergo multiple scattering  inside the large nucleus and thus lead to induced gluon radiation \cite{Guo:2000nz,Wang:2001ifa,Wang:2002ri} when passing through the nucleus to form the observed hadron or jet. While such effects can be studied by varying  jet shape parameters, the information about soft radiation at wide angles from the jet is often lost.  The main idea advocated in this paper is to study the properties of the observed radiation in Fig.~\ref{fig:process}, quantified by distributions in the configuration space $(\tau_1,P_{J_T},y)$, as a probe of nuclear dynamics. 
In particular, the 1-jettiness $\tau_1$ global event shape is sensitive to soft radiation at wide angles from the jet and nuclear beam directions. Thus, 1-jettiness allows one to study jet shapes while simultaneously providing sensitivity to wide-angle soft radiation.

For processes with $N$ final state jets, the appropriate event shape is called $N$-jettiness ($\tau_N$)\cite{Stewart:2010tn}, corresponding to a generalization of $\tau_1$ for $N$-jet events. N-jettiness has been studied previously in the context of implementing jet vetoes in hadron collider processes. New physics analyses typically classify data by the number of hard jets observed in the final state. Such jet binning is an effective way to enhance signals over background processes which are often accompanied by additional hard jets. Vetoing additional jets restricts the phase space for additional radiation, giving rise to large jet-veto Sudakov  logarithms that can spoil the convergence of perturbation theory.  The $N$-jettiness framework, first introduced in Ref.\cite{Stewart:2010tn}, allows for vetoes on additional jets in an inclusive manner that facilitates resummation of the  jet-veto logarithms. In this framework, the jet-veto logarithms correspond to Sudakov  logarithms of the form $\sim \alpha_{s}^n\ln^{m} (\tau_N/Q)$, where $m\leq 2n$ and $Q$ denotes the hard scale in the process. Within this context,  numerical results have been obtained for beam thrust ($0$-jettiness) distributions for Drell-Yan processes \cite{Stewart:2009yx,Stewart:2010pd} and Higgs production \cite{Berger:2010xi},  threshold resummation in gauge boson production with two final-state jets \cite{Liu:2012zg}, and the jet mass spectrum for Higgs production with one final-state jet~\cite{Jouttenus:2013hs}.

In this work, we apply the technology of the $N$-jettiness formalism, previously studied for physical processes in hadronic collisions, to electron-nucleus scattering. In this case, the 1-jettiness event shape $\tau_1$  for the process in Eq.(\ref{process}) is defined as
\bea
\label{1-jettiness}
\tau_1 &=& \sum_k \text{min} \Big \{ \frac{2q_A\cdot p_k}{Q_a}, \frac{2q_J\cdot p_k }{Q_J}\Big \},
\eea
where the sum is over all final state particles (except the final state electron) with momenta denoted by $p_k$. The null four-vectors $q_A, q_J$ denote reference vectors along the nuclear beam and jet directions respectively. The choice of $Q_a$ and $Q_J$ is not unique, so long as they are of the same order as the hard scale in the process. Different choices of $Q_a$, $Q_J$ correspond to different definitions of $\tau_1$, and lead to different geometric shapes for the beam and final-state jets~\cite{Jouttenus:2011wh,Jouttenus:2013hs}. One can also appropriately choose $Q_J$ as a function of the jet algorithm parameters to produce jets that look very close to the jets arising from commonly used jet algorithms. For example, in Ref. \cite{Jouttenus:2013hs}, $Q_J$ was chosen to depend on the jet size parameter $R$ to produce jets consistent with the anti-$k_T$ jet algorithm. Thus, varying the choices for $Q_J$ corresponds to performing a jet shape analysis.   One the other hand, keeping $Q_J$ fixed while varying $\tau_1$ corresponds to  controlling the amount of radiation near the boundary or far away from the jet while keeping the jet algorithm parameters fixed. A jet shape analysis can also be performed by being differential in an extra jet shape parameter, such as the jet mass. In this way, 1-jettiness gives us the flexibility to study jet shapes while also providing  sensitivity to soft radiation at wide angles from the jet.

In this paper, we work with specific choices for $Q_a, Q_J$ and leave a jet shape analysis in this context for future work.
For specific choices of $Q_{a}$ and $Q_J$, the reference vectors $q_A$ and $q_J$ can be determined experimentally by a minimization condition \cite{Thaler:2011gf} such that the optimal choice for $q_A$ and $q_J$ minimizes the value of $\tau_1$ in Eq.(\ref{1-jettiness}). Such an analysis is similar to that employed for finding the thrust axis for the  thrust event shape in $e^+e^-$ colliders and does not rely on any jet algorithm. Alternatively, $q_A$ can be chosen along the beam axis and $q_J$ can be determined by employing a standard jet algorithm. However, as we discuss below, in the region $\tau_1 \ll P_{J_T}$, the computation of $\tau_1$ is insensitive to the details of the jet algorithm, up to power suppressed terms \cite{Stewart:2010tn}. This feature gives analytically simpler expressions compared to methods that depend on the detailed properties of the jet algorithm, allowing for an easier implementation of higher order corrections for increased precision.

From the definition in Eq.(\ref{1-jettiness}), it becomes clear that energetic particles at wide angles from the beam and jet reference vectors $q_A$ and $q_J$ make the largest contributions to $\tau_1$. On the other hand, energetic radiation closely aligned with either $q_A$ or $q_J$ and soft radiation make relatively small contributions to $\tau_1$. Thus, the region of small $\tau_1$, quantified by the condition in Eq.(\ref{pscond}), corresponds to a single narrow jet with only soft radiation between the beam and jet directions, as illustrated in Fig. \ref{fig:process}.

In this paper, for the computation of $\tau_1$ in Eq.(\ref{1-jettiness}), we make the choices
\bea
\label{choices}
q_A = x_A P_A, &&\qquad q_J= (P_{J_T}\cosh y, \vec{P}_{J_T},P_{J_T}\sinh y), \nn \\
Q_a=x_A A Q_e,&& \qquad  Q_J = 2P_{J_T}\cosh y,
\eea
where $x_A$ denotes the nucleus momentum fraction carried by the initial parton that enters the hard interaction. The value of $x_A$ can be determined from momentum conservation in terms of the electron energy, $P_{J_T}$, and $y$ and is given later on in Eq.(\ref{xA}). Note that the reference vector $q_J$ 
\bea
\label{jet-ref}
q_J^\mu&=& (P_{J_T}\cosh y, \vec{P}_{J_T},P_{J_T}\sinh y),
\eea
is defined without any explicit reference to a particular jet algorithm and is simply a massless vector constructed for each value of  $P_{J_T}$ and $y$ in Eq.(\ref{obs}). As we explain below, such a definition without reference to a jet algorithm is consistent in the resummation region  $\tau_1 \ll P_{J_T}$, where relevant corrections are power suppressed in $\tau_1/P_{J_T}$. The choices for $Q_a$ and $Q_J$ correspond to twice the energy of the initial parton entering the hard interaction and twice the energy of the final-state jet respectively.

Experimentally, the jet reference vector $q_J^\mu$ can be obtained by applying any standard jet algorithm to obtain a leading jet with momentum $K_J$ and then construct the massless vector  $q_J^\mu= (K_{J_T}\cosh y_K, \vec{K}_{J_T},K_{J_T}\sinh y_K)$.  As we explain below, in the resummation region we will find that $q_J^\mu= (K_{J_T}\cosh y_K, \vec{K}_{J_T},K_{J_T}\sinh y_K)\simeq(P_{J_T}\cosh y, \vec{P}_{J_T},P_{J_T}\sinh y)$ up to power corrections, justifying the definition in Eq.(\ref{jet-ref}). Note that the only information from the jet algorithm used to compute $\tau_1$, is the jet reference vector $q_J$ which only depends on the transverse momentum ($K_{J_T}$) and rapidity ($y_K$) of the leading jet; i.e. the energy and direction of the leading jet.  In particular, it does not depend on the mass of the leading jet  which is sensitive to how soft radiation is clustered. Different jet algorithms will in general give different results for the jet reference vector  $q_J^\mu$. The result extracted for $K_{J_T}$ and $y_K$ depends on which final state particles are grouped into the jet by the jet algorithm in question. However, by restricting to the region $\tau_1\ll P_{J_T}$, this jet algorithm dependence becomes power suppressed \cite{Stewart:2010tn}. This can be understood by recalling that the  limit $\tau_1\to 0$ corresponds to an infinitely narrow jet with any additional wide-angle radiation  being  restricted to be soft (of energy $E \sim \tau_1$), as shown in Fig.~\ref{fig:process}. In this region of phase space, different jet algorithms will find the same energy and direction for the narrow jet, up to power corrections. In particular, they same values for $K_{J_T}$ and $y_K$ will be found so that different jet algorithms will yield the same $q_J$ in the resummation region $\tau_1\ll P_{J_T}$.  In other words, different jet algorithms will give the same result for $q_J$, for events characterized by well-separated  narrow jets.  Any differences in the jet algorithms are associated with how they treat  wide-angle soft radiation, which only affects the mass but has has little impact in determining the energy and direction of the leading jet, used to obtain the reference vector $q_J$. 

In the theoretical calculation of the observable in Eq.(\ref{obs}), we define the jet momentum as 
\bea
\label{pjet}
P_J &=& \sum_k p_k \>\theta (\frac{2q_A\cdot p_k}{Q_a} - \frac{2q_J\cdot p_k}{Q_J} ),
\eea
where the sum is over all final state particles (except the scattered electron) with momenta denoted by $p_k$. This definition of the jet momentum is closely tied with the definition of $\tau_1$ in Eq.(\ref{1-jettiness}). In the calculation of $\tau_1$, all final state particles ($p_k$) are associated with either  the $q_A$  or  $q_J$ directions as determined by the minimization condition in Eq.(\ref{1-jettiness}). The jet momentum is then defined as the sum of the particle momenta ($p_k$) associated with the $q_J$ direction, selected by the theta function condition in Eq.(\ref{pjet}). The transverse momentum $P_{J_T}$ and rapidity $y$ of the jet, appearing in Eq.(\ref{obs}), just correspond to the magnitude of the transverse momentum component and the rapidity of four-momentum $P_J$ of Eq.(\ref{pjet}). Note that in the region of small $\tau_1$, the total jet momentum $P_J$ as defined in Eq.(\ref{pjet}), will have the same energy and direction as the leading jet obtained by a standard jet algorithm up to power corrections in $\tau_1/P_{J_T}$; in particular $P_{J_T}\simeq K_{J_T} $ and $y\simeq y_K$. There can still be differences in the jet masses of $P_J$ and $K_J$ which depend on how wide angle soft radiation is clustered; however this does not affect the extraction of $q_J$ since it only depends on the energy and direction of the leading jet. Thus, due to these properties of the jet configurations in the resummation region $\tau_1 \ll P_{J_T}$, one can simply use the definition of $q_J$ in Eq.(\ref{jet-ref}) without explicit reference to any jet algorithm. The jet algorithm dependence will become important in the region $\tau_1 \sim P_{J_T}$ where power corrections cannot be ignored. Since the focus of this paper is on the resummation region $\tau_1 \ll P_{J_T}$, we use the definition of $q_J$ in Eq.(\ref{jet-ref}) in all calculations.

As discussed earlier, different choices of $Q_a,Q_J$ in Eq.(\ref{pjet}) can be made to change the geometric properties of the jet. For example, as one changes $Q_J$ in Eq.(\ref{pjet}), the set of particles that are grouped into the jet will change. This property can be exploited to perform a jet shape based analysis. In particular, one can study the dependence of $P_{J_T}$ as a function of $Q_J$ for a fixed value of $\tau_1$. By choosing $Q_J$ as a function of a jet size parameter $R$ \cite{Jouttenus:2013hs},  one can study the energy contained in the jet as a function of its cone size.  This allows us to probe energy loss near the boundary of the jet while still retaining information on wide-angle soft radiation through the value of $\tau_1$.

The dynamics of the process in Eq.(\ref{process}), in the restricted region $\tau_1 \ll P_{J_T}$, is dominated by energetic collinear emissions ($E \sim P_{J_T}$) along the nuclear beam and final state jet directions and soft emissions ($E\sim \tau_1$) in all directions. A convenient framework for such processes is given by the Soft-Collinear Effective Theory (SCET) \cite{Bauer:2000ew,Bauer:2000yr,Bauer:2001ct,Bauer:2001yt,Bauer:2002nz,Beneke:2002ph}, which is a Lagrangian and operator based formulation of the soft-collinear limit of QCD. The SCET naturally separates the physics of the disparate scales $\tau_1 \ll P_{J_T}$. A resummation of the Sudakov logarithms  $\sim \alpha_s^{n} \ln ^{m} (\tau_1/P_{J_T})$ with $m\leq 2n$, associated with the restricted radiation or equivalently a veto on additional jets or hard radiation, naturally arises through solutions to the renormalization group (RG) equations in the SCET. For the process in Eq.(\ref{process}), the SCET framework has a well-defined power counting in the small parameter $\lambda$
\bea
\lambda^2 \sim \frac{\tau_1}{P_{J_T}}.
\eea 

In the region of $\tau_1\sim P_{J_T}$, corresponding to allowing hard radiation or additional jets between the nuclear beam and jet directions, resummation effects are no longer important but power corrections can no longer be neglected. In addition,  the jet algorithm dependence is no longer suppressed.  The regions $\tau_1\ll P_{J_T}$ and $\tau_1\sim P_{J_T}$ can be smoothly connected via a matching calculation. In this work, we only focus on the resummation region $\tau_1\ll P_{J_T}$, leaving the matching calculation for future work. 

A factorization framework based on the SCET, applicable in the region $\tau_1 \ll P_{J_T} $, was first derived  for the observable in Eq.(\ref{obs}) in Ref.~\cite{Kang:2012zr}. In that work, numerical results at the next-to-leading logarithmic (NLL) accuracy were derived for the case of a proton target and the impact of non-perturbative effects in the region $\tau_1 \sim \Lambda_{QCD}$ were studied. In this work, we extend the numerical results to include a wide range of nuclear targets. In particular, we give numerical results for the nuclear targets: Proton, Carbon (C), Calcium (Ca), Iron (Fe), Gold (Au), and Uranium (Ur). In addition, we extend resummation to the next-to-next-to-leading logarithmic (NNLL) level of accuracy. This is the first time that NNLL resummation has been performed for a DIS event shape\footnote{After the first version of this paper appeared, Ref.~\cite{Kang:2013nha} appeared and it also studied the 1-jettiness DIS event shape and presented results at the NNLL level of accuracy. Their analysis was restricted to the proton target. They studied three different versions of 1-jettiness for DIS which they denoted as $\tau_a,\tau_b,$ and $\tau_c$. These different versions correspond to different choices for the reference vectors, used to define the 1-jettiness event shape, and have correspondingly different factorization structures. The event-shape $\tau_a$ is equivalent to  $\tau_1$, first studied in Ref.~\cite{Kang:2012zr} and the focus of this paper.  $\tau_b$ was shown to be equivalent to the thrust distribution studied in Ref. \cite{Antonelli:1999kx} and  $\tau_c$  was a new definition of 1-jettiness that is naturally conducive to analysis in the target rest frame.}.

The  factorization formula  for the observable in Eq.(\ref{obs}) has the schematic form \cite{Kang:2012zr}
\bea
\label{schem-1}
\frac{d^3\sigma}{dy dP_{J_T} d\tau_1} &\sim &H \otimes B \otimes J \otimes {\cal S},  
\eea
where $H$, $B$, $J$, and ${\cal S}$ denote the hard function,  the nuclear beam function, the jet function, and the soft function respectively. The hard function captures the physics of the hard partonic interaction that initiates the final state jet. Similarly, the jet function describes the dynamics of collinear energetic radiation in the final state jet and the soft function describes the low energy radiation throughout the event. The beam function \cite{Fleming:2006cd,Stewart:2009yx} $B$ is a nuclear matrix element and encodes the physics of parton correlations in the initial nucleus, collinear radiation from the initial state, and the beam remnants. The various objects in Eq.(\ref{schem-1}) have well defined field-theoretic definitions and  correspond to the various parts shown schematically in Fig. \ref{fig:process}. 

\begin{figure}
\includegraphics{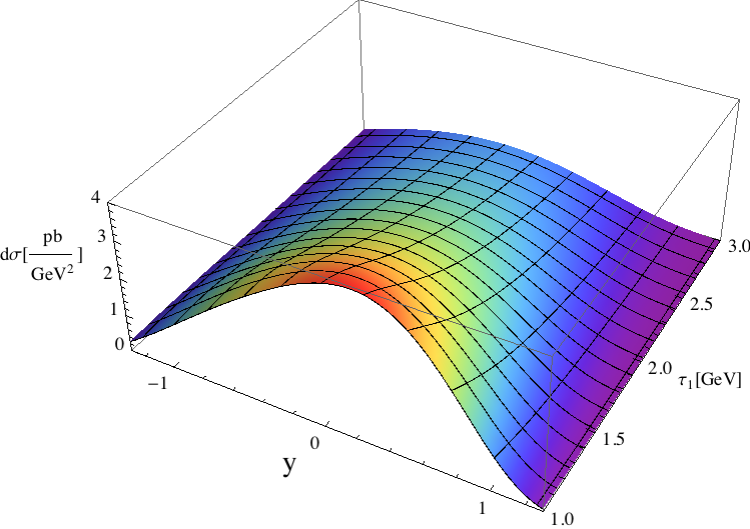}
\caption{Cross-section differential in $\tau_1$ and y  with NNLL resummation for a proton target, at $P_{J_T}=20$ GeV and center of mass energy of 90 GeV.}
\label{3dplot}
\end{figure}

In the nuclear beam function $B$, one can separate the physics of perturbative collinear initial state radiation from the non-perturbative dynamics of the initial state nucleus by performing an operator product expansion (OPE). At leading order in the OPE, the beam function can be written as a convolution between a perturbatively calculable coefficient ${\cal I}$ and the standard nuclear PDF $f_A$ 
\bea
\label{schem-2}
B \sim {\cal I} \otimes f_A .
\eea
The OPE is an expansion in the $Q_s^2(A)/t_a$, where $Q_s(A)$ is a dynamical nuclear scale and $t_a \sim \tau_1 P_{J_T}$ denotes the virtuality squared of the initial state parton that enters the hard interaction after being taken off-shell by initial-state radiation. The physics of these perturbative collinear emissions from the incoming parton, after absorbing the non-perturbative collinear emissions into the PDF,  is contained in the coefficient ${\cal I}$. The dependence of the nuclear scale $Q_s^2(A)$ on the atomic weight $A$ of the nucleus is typically parameterized as \cite{Luo:1994np,Kang:2011bp,Dusling:2009ni}
\bea
\label{nucscale}
Q_s^2(A) \sim A^\alpha \Lambda_{QCD}^2,
\eea
where the parameter $\alpha$ determines the scaling of $Q_s(A)$ with the the atomic weight of the nucleus. Note that for the simplest case of a proton target ($A=1$), the nuclear scale is just $Q_s^2(A=1) \sim  \Lambda_{QCD}^2\sim 1/R_N^2$, where $R_N$ is the nucleon radius. The power corrections in $Q_s^2(A)/(\tau_1 P_{J_T})$ can allow one to extract information on higher twist parton correlations in the nucleus and the nuclear modification of gluon radiation. Note that the size of the power corrections will increase for heavier nuclear targets as determined by the scaling with the atomic weight  in Eq.(\ref{nucscale}). The size of these power corrections for a given nuclear target will also increase at smaller values of $\tau_1$ and $P_{J_T}$. Thus, by analyzing the dependence of data on the $A,\tau_1$, and $P_{J_T}$, one can extract information on the size and properties of the nuclear-dependent power corrections. These power corrections will manifest themselves as deviations from the leading twist results of Eqs.(\ref{schem-1}) and (\ref{schem-2}) that have increased effects for heavier nuclear targets and smaller values of $\tau_1$ and $P_{J_T}$. Also, note that while the jet algorithm dependence is suppressed in powers of $\tau_1/P_{J_T}$, the nuclear medium induced effects are suppressed by $Q_s^2(A)/(\tau_1 P_{J_T})$. Thus, for a fixed $P_{J_T}$, by going to smaller values of $\tau_1$ we can reduce the jet algorithm dependence while increasing the nuclear medium effects. 

In the region $\Lambda_{QCD} \ll \tau_1 \ll P_{J_T}$, where $\tau_1$ is perturbative, the functions $H,J,{\cal I},$ and ${\cal S}$ are all perturbatively calculable and are independent of the properties of the initial state nucleus. Thus, at leading twist, the only dependence on the nuclear target comes from the nuclear PDF $f_A$ and the observable in Eq.(\ref{obs}) becomes a direct probe of the nuclear PDFs. In the region where $\tau_1 \sim \Lambda_{QCD}$, the soft function $S$ becomes non-perturbative. This can be understood by recalling that the soft function describes the dynamics of soft radiation with energy $E\sim \tau_1$. In this region, a non-perturbative model must be employed for the soft function and its parameters can be extracted by a comparison with data. Note that since the soft function is independent of the nuclear target, it is a universal function. One can exploit this universality to extract the non-perturbative soft function from data for the proton target and then use it as a known quantity for processes with other nuclear targets.

For the purposes of illustration, in Fig.~\ref{3dplot} we show the differential cross-section in Eq.(\ref{obs}) as a function of $\tau_1$ and the jet rapidity ($y$) for a proton target at $P_{J_T}=20$ GeV and a center of mass energy of 90 GeV. This result includes resummation of the jet-veto Sudakov logarithms at the NNLL level of accuracy. Studying such distributions in the configuration space of $\{\tau_1,P_{J_T},y\}$ for a wide range of nuclear targets and center of mass energies, can provide detailed information on the structure and dynamics of nuclei. This paper is a first step towards such a program of exploring nuclear physics in exclusive jet production using a global event shape.

In the rest of the paper, we give details of the formalism described in this section.  In section \ref{kin}, we describe the kinematics of the process in Eq.(\ref{process}). We also describe the result for a naive tree-level parton model calculation and discuss how it will be modified by perturbative and non-perturbative effects. In section \ref{factorization}, we give details of the factorization formula shown schematically in Eqs.(\ref{schem-1}) and (\ref{schem-2}), describe the framework used for the soft function in the non-perturbative region, and discuss power corrections. In section \ref{numerical}, we give numerical results and plots. We make concluding remarks in section \ref{conclusions}. Various useful formulae and field-theoretic definitions are collected in the appendices at the end of this manuscript. The reader not interested in the technical details of the factorization and resummation framework, can skip section \ref{factorization} and go directly to section \ref{numerical} for the numerical results.

\section{Kinematics}
\label{kin}

We carry out our analysis  in the center of mass frame defined by the electron momentum and the \textit{average} nucleon momentum in the nucleus. The electron and nucleus momentum, $p_e$ and $P_A$ respectively, take the form \footnote{Note that in the earlier work of Ref. \cite{Kang:2012zr}, the framework was set up in the electron-nucleus center of mass frame. This differs from the center of mass frame of the electron and average nucleon momentum in the nucleus considered here. The frame defined by Eq.(\ref{kin1}) is the one typically used in the experimental analysis, allowing for a more direct comparison.}
\bea
\label{kin1}
p_e^\mu &=& (p_e^0,\>\vec{p}_e), \qquad P_A^\mu = A (p_e^0,\>-\vec{p}_e),
\eea
where $A$ is the atomic weight of the nucleus and the electron momentum satisfies the on-shell condition $p_e^2=0$ so that the nucleus is also treated as a massless particle $P_A^2=0$. We introduce the quantity $Q_e$ which is related to the electron energy as
\bea
p_e^0 = |\vec{p}_e| =\frac{Q_e}{2},
\eea
so that the hadronic Mandelstam invariant $s$ is given by
\bea
s=(p_e +P_A)^2 = A \>Q_e^2.
\eea
We introduce the light-cone vector $n_A^\mu$  and its conjugate $\bn_A^\mu$ so that we can write the electron and nucleus momenta as
\bea
P_A^\mu &=& A \frac{Q_e}{2}n_A^\mu, \qquad n_A^\mu = (1,0,0,1), \nn \\
p_e^\mu &=& \frac{Q_e}{2}\bar{n}_A^\mu, \qquad \bar{n}_A^\mu = (1,0,0,-1).
\eea
The light-cone vectors satisfy $n_A^2=\bn_A^2=0$ and $\bn_A \cdot n_A=2$.  The final state jet momentum ($P_J$) is given in Eq.(\ref{jet-ref}). $P_{J_T}= |\vec{P}_{J_T}|$ and $y$  denote the transverse momentum and rapidity of the jet respectively. We denote the light-cone four momentum vector along the jet direction and its conjugate as  $n_J$ and  $\bn_J$ respectively, such that $n_J^2=\bn_J^2=0, \bn_J \cdot n_J=2$ and $\vec{n}_J=-\vec{\bar{n}}_J$. 

\OMIT{
\begin{figure}
\subfigure [NNLL: Proton \& Uranium] { \label{fig:subfig1}\includegraphics[scale=0.8]{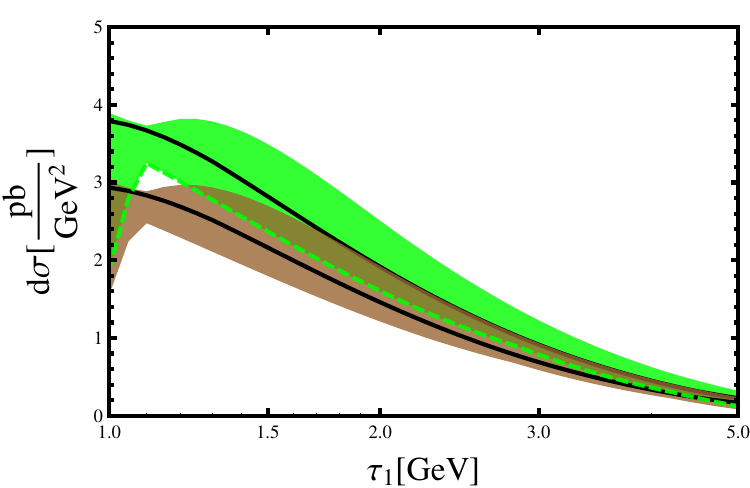}}
\subfigure [NNLL: Ratio of Uranium to Proton] { \label{fig:subfig1}\includegraphics[scale=0.8]{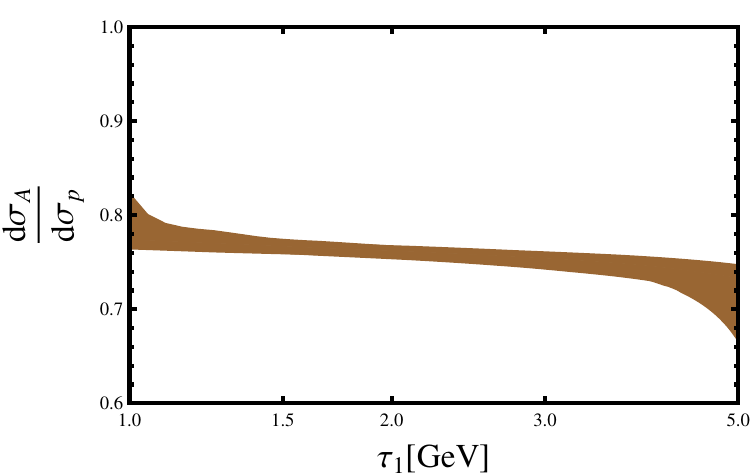}}
\caption{Top Graph: 1-jettiness distribution at NNLL level of accuracy for the proton (upper green band) and Uranium (lower brown band) for $Q_e=90$ GeV and $y=0$. Bottom Graph: Ratio of the Uranium to Proton distributions. The scale variation seems dramatically reduced in the ratio so that the overall theoretical uncertainty can be reduced by measuring ratios. }
\end{figure}}

\section{Factorization}
\label{factorization}

In this section we give the factorization formula for the process in Eq.(\ref{process}) in the region $\tau_1\ll P_{J_T}$. This formula is derived using an effective field theory approach as described by the SCET. However, before going into the details of the factorization framework, for illustration and establishing the normalization of the cross-section, we first give the lowest order result for the observable in Eq.(\ref{obs}), using the parton model. The lowest order parton model result
\bea
\label{fac-tree}
\frac{d^3\sigma^{(0)}}{dy dP_{JT} d\tau_1} &=&\sigma_0\> \delta(\tau_1) \sum_{q} e_q^2\>\frac{1}{A} f_{q/A}(x_A,\mu),  
\eea
is obtained from tree-level partonic process convoluted with the nuclear PDF. $\sigma_0$ is the tree-level partonic cross-section differential in $P_{J_T}$ and $y$
\bea
\label{sigma0}
\sigma_0 &\equiv& \frac{d\hat{\sigma}^{(0)}}{dP_{J_T}dy} = \frac{4\pi \alpha_{em}^2}{Q_e^3 e^y} \frac{\hat{s}^2 + \hat{u}^2}{\hat{t}^2},
\eea
and the  partonic Mandelstam variables $\hat{s},\hat{t}$, and $\hat{u}$ take the form
\bea
\label{stu}
\hat{s} &=& (p_e + x_A P_A)^2=x_A A Q_e^2, \nn \\
\hat{t}&=& (x_A P_A-P_J)^2 = -x_AA Q_eP_{J_{T}} e^{-y}, \nn \\
\hat{u}&=& (p_e -P_J)^2 =-Q_eP_{J_T} e^{y}.\nn \\
\eea
From the momentum conservation condition $\hat{s}+\hat{t}+\hat{u}=0$, the momentum fraction $x_A$  is given by
\bea
\label{xA}
x_A &=& \frac{e^y P_{J_T}}{A(Q_e-e^{-y}P_{J_T})}.
\eea
Note that from Eqs.(\ref{xA}) and (\ref{stu}), the dependence on the atomic weight $A$ completely cancels out in $\hat{s},\hat{t},\hat{u}$. Thus, for all nuclear targets, $\sigma_0$ is independent of $A$ and is equal to the partonic cross-section for the case of $A=1$. In other words, for the kinematics given by Eq.(\ref{kin1}), the $A$-dependence is isolated into the factor $\frac{1}{A} f_{q/A}(x_A,\mu)$ in Eq.(\ref{fac-tree}), the effective nuclear parton distribution per nucleon. 

As seen in Eq.(\ref{fac-tree}), this lowest order parton-model calculation gives a cross-section that is proportional to $\delta (\tau_1)$. This simply corresponds to the fact that at the lowest order the final state involves a jet made up of a single quark recoiling against the final state lepton. A calculation of the 1-jettiness in Eq.(\ref{1-jettiness}), for this configuration, trivially gives zero since the only final state particle that contributes is the quark which is exactly in the direction of the reference vector $q_J$. However, this parton model result is not an adequate description since important non-perturbative effects already come into play once $\tau_1 \sim \Lambda_{QCD}$. Recall that the soft radiation, schematically shown in Fig. \ref{fig:process}, has energy $E\sim \tau_1$ and will give rise to non-perturbative effects when $\tau_1 \sim \Lambda_{QCD}$. These non-perturbative effects will smear the $\delta(\tau_1)$ distribution in a way that cannot be captured by the naive parton-model calculation. A rigorous analysis requires working within a factorization framework that properly treats the physics associated with the scales $\tau_1\ll P_{J_T}$ and incorporates non-perturbative effects. 

As described earlier, the region $\tau_1 \ll P_{J_T}$ is dominated by configurations that correspond to a narrow jet with low-energy radiation between the nuclear beam and jet directions. The dynamics of this region of phase is dominated by collinear emissions along the jet and nuclear beam directions and soft emissions in all directions. The corresponding physics can be described by formulating the problem in terms of the SCET. The relevant degrees of freedom are the collinear modes along the nuclear beam and jet directions and the soft emissions with momentum scalings and virtuality given by
\bea
\text{beam-collinear}: (n_A\cdot p, \bn_A \cdot p, p^{\perp A}) &&\sim P_{J_T} (\lambda^2,1,\lambda); \>\>\>\>\>\> p^2\sim \tau_1P_{J_T}, \nn \\
\text{jet-collinear}: (n_J\cdot p, \bn_J \cdot p, p^{\perp J}) &&\sim P_{J_T} (\lambda^2,1,\lambda);\>\>\>\>\>\> p^2\sim \tau_1P_{J_T},\nn \\
\text{soft}: (n_A\cdot p, \bn_A \cdot p, p^{\perp A}) &&\sim P_{J_T} (\lambda^2,\lambda^2,\lambda^2); \>\> p^2\sim \tau_1^2,\nn \\
\eea
where $p$ denotes a generic four-momentum vector, the power counting parameter $\lambda^2 \sim \tau_1/P_{J_T}$, 
and $p^{\perp A},p^{\perp J}$ denote momentum components perpendicular to the beam and jet directions respectively. The beam-collinear modes describe the dynamics of physics along the beam direction, including the beam remnants. Similarly, the jet-collinear modes describe the dynamics of the final state jet. The typical virtuality $p^2\sim \tau_1 P_{J_T}$ of the beam and jet collinear modes is roughly the order of the invariant mass of the final state  beam and jet respectively. The soft modes describe the dynamics of soft radiation of virtuality $p^2\sim \tau_1^2$ that is present throughout the event. In the region $\tau_1 \sim \Lambda_{QCD}$, the soft radiation becomes non-perturbative.  Even at small perturbative values of $\tau_1$ where $\tau_1\ll P_{J_T}$, calculations in fixed order perturbation theory are not reliable due to the presence of large Sudakov logarithms  $\sim \alpha_s^n \ln ^{2n}(\tau_1/P_{J_T})$ that can spoil the convergence of perturbation theory.

A resummation of large logs and an incorporation of non-perturbative effects can be accomplished via a factorization framework in the SCET. This was recently done, for the observable under consideration, in Ref. \cite{Kang:2012zr}. Using the by now standard techniques in the SCET, the factorization formula for the for the kinematics of Eq.(\ref{kin1}), is given by
\bea
\label{factor-1}
\frac{d^3\sigma}{dy dP_{JT} d\tau_1} &=&\frac{\sigma_0}{A} \sum_{q,i} e_q^2 \int_0^1 dx \int ds_J \int dt_a  
\nn \\
&&\times H(x A Q_e P_{J_T}e^{-y}, \mu; \mu_H)\delta \big [ x- \frac{e^y P_{J_T}}{A(Q_e-e^{-y}P_{J_T})}\big ]  
\nn \\
&&\times J^q(s_J, \mu;\mu_J)B^q(x,t_a,\mu;\mu_B) \\
&&\times {\cal S}\left(\tau_1 - \frac{t_a}{Q_a}-\frac{s_J}{Q_J}, \mu;\mu_S\right),\nn
\eea
where the nuclear quark beam function ($B^q$), up to power corrections, is given in terms of the nuclear PDF ($f_{i/A}$) as \cite{Stewart:2009yx}
\bea
\label{beam}
B^q(x,t_a,\mu;\mu_B) &=& \int_x^1 \frac{dz}{z} {\cal I}^{qi}\left(\frac{x}{z}, t_a, \mu;\mu_B\right) f_{i/A}(z,\mu_B),
\eea
where the ${\cal I}^{qi}$ are a perturbatively calculable matching coefficients and the index $i$ runs over the initial parton species in the nucleus. The one-loop matching of the quark (and gluon) beam functions were computed in Refs. \cite{Stewart:2009yx, Stewart:2010qs, Mantry:2009qz,Berger:2010xi} and are given in appendix \ref{nlo}. Note that the argument of the hard function is independent of $A$, since the $A$-dependence cancels out in the combination $x A$ when $x$ is evaluated at its value determined by the delta function in Eq.(\ref{factor-1}).  The soft function appearing in Eq.(\ref{factor-1}) is defined in terms of the generalized hemisphere soft function \cite{Jouttenus:2011wh} as
\bea
\label{soft-1}
{\cal S}\left(\tau_1, \mu;\mu_S\right) &=& \int dk_a \int dk_J \>\delta (\tau_1-k_a-k_J) \>{\cal S} (k_a,k_J,\mu;\mu_S). \nn \\
\eea
The generalized hemisphere soft function ${\cal S} (k_a,k_J,\mu;\mu_S)$, appearing on the RHS above, is a function of two kinematic arguments $k_a,k_J$, corresponding to the contribution to $\tau_1$ of soft radiation grouped with the nuclear beam and jet directions respectively, as determined by the 1-jettiness algorithm used to calculate $\tau_1$ in Eq.(\ref{1-jettiness}). It is also known \cite{Jouttenus:2011wh}  at the one-loop level in fixed-order perturbation theory.

Eqs.(\ref{factor-1}) and (\ref{beam}) are detailed versions of the schematic formulae in Eqs.(\ref{schem-1}) and (\ref{schem-2}) respectively. The intuitive role of the hard ($H$), nuclear beam ($B^q$), jet ($J^q$), and soft (${\cal S}$)  functions were discussed in section \ref{introduction}. 
All of these objects have well-defined field-theoretic definitions. These definitions are given in appendix \ref{QFT-def} for completeness.  Furthermore, the functions $H$, ${\cal I}^{qi}$, $J^q$, and ${\cal S}$ are independent of the nuclear target and this universality can be exploited in nuclear studies. The argument $s_J$ of the jet function in Eq.(\ref{factor-1}) is a measure of the virtuality of the parton initiating the final state jet. Similarly, the argument of the beam function $t_a$ is a measure of the virtuality of the initial parton entering the hard scattering. Eq.(\ref{beam}), describes the process by which the initial state parton goes off-shell by an amount $p^2\sim t_a$ via initial state radiation (along the nuclear beam direction) which shifts the initial momentum fraction  from $z$ to $x$ as seen in Eq.(\ref{beam}). The perturbative coefficient ${\cal I}^{qi}$ captures the physics of the perturbative  initial state radiation. The convolution structure between the hard, beam, jet, and soft functions in Eq.(\ref{factor-1}) captures the dynamics of the interplay between the  soft-collinear factored sectors. 

The hard, beam, jet, and soft scales $\mu_H,\mu_B,\mu_J$, and $\mu_S$ respectively  are of typical size
\bea
\label{scales}
\mu_H \sim P_{J_T}, \qquad \mu_B \sim \mu_J \sim \sqrt{\tau_1 P_{J_T}}, \qquad \mu_S \sim \tau_1.
\eea
All objects in the factorization forumla are evaluated at a common scale $\mu$. Their evolution from their natural scales in Eq.(\ref{scales}) to the scale $\mu$ are determined by their respective renormalization group (RG) equations. The RG evolution between the various scales allows for a resummation of logarithms associated between the scales $P_{J_{T}},\tau_1,$ and $\Lambda_{QCD}$. The evolution equations for the various objects are given by
\bea
\label{evol}
H(Q^2, \mu; \mu_H) &=& U_H (Q^2,\mu, \mu_H) H(Q^2,\mu_H) ,\nn \\
{\cal I}^{qi}\left(\frac{x}{z}, t_a, \mu;\mu_B\right)&=& \int dt_a'\> U_B(t_a-t_a',\mu,\mu_B){\cal I}^{qi}\left(\frac{x}{z}, t_a', \mu_B\right)\nn \\
J^q(s_J, \mu;\mu_J) &=& \int  ds_J' \> U_J(s_J-s_J',\mu,\mu_J) J^q(s_J',\mu_J), \nn \\
{\cal S} (k_a,k_J,\mu;\mu_S) &=& \int dk_a' \int dk_J' \>U_S(k_a-k_a',k_J-k_J',\mu,\mu_S)\> {\cal S} (k_a,k_J,\mu_S), \nn \\
\eea
where $U_H(Q^2,\mu,\mu_0),U_B(t_a,\mu,\mu_0),U_J(s_J,\mu,\mu_0)$ and $U_S(k_a,k_J,\mu,\mu_0)$ are the RG evolution factors, from the scale $\mu_0$ to the scale $\mu$, of the hard, beam, jet, and soft functions respectively. The nuclear PDF $f_{i/A}$ in Eq.(\ref{beam}) is evaluated at the scale $\mu$ as determined by the standard DGLAP evolution equations.  A collection of useful formulae that determine the various RG evolution equations is given in appendix \ref{resum}.

\subsection{Factorization in position space}
The beam, jet, and soft functions that appear in Eqs.(\ref{factor-1}), (\ref{beam}) and (\ref{soft-1}) depend on variables in momentum space. One can also rewrite the factorization formula in terms of position space quantities. This can often simplify its implementation since the RG evolution equations become multiplicative instead of  the convolution structure seen in Eq.(\ref{evol}). The momentum and position space functions are related via Fourier transforms as
\bea
\label{mom-pos}
{\cal I}^{qi}(\frac{x_a}{z_a},t_a,\mu;\mu_B) &=& \int \frac{dy_{t_a}}{2\pi}\> e^{iy_{t_a} t_a} {\cal I}^{qi}(\frac{x_a}{z_a},y_{t_a},\mu;\mu_B), \nn \\
J(s_J,\mu;\mu_J) &=& \int \frac{dy_J}{2\pi} \> e^{iy_Js_J} J(y_J,\mu;\mu_J),\nn \\
S(k_a,k_J,\mu;\mu_S) &=& \int \frac{dy_{k_a} dy_{k_J}}{4\pi^2} \> e^{iy_{k_a}k_a+iy_{k_J}k_J} S(y_{k_a},y_{k_J},\mu;\mu_S),\nn \\
\eea
where the position-space quantities appear on the RHS above and the variables $y_{t_a},y_J,y_{k_a},y_{k_J}$ are the position space analogs of $t_a,s_J,k_a,k_J$ respectively. Note that we use the same notation for a given function and its Fourier transform in order to avoid to much clutter in notation. A given function and its Fourier transform are distinguished by looking at their arguments. The corresponding position space RG evolution equations are multiplicative and given by
\bea
{\cal I}^{qi}(\frac{x_a}{z_a},y_{t_a},\mu;\mu_B)&=&  U_B(y_{t_a},\mu,\mu_B){\cal I}^{qi}(\frac{x_a}{z_a},y_{t_a},\mu_B),\nn \\
J(y_J,\mu;\mu_J) &=& U_J(y_J,\mu,\mu_J)J(y_J,\mu_J)\nn \\
S(y_{k_a},y_{k_J},\mu;\mu_S) &=& U_S(y_{k_a},y_{k_J},\mu,\mu_S)S(y_{k_a},y_{k_J},\mu_S),\nn \\
\eea
where $U_B(y_{t_a},\mu,\mu_0),U_J(y_J,\mu,\mu_0)$, and $U_S(y_{k_a},y_{k_J},\mu,\mu_S)$ are the Fourier transforms of $U_B(t_a,\mu,\mu_0),U_J(s_J,\mu,\mu_0)$ and $U_S(k_a,k_J,\mu,\mu_0)$ respectively.
The factorization formula in terms of position-space quantities is given by 
\bea
\label{fac-pos-3}
\frac{d^3\sigma}{dy dP_{JT} d\tau_1} &=&\sigma_0 \>U_H (\xi^2,\mu, \mu_H)H(\xi^2, \mu_H) \nn \\
&&\times \sum_{q,i} e_q^2  \int_0^1 dx\int_x^1 \frac{dz}{z} \> \delta \big [ x- \frac{e^y P_{J_T}}{A(Q_e-e^{-y}P_{J_T})}\big ]
\nn \\
&&\times \int \frac{dy_\tau}{2\pi}   e^{iy_\tau \tau_1}U_J(\frac{y_\tau}{Q_J},\mu,\mu_J) U_S(y_\tau,y_\tau,\mu,\mu_S) U_B(\frac{y_\tau}{Q_a},\mu,\mu_B)
\nn \\
&&\times J^q(\frac{y_\tau}{Q_J}, \mu_J){\cal I}^{qi}\left(\frac{x}{z}, \frac{y_\tau}{Q_a},\mu_B\right) {\cal S}\left(y_\tau,y_\tau, \mu_S\right) \frac{1}{A} f_{i/A}(z,\mu_B).
\eea
where we have defined
\bea
\label{xi}
\xi^2 &\equiv& \frac{P_{J_T}^2}{1-e^{-y}P_{J_T}/Q_e}.
\eea

\subsection{Non-perturbative soft function}
\label{npsoft}

In the region where $\tau_1\sim \Lambda_{QCD}$, the soft function becomes non-perturbative since now $\mu_S \sim \Lambda_{QCD}$ as seen in Eq.(\ref{scales}).
In this region, the soft function cannot be computed using perturbative techniques. 	
In this case, a soft function model can be introduced for phenomenological purposes and the parameters of the model can be extracted from data. As seen in Eq.(\ref{QFT-definition}), the field-theoretic definition of the soft function is independent of the nuclear target. This universality can be exploited to extract the soft function from data collected with a proton target and used an a known quantity in for analysis with other nuclear targets.

We treat non-perturbative effects with a phenomenological model for the soft function. In particular, we write the momentum-space generalized hemisphere soft function, that appears in Eq.(\ref{soft-1}), as a convolution  \cite{Ligeti:2008ac,Hoang:2007vb} of the partonic soft function (${\cal S}_{\text{part}.}$) and a model function ($S_{\text{mod}.}$) as
\bea
\label{softmod}
{\cal S} (k_a,k_J,\mu_S) &=&  \int dk_a' \int dk_J' \>{\cal S}_{\text{part}.} (k_a-k_a',k_J-k_J',\mu_S) S_{\text{mod}.}(k_a',k_J').
\eea
The   model function satisfies the normalization condition
\bea
\label{smodnorm}
\int dk_a' dk_J'\> S_{\text{mod}.}(k_a',k_J') &=&1.
\eea
The partonic soft function ${\cal S}_{\text{part}.}$ is simply the result of the perturbative computation of the soft function. The model function $S_{\text{mod}.}(k_a',k_J')$ is typically chosen to peak around $k_{a,J}'\sim \Lambda_{QCD}$, so that as expected for $\tau_1 \gg \Lambda_{QCD}$, the soft function reduces entirely to ${\cal S}_{\text{part}.}$ up to power corrections in $\Lambda_{QCD}/\tau_1$. This can be seen by noting that  the since dominant contribution of $S_{\text{mod}.}$ comes from its peak  region $k_{a,J}'\sim \Lambda_{QCD}$ and the typical scaling of the soft momenta in the perturbative region is $k_{a,J}\sim \tau_1\gg \Lambda_{QCD}$, an OPE of the partonic soft function can be performed in the limit $k_{a,J} \gg k'_{a,J}$ to get
\bea
\label{softmod-ope}
{\cal S} (k_a,k_J,\mu_S) &=&  {\cal S}_{\text{part}.} (k_a,k_J,\mu_S) + {\cal O} (\frac{\Lambda_{QCD}}{\tau_1}),
\eea
where the normalization condition in Eq.(\ref{smodnorm}) was used to obtain the first term above. Thus, as expected, in the perturbative region $\tau_1\gg \Lambda_{QCD}$ the model soft function of Eq.(\ref{softmod}) reduces to the perturbative result ${\cal S}_{\text{part}.}$ and the model dependence arising  through $S_{\text{mod}.}$ is power suppressed.

The scale dependence of the soft function in Eq.(\ref{softmod}) is contained entirely in ${\cal S}_{\text{part}.}$. There is no scale dependence in the model function $S_{\text{mod}.}$. Since ${\cal S}_{\text{part}.}$ is just the perturbative soft function, the convolution structure Eq.(\ref{softmod}) correctly reproduces the perturbative scale dependence of the soft function.

The soft function in position space ${\cal S}(y_\tau,y_\tau,\mu)$, that appears in Eq.(\ref{fac-pos-3}) and is related to the momentum space soft function via Eq.(\ref{mom-pos}), is correspondingly modeled  using Eq.(\ref{softmod}) as
\bea
\label{pos-soft-2}
{\cal S}\left(y_\tau,y_\tau, \mu_S\right) &=&  \int dk_a \int dk_J \int dk_a' \int dk_J'\>e^{-iy_\tau (k_a + k_J)} \nn \\
&\times&{\cal S}_{\text{part}.}(k_a-k_a',k_J-k_J',\mu_S){\cal S}_{\text{mod}.}(k_a',k_J').
\eea
We can further simplify  by writing the momentum-space partonic soft function that appears above in terms position-space partonic soft function as
\bea
\label{pos-soft-3}
{\cal S}_{\text{part}.}(k_a-k_a',k_J-k_J',\mu_S) &=& \int \frac{dy_{k_a}  dy_{k_J}}{4\pi^2} e^{iy_{k_a}(k_a-k_a')+iy_{k_J}(k_J-k_J')} {\cal S}_{\text{part}.}(y_{k_a},y_{k_J},\mu_S). \nn \\
\eea
Combining Eqs.(\ref{pos-soft-2}) and (\ref{pos-soft-3}), the convolution in Eq.(\ref{softmod}) becomes a simple product in position space
\bea
\label{pos-soft-4}
{\cal S}\left(y_\tau,y_\tau, \mu_S\right) &=& {\cal S}_{\text{part}.}(y_{\tau},y_{\tau},\mu_S) {\cal S}_{\text{mod}.}(y_\tau,y_\tau), \nn \\
\eea
where the position-space model function ${\cal S}_{\text{mod}.}(y_\tau,y_\tau)$ given by
\bea
\label{mod-pos}
 {\cal S}_{\text{mod}.}(y_\tau,y_\tau)&=& \int dk_a' \int dk_J' \> e^{-iy_{\tau}(k_a'+k_J')}{\cal S}_{\text{mod}.}(k_a',k_J'). \nn \\
 \eea
Further simplifications can be made by changing the variables  of integration in Eq.(\ref{mod-pos}). In particular, we introduce new integration variables $u,\zeta$, defined as
\bea
u=k_a'+k_J', \qquad \zeta=k_a'-k_J',
\eea
to rewrite the position-space model soft function $ {\cal S}_{\text{mod}.}(y_\tau,y_\tau)$ as
 \bea
 \label{smody}
{\cal S}_{\text{mod}.}(y_\tau,y_\tau)&=& \int du \> e^{-iy_{\tau}u} \int_{-u}^u \frac{d\zeta}{2} \>{\cal S}_{\text{mod}.}(\frac{u+\zeta}{2},\frac{u-\zeta}{2}).
\eea
The integration over the $\zeta$ variable can be perfomed to define   a new single-variable function $F_{\text{mod}.}(u)$ and its position space version $F_{\text{mod}.}(y)$ as
\bea
\label{fmoduy}
F_{\text{mod}.}(u) &=& \int_{-u}^u \frac{d\zeta}{2} \>{\cal S}_{\text{mod}.}(\frac{u+\zeta}{2},\frac{u-\zeta}{2}), \qquad
F_{\text{mod}.}(y)=\int du \> e^{-iy_{\tau}u} \>F_{\text{mod}.}(u).
\eea
Using Eq.(\ref{fmoduy}) in Eq.(\ref{smody}), the position-space model soft function ${\cal S}(y_\tau,y_\tau,\mu_S)$ takes the form
\bea
{\cal S}(y_\tau,y_\tau,\mu_S)&=&{\cal S}_{\text{part}.}(y_{\tau},y_{\tau},\mu_S)F_{\text{mod}.}(y_\tau).  
\eea
Using the above relation for the soft function that appears in Eq.(\ref{fac-pos-3}), the factorization formula in terms of position-space quantities, including a parameterization of soft non-perturbative effects, takes the form
\bea
\label{fac-pos-4}
\frac{d^3\sigma}{dy dP_{JT} d\tau_1} &=&\sigma_0 \>U_H (\xi^2,\mu, \mu_H)H(\xi^2, \mu_H) \nn \\
&&\times \sum_{q,i} e_q^2  \int_0^1 dx\int_x^1 \frac{dz}{z} \> \delta \big [ x- \frac{e^y P_{J_T}}{A(Q_e-e^{-y}P_{J_T})}\big ]\frac{1}{A}f_{i/A}(z,\mu_B) 
\nn \\
&&\times \int \frac{dy_\tau}{2\pi}   e^{iy_\tau \tau_1}U_J(\frac{y_\tau}{Q_J},\mu,\mu_J) U_S(y_\tau,y_\tau,\mu,\mu_S) U_B(\frac{y_\tau}{Q_a},\mu,\mu_B)
\nn \\
&&\times J^q(\frac{y_\tau}{Q_J}, \mu_J){\cal I}^{qi}\left(\frac{x}{z}, \frac{y_\tau}{Q_a},\mu_B\right) {\cal S}_{\text{part}.}\left(y_\tau,y_\tau, \mu_S\right)F_{\text{mod}.}\left(y_\tau \right).\nn \\
\eea



\subsection{Power corrections}
\label{powercorrections}

The factorization formula of Eqs.(\ref{factor-1}) and (\ref{beam}) and its equivalent form in terms of position-space quantities in Eq.(\ref{fac-pos-3}), is valid at leading order in the power counting of the SCET. Several types of power corrections can arise and we discuss their impact on the $\tau_1$-distributions.  The  sizes of the power corrections, in the effective theory language, are characterized by ratios between the scales $\mu_H, \mu_B,  \mu_J,\mu_S$, and $Q_s(A)$. The scalings of $\mu_H,\mu_B,\mu_J,$ and $\mu_S$ are given in Eq.(\ref{scales}).   $Q_s(A)$ is a dynamical scale, often referred to as the saturation scale \cite{Gelis:2010nm}, associated with multiple scatterings in the nuclear medium. It depends on the atomic weight ($A$) of the nucleus and its size is typically given by Eq.(\ref{nucscale}),
where the value of $\alpha$ determines the power law dependence. If there is no color exchange between the nucleons in the nucleus, $\alpha \sim 1/3$ \cite{Luo:1994np,Kang:2011bp,Dusling:2009ni} corresponding to the path length available for the jet parton to have  multiple scattering in the nucleus. For the simplest case of the proton,  Eq.(\ref{nucscale}) gives $Q_s^2(A=1) \sim \Lambda_{QCD}^2$ as expected. 

A systematic analysis can be performed in the SCET to derive the operator structure of the various power corrections. We leave such an analysis for future works, where we will study in detail how the multiple scattering induced gluon radiation in the final state will alter the radiation pattern, in particular the $\tau_1$ distribution. At the moment we discuss nuclear-dependent power corrections, that depend on $Q_s(A)$, and how they may be  probed through measurements of $\tau_1$-distributions.  As seen in Eq.(\ref{bf}), the beam function is nuclear matrix element and is the only source of nuclear target dependence in the factorization formula of Eq.(\ref{factor-1}). An operator product expansion (OPE) in $Q_s(A)/t_a$ can be performed on the beam function where the leading term is given by a perturbative function convoluted with the standard PDFs as shown in Eq.(\ref{beam}). However, higher order terms in the OPE lead to the more general form of the beam function
\bea
\label{beam-power}
B^q(x,t_a,\mu;\mu_B) &=& \int_x^1 \frac{dz}{z} {\cal I}^{qi}\left(\frac{x}{z}, t_a, \mu;\mu_B\right) f_{i/A}(z,\mu_B) + {\cal O} \Big (\frac{Q_s^2(A)}{t_a} \Big ),
\eea
where the power corrections in $Q_s^2(A)/t_a$ are associated with higher twist nuclear matrix elements.  Recall that the beam function argument $t_a \sim \mu_B^2 \sim  \tau_1 P_{J_T}$, gives the virtuality of the initial parton that goes off-shell via initial state radiation before entering the hard interaction, as explained in section \ref{factorization}. Thus, the power correction to the beam function has a scaling
\bea
\label{Apower}
\frac{Q_s^2(A)}{t_a} \sim \frac{A^{\alpha} \Lambda_{QCD}^2 }{\tau_1 P_{J_T}}.
\eea
Note that this power correction has a dependence on the nuclear atomic weight through the factor of $A^\alpha$. Thus,  for heavier nuclei, the effect of these of power corrections is expected to be larger.  A detailed study of $\tau_1$ distributions over a wide range of nuclear targets can probe these nuclear-dependent power corrections. In particular, these power corrections will lead to deviations from prediction of the leading twist factorization given by Eqs.(\ref{factor-1}) and (\ref{beam}) and these deviations are expected to be larger for heavier nuclei. Also, note that the scaling of this power correction goes like $\sim 1/(\tau_1 P_{J_T})$ compared to the typical scaling of $1/P_{J_T}^2$ associated with power corrections to the hard function. This corresponds to the fact that this power correction is probing multiple scattering or nuclear modification at the beam scale $\mu_B \sim \sqrt{\tau_1P_{J_T}} $. 

Power suppressed nuclear effects can also arise from multiple hard scatterings in the nuclear medium. These will arise as power corrections to the hard function and after the soft-collinear decoupling will give rise to higher twist nuclear beam functions, which will then be matched onto higher twist nuclear parton correlation functions. However, these types of nuclear-dependent power corrections have an additional suppression  of $\sim 1/P_{J_T}^2$. Thus, the dominant nuclear power corrections will arise from Eq.(\ref{beam-power}).

From Eq.(\ref{Apower}) we see that the nuclear-medium-induced power corrections get larger for smaller values of $\tau_1$. On the other hand, the jet algorithm dependence is suppressed by powers of $\tau_1/P_{J_T}$. Thus, the 1-jettiness formalism has the advantage that in the region of small $\tau_1$, one can study the enhanced nuclear-medium-induced power corrections without much sensitivity to uncertainties typically associated with implementing the details of a jet algorithm.

\section{Numerical Results}
\label{numerical}

In this section, we present numerical results for the differential cross-section in Eq.(\ref{obs}). We present results for a range of nuclear targets: Proton, Carbon (C), Calcium (Ca), Iron (Fe), Gold (Au), and Uranium (Ur). The results are at leading order in the SCET power counting parameter $\lambda^2 \sim \tau_1/P_{J_T}$ and include a resummation of large logarithms in $\tau_1/P_{J_T}$ up to the next-to-next-to-leading logarithmic (NNLL) level of accuracy, using the convention in Table 1 of  Ref. \cite{Berger:2010xi} for determining the order of resummation. In the region $\tau_1 \gg \Lambda_{QCD}$, the numerical results are determined entirely in terms of perturbatively calculable functions and the nuclear PDFs. While the nuclear-size enhanced power corrections discussed in Sec.~\ref{powercorrections} will be left for future work, we  study in detail the nuclear modification coming from the leading twist nuclear PDFs.  For the purpose of generating numerical results, we use the EPS09 nuclear PDF sets from the analysis of Ref. \cite{Eskola:2009uj}.  We also give results in the region where $\tau_1\sim \Lambda_{QCD}$ where the soft function becomes non-perturbative.  In this region, we use a phenomenological model for the non-perturbative soft function, as described in section \ref{npsoft}, and show that while different model parameter choices lead to different predictions in the $\tau_1 \sim \Lambda_{QCD}$ region, they all converge to the perturbative result for $\tau_1 \gg \Lambda_{QCD}$ as required.  Eqs.(\ref{factor-1}) and (\ref{beam}), corresponding to the detailed version of the schematic formulae given in Eqs.(\ref{schem-1}) and (\ref{schem-2}) respectively, give the master factorization formula for the leading-twist numerical results presented in this section. Power corrections will appear  in the data as deviations from the leading twist predictions. The scaling of such deviations with $\{ A, \tau_1, P_{J_T}\}$ were discussed in section \ref{powercorrections} and are expected be larger for heavier nuclei. Thus, in addition to probing nuclear PDFs, the leading twist numerical results presented in this section can serve as a baseline to probe nuclear power corrections.

\subsection{Nuclear PDFs and master formula}

In order to generate numerical results, the nuclear PDFs $f_{i/A}(Z,\mu)$, appearing in the factorization formula in Eqs.(\ref{factor-1}) and (\ref{beam}), must be modeled and extracted from data. In the factorization formula, no assumption is made about the form of the nuclear PDF. The parametric form of  nuclear PDFs, their connection to nuclear structure, and the extraction from data is still an active  area of research~\cite{Eskola:2009uj,deFlorian:2003qf,deFlorian:2011fp,Hirai:2007sx,Kovarik:2010uv,Owens:2012bv}. In this work, we use the EPS09 nuclear PDFs from the analysis of Ref. \cite{Eskola:2009uj} to generate numerical results and plots. Such an analysis can be repeated for different parameterizations of the nuclear PDFs and it will be interesting to study the resulting differences. We leave such a comparative study for future work and limit our analysis to only working with the PDF sets in Ref. \cite{Eskola:2009uj}.  Before presenting the numerical results, we describe the form of these nuclear PDFs   and how they can be incorporated into the factorization formula in Eqs.(\ref{factor-1}) and (\ref{beam}).

The momentum fraction $z$,  appearing in the nuclear PDFs $f_{i/A}(z,\mu)$ in  Eq.(\ref{beam}), is such that at $z=1$ the initial parton $i$ carries the entire momentum of the nucleus. Typically, models of the nuclear PDF are such that the momentum of a parton in the nucleus does not exceed that of the nucleon in which it is bound. In its implementation, this corresponds to the assumption that the  nuclear PDF  falls of rapidly for $z\gtrsim 1/A$, corresponding to the intuitive expectation that average nucleon momentum in the nucleus is about a factor of $1/A$ smaller than the total nucleus momentum. The simplest way to incorporate this picture is to view the nuclear PDF as a sum of  free-nucleon PDFs in the nucleus, modified by nuclear correction factors.   After incorporating isospin symmetry, so that the $u$ and $d$ quarks of the proton PDF are the same as the $d$ and $u$ quarks of the neutron PDF respectively, the nuclear PDFs take the form  \cite{Eskola:2009uj}
\bea
\label{EPS09}
f_{u/A}^{EPS09}(x,\mu) &=&   \frac{Z}{A} \>R_u^A (x,\mu) \>f_{u/p}(x,\mu) + \frac{A-Z}{A} \>R_d^A(x,\mu) \>f_{d/p}(x,\mu) , \nn \\
f_{d/A}^{EPS09}(x,\mu) &=& \frac{Z}{A}\> R_i^A (x,\mu)\> f_{d/p}(x,\mu) + \frac{A-Z}{A}\> R_u^A(x,\mu) \> f_{u/p}(x,\mu) , \nn \\
f_{{s,c,b}/A}^{EPS09}(x,\mu) &=& R_{s,c,b}^A(x,\mu)\> f_{{s,c,b}/p}(x,\mu), \nn \\
f_{g/A}^{EPS09}(x,\mu) &=& R_g^A(x,\mu)\> f_{g/p}(x,\mu),
\eea
where the  $f_{i/p}(x,\mu)$ are the standard free-proton PDFs, the $R_{i}^A(x,\mu)$ denote nuclear correction factors arising from nuclear effects on a proton bound in the nucleus, and the nuclear PDFs $f_{u/A}^{EPS09}(x,\mu)$ are defined with an overall normalization factor of $1/A$ to give the effective nuclear PDF per nucleon. Note that the $f_{i/A}^{EPS09}(x,\mu)$  vanish for  $x>1$ since they are given by linear combinations of the proton PDFs $f_{i/p}(x,\mu)$.  The argument $x$ in Eq.(\ref{EPS09}) corresponds to the parton momentum fraction of the average nucleon momentum in the nucleus. On the other hand, the momentum fraction $z$ in Eq.(\ref{beam}), corresponds to the parton momentum fraction of entire nucleus.   As result,  the EPS09  PDFs  $f_{i/A}^{EPS09}(z,\mu)$ \cite{Eskola:2009uj}  are related to the PDFs $f_{i/A}(z,\mu)$ in the factorization formula in Eqs.(\ref{factor-1}) and (\ref{beam}) by
\bea
\label{pdftoeps09}
\frac{1}{A}f_{i/A}(z,\mu) &=& f_{i/A}^{EPS09}( A \>z,\mu).
\eea
Since $f_{i/A}^{EPS09}( A \>z,\mu)$ vanishes for $A\>z>1$, the upper limit of the range of integration for $z$, becomes $1/A$.  Using the relation in Eq.(\ref{pdftoeps09}),  the factorization formula in Eqs.(\ref{factor-1}) and (\ref{beam}) can be brought to the relatively simple form
\bea
\label{factorization-EPS09}
d\sigma_A(\tau_1,P_{J_T},y)\equiv \frac{d^3\sigma}{dy dP_{J_T} d\tau_1}\Big |_{EPS09} &=&\sigma_0 \sum_{q,i} e_q^2 \int_{x_*}^{1} \frac{dx}{x} \int ds_J \int dt_a  
\nn \\
&&\times H(\xi^2, \mu; \mu_H)J^q(s_J, \mu;\mu_J){\cal I}^{qi}\left(\frac{x_*}{x}, t_a, \mu;\mu_B\right)
\nn \\
&&\times {\cal S}\left(\tau_1 - \frac{t_a}{Q_a}-\frac{s_J}{Q_J}, \mu;\mu_S\right) f_{i/A}^{EPS09}(x,\mu_B), \nn \\
\eea
where the subscript EPS09 on the differential cross-section indicates that the factorization formula has been written in terms of the EPS09 nuclear PDFs.  Note that the dependence of the cross-section on the nuclear target is contained entirely in the nuclear PDF $f_{i/A}^{EPS09}(x,\mu)$, as seen from Eqs.(\ref{factorization-EPS09}), (\ref{xstar}), (\ref{xi}), and (\ref{sigma0}).  The $A$-dependence completely cancels out in the rest of the cross-section. The cross-section in Eq.(\ref{factorization-EPS09}), for electron-nucleus scattering, has the simple interpretation of  electron-proton scattering at a center of mass energy of $s=Q_e^2$ where the proton PDF has been dressed for nuclear corrections via the replacement $f_{i/p} \to f_{i/A}^{EPS09}$. This result is a consequence of the kinematics of Eq.(\ref{kin1}), the relation in Eq.(\ref{pdftoeps09}), and the property that $f_{i/A}^{EPS09}(A\>z,\mu)$ vanishes for $A\>z >1$.
\begin{figure}
\subfigure [$R_u^{\text{Ur(V)}}$] { \label{fig:subfig1}\includegraphics[scale=0.4]{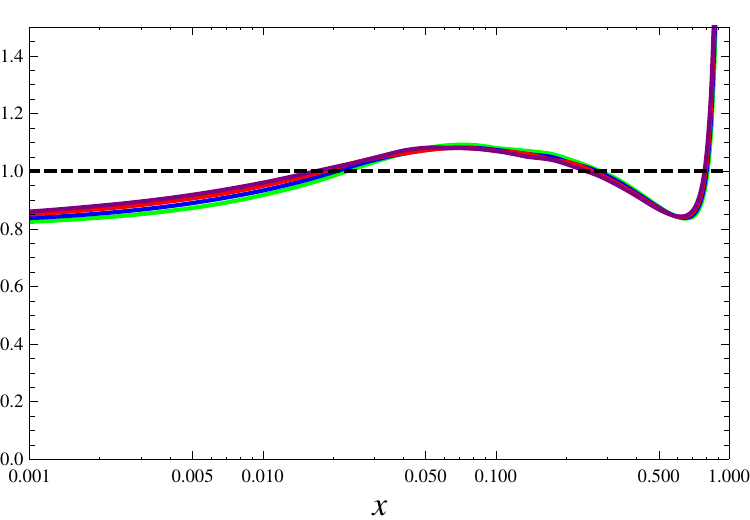}}
\subfigure [$R_d^{\text{Ur(V)}}$] { \label{fig:subfig2}\includegraphics[scale=0.4]{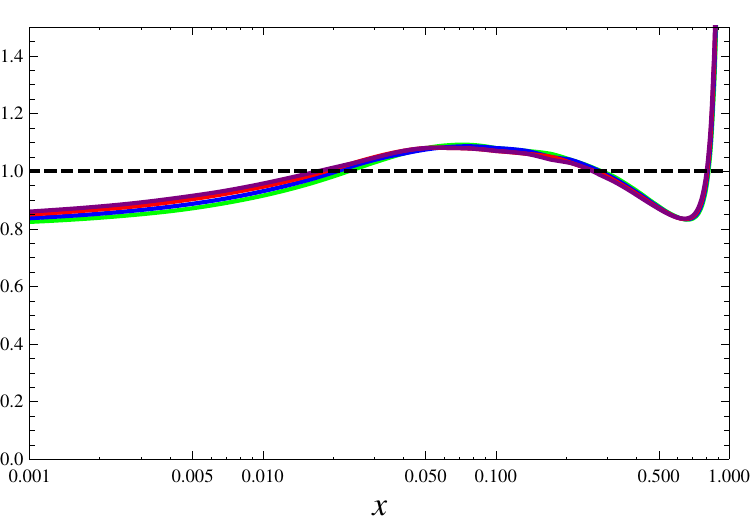}}
\subfigure [$R_u^{\text{Ur(S)}}$] { \label{fig:subfig3}\includegraphics[scale=0.4]{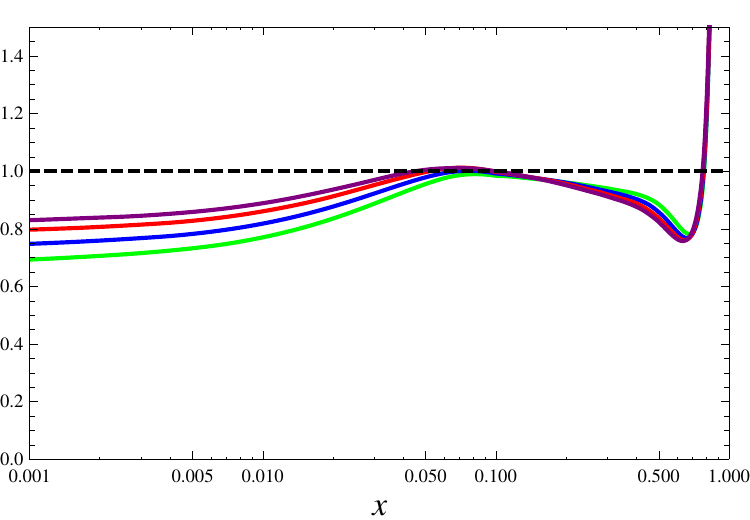}}
\subfigure [$R_d^{\text{Ur(S)}}$] { \label{fig:subfig4}\includegraphics[scale=0.4]{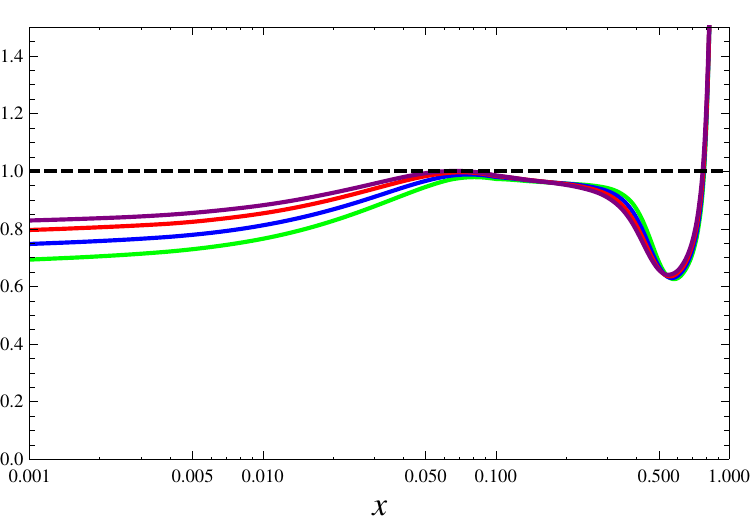}}
\subfigure [$R_s^{\text{Ur}}$] { \label{fig:subfig5}\includegraphics[scale=0.4]{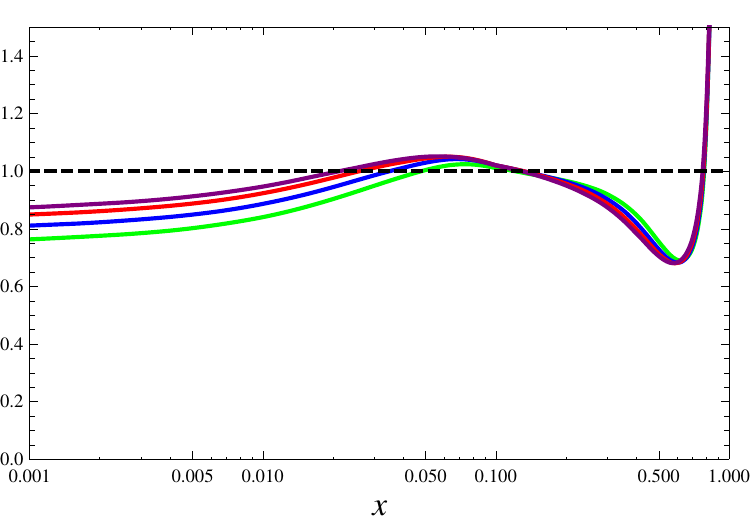}}
\subfigure [$R_g^{\text{Ur}}$] { \label{fig:subfig8}\includegraphics[scale=0.4]{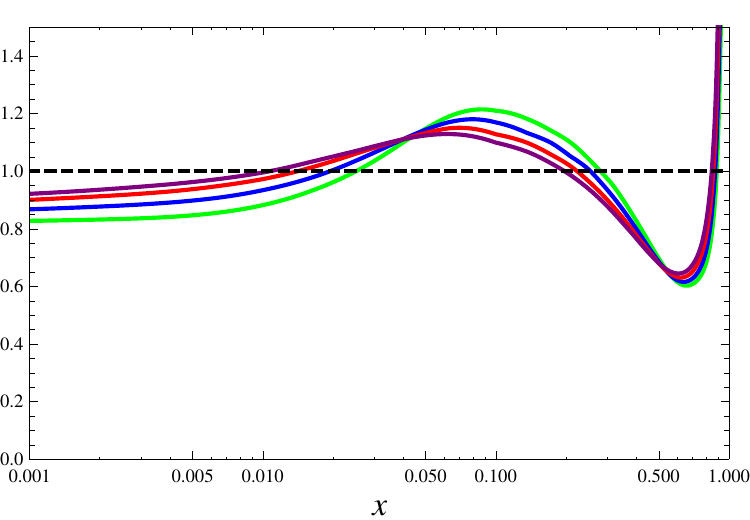}}
\caption{Nuclear correction factors $R_i^{\text{Ur}}(x,\mu)$ for the NLO nuclear PDF for a Uranium target as defined in Eq.(\ref{EPS09}). The subscript $i$ runs over the parton species $i=\{u,d,s,g\}$. For the $u$ and $d$ quarks, separate  $R$-factors are given for the valence (V) and sea quarks (S). The different curves in each graph correspond to different values for the scale $\mu$. By looking at the region of small Bjorken-$x$, the different curves from the bottom to the top correspond to $\mu=3$ GeV (Green), $\mu=5$ GeV (Blue), $\mu=10$ GeV (Red), and $\mu=20$ GeV (Purple). These plots were generated using publicly available code for the EPS09 PDF set \cite{Eskola:2009uj}.}
\label{Rfactors}
\end{figure}

The lower limit of integration $(x_*)$, over the argument of the nuclear PDF $f_{i/A}^{EPS09}(x,\mu_B)$ in Eq.(\ref{factorization-EPS09}), is given by
\bea
\label{xstar}
x_* &=& \frac{e^y P_{J_T}}{Q_e-e^{-y}P_{J_T}}.
\eea
The corresponding range  of integration  $[x_*,1]$ is then determined by the choice of the kinematic variables $\{Q_e,P_{J_T},y\}$, defined in section \ref{kin}. Thus, one can access smaller values of  Bjorken-$x$ by increasing $Q_e$ and decreasing $P_{J_T}$ and $y$. 
\begin{figure}
\includegraphics[scale=0.8]{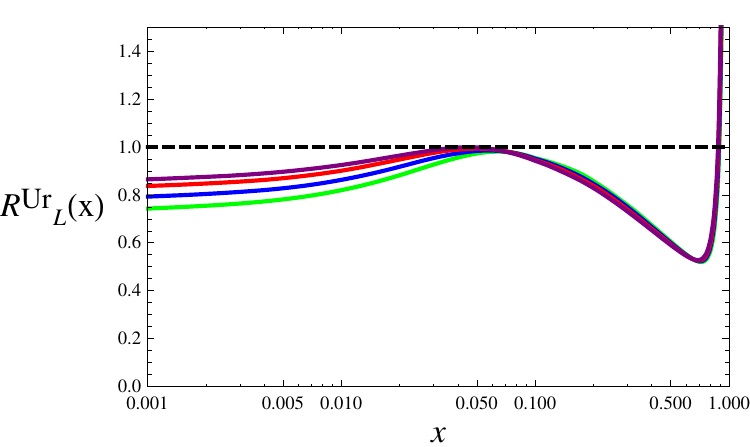}
\caption{Luminosity ratio for Uranium to proton using NLO PDFs for  $\mu=3$ GeV (Green), $\mu=5$ GeV (Blue), $\mu=10$ GeV (Red), and $\mu=20$ GeV (Purple).}
\label{lumratio}
\end{figure}

Different regions in Bjorken-$x$ are sensitive to different types of nuclear effects. For example, shadowing suppresses  the number density of partons in the region of small Bjorken-$x$, anti-shadowing enhances the parton density at  values of Bjorken-$x$ about $\sim 0.1$ , the EMC effect suppresses the parton density at intermediate values of Bjorken-$x$ $(> 0.2)$, and the effect from Fermi motion  of the nucleons enhances the parton density at Bjorken-$x$ values close to one. In Fig.~\ref{Rfactors}, we show numerical results for the nuclear correction factors $R_i^{A}$ for the NLO PDFs in Eq.(\ref{EPS09}) for the case of a Uranium target ($A=238, Z=92$). From these results, generated using the publicly available code for the EPS09 PDF set \cite{Eskola:2009uj}, we see that the shape of the $R_i^{A}$ factors clearly show the distinct regions in Bjorken-$x$ that are dominated by shadowing, anti-shadowing, the EMC-effect, and Fermi motion.
These different regions can be probed by appropriate choices for the kinematic variables $Q_e,P_{J_T},$ and $y$ to determine the lower limit $x_*$ of the Bjorken-$x$ integration, as determined by Eq.(\ref{xstar}).

In order to illustrate the effects of the nuclear correction factors $R_i^A(x,\mu)$ on the cross-sections, in Fig.~\ref{lumratio} we first consider the ratio of the luminosity functions, that appear in the tree-level cross-sections (see Eqs.(\ref{fac-tree}) and (\ref{pdftoeps09})), for a nucleus $A$ compared to the case of a proton target
\bea
\label{RL}
R_L^A(x,\mu) &=& \frac{\sum_q e_q^2 f_{q/A}^{EPS09}(x,\mu)}{\sum_q e_q^2 f_{q/p}(x,\mu)}.
\eea
The luminosity functions appearing in the ratio above are identical to the luminosity functions for fully-inclusive deep inelastic scattering at tree-level. Using Eq.(\ref{EPS09}) for the nuclear PDF $f_{q/A}^{EPS09}(x,\mu)$, one can study the effect of the nuclear correction factors $R_i^A(x,\mu)$ on the ratio of the tree-level cross-section for a nucleus $A$ compared to that of a proton target, through luminosity ratio in Eq.(\ref{RL}). In Fig.~\ref{lumratio},  we show this luminosity ratio for the case of a Uranium target ($A=238$). We see that it captures the qualitative features of shadowing, anti-shadowing, EMC effects, and Fermi motion as seen in the nuclear correction $R$-factors in Fig.~\ref{Rfactors}. However, the anti-shadowing region is completely washed out due to the isospin effect so that $R_L^A \lesssim 1$  in the anti-shadowing region.

\subsection{Distributions in $\tau_1, P_{J_T},$ and $y$}

Once higher order perturbative effects, resummation effects, and non-perturbative effects from soft radiation are included, the cross-section is more complicated and is given by Eq.(\ref{factorization-EPS09}). In this case, a simple comparison of the the tree-level luminosity ratio in Eq.(\ref{RL}) is no longer sufficient. Instead, a comparison of the predictions from the cross-section formula in Eq.(\ref{factorization-EPS09}) for different nuclear targets must be carried out and is the focus of the rest of this section. In particular, we give numerical results  for a variety of nuclear targets and kinematic configurations in $\{Q_e,\tau_1, P_{J_T},y\}$ and discuss their implications.

Theoretical uncertainties to the factorization formula in Eq.(\ref{factorization-EPS09}) will arise from a truncation of the perturbative series in the calculation of the hard ($H$), jet ($J$), beam (${\cal I}$), and soft ($\cal{S}$) functions, higher order resummation effects not included at a given level of resummation accuracy, and non-perturbative effects in the soft function (${\cal S}$). In addition, theoretical predictions will be affected by the standard PDF uncertainties. 
 Corrections to Eq.(\ref{factorization-EPS09}) will also arise from the power corrections discussed in section \ref{powercorrections}.  If one is interested in probing these power corrections, the uncertainties mentioned for the leading twist formula of Eq.(\ref{factorization-EPS09}) must be sufficiently under control.
 
In order to isolate nuclear effects we will compute the ratio 
\bea
\label{ratio}
R_A (\tau_1,P_{J_T},y) &=& \frac{d\sigma_A(\tau_1,P_{J_T},y)}{d\sigma_p(\tau_1,P_{J_T},y)},
\eea
which compares distributions in $\tau_1,P_{J_T},$ and $y$ for a  nuclear target with atomic weight $A$ to that of a proton target.  In addition to the isolation of nuclear-dependent effects, the ratio $R_A$ has the advantage that many of the uncertainties in the calculation of $d\sigma_A$ and $d\sigma_p$, as determined by Eq.(\ref{factorization-EPS09}), cancel in the ratio. In particular, we will show that the perturbative uncertainties associated with resummation and the calculation of the hard, beam, jet, and soft functions in fixed-order perturbation theory largely cancel in the ratio, leading to much smaller overall uncertainty for $R_A$. We also show that in the region $\tau_1\sim \Lambda_{QCD}$ where the soft function ${\cal S}$ is non-perturbative, the dependence on the phenomenological model implemented to describe ${\cal S}$  largely cancels in the ratio $R_A$. This can be understood as a consequence of the fact that the soft function ${\cal S}$ in Eq.(\ref{factorization-EPS09}) is universal and independent of the nuclear target.

In order to estimate the perturbative uncertainty, we employ a standard scale variation procedure. As seen in Eq.(\ref{factorization-EPS09}), the cross-section depends on a 
hard function, beam function coefficient, jet function, and soft function which naturally live at the scales $\mu_H,\mu_B,\mu_J,$ and $\mu_S$ respectively. The typical size of these scales are given in Eq.(\ref{scales}). All of these objects are evaluated at the common scale $\mu$ using their  renormalization group equations  to evolve from their natural scales. We perform a scale variation analysis similar to that carried out in Ref.\cite{Stewart:2010pd}. The nuclear PDFs are evaluated at the beam scale $\mu_B$ corresponding to the scale at which the beam function is matched on to the nuclear PDF as shown in Eq.(\ref{beam}), or more schematically in Eq.(\ref{schem-2}). We compute the cross-sections by choosing $\mu=\mu_H$ and  make four independent choices for the relative values of the scales $\mu_H,\mu_B,\mu_J,$ and $\mu_S$
\bea
\label{scalevar}
&&(a)\> \mu=\mu_H = r \sqrt{\xi^2}, \>\mu_B=r \sqrt{Q_a \tau_1},\> \mu_J = r \sqrt{Q_J\tau_1},\> \mu_S = r\> \tau_1, \nn \\
&& (b) \> \mu=\mu_H = \sqrt{\xi^2}, \>\mu_B= \sqrt{Q_a \tau_1},\> \mu_J = \sqrt{Q_J\tau_1},\> \mu_S = r^{-\frac{1}{4}\ln \frac{\tau_1}{\xi}}\> \tau_1, \nn \\
&& (c) \> \mu=\mu_H = \sqrt{\xi^2}, \>\mu_B=  r^{-\frac{1}{4}\ln \frac{\tau_1}{\xi}}\sqrt{Q_a \tau_1},\> \mu_J =  \sqrt{Q_J\tau_1},\> \mu_S = \> \tau_1, \nn \\
&&(d) \> \mu= \mu_H = \sqrt{\xi^2}, \>\mu_B=  \sqrt{Q_a \tau_1},\> \mu_J =r^{-\frac{1}{4}\ln \frac{\tau_1}{\xi}} \sqrt{Q_J\tau_1},\> \mu_S = \> \tau_1, 
\eea
where $\xi$ is given in Eq.(\ref{xi}) and $r$ denotes the scale variation parameter. For each of these choices, the scale variation parameter $r$ is varied in the range $r=\{1/2,2\}$.  One can estimate the perturbative uncertainty by adding in quadrature the uncertainty associated with the variation of each of the scales $\mu_H,\mu_B,\mu_J,$ and $\mu_S$ or by analyzing the covariance matrix. However, for simplicity, in this work we estimate the perturbative uncertainty as the envelope  \cite{Stewart:2010pd,Berger:2010xi} of the independent scale variations in Eq.(\ref{scalevar}). These two methods are expected to give similar results and a more detailed discussion can be found in \cite{Stewart:2011cf}.
\begin{figure}
\includegraphics[scale=0.8]{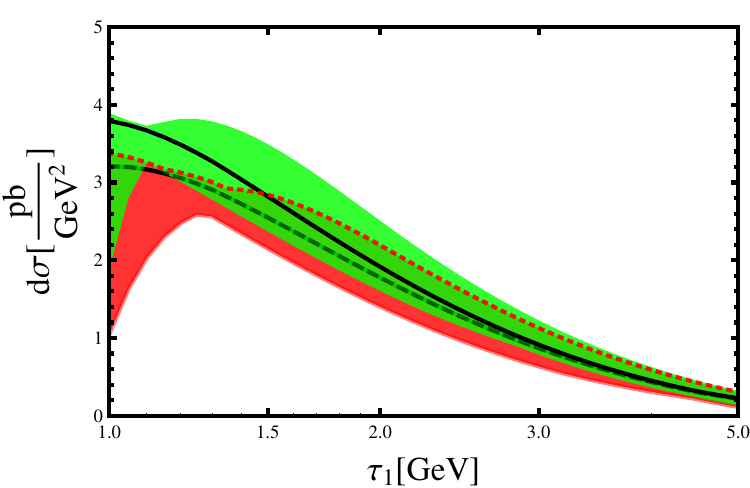}
\caption{$\tau_1$ distribution for a proton target with NLL$^\prime$ (lower red band) and NNLL (upper green band) resummation for $Q=90$ GeV, $P_{J_T}=20$ GeV and $y=0$. A more detailed description is given in the text.}
\label{protontau1}
\end{figure}

\begin{figure}
\subfigure [Proton] { \label{fig:subfig1}\includegraphics[scale=0.5]{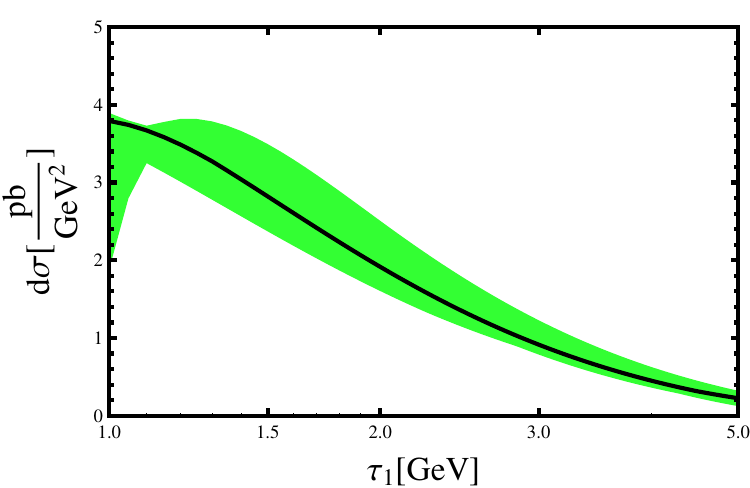}}
\subfigure [C and Proton] { \label{fig:subfig1}\includegraphics[scale=0.5]{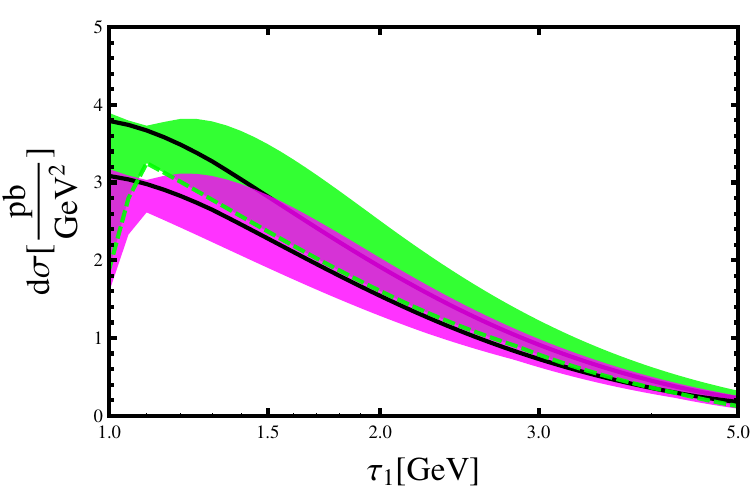}}
\subfigure [Ca and Proton] { \label{fig:subfig1}\includegraphics[scale=0.5]{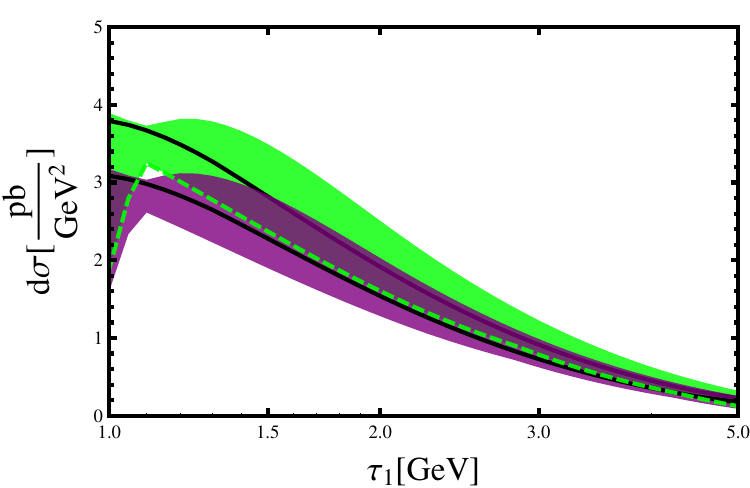}}
\subfigure [Fe and Proton] { \label{fig:subfig1}\includegraphics[scale=0.5]{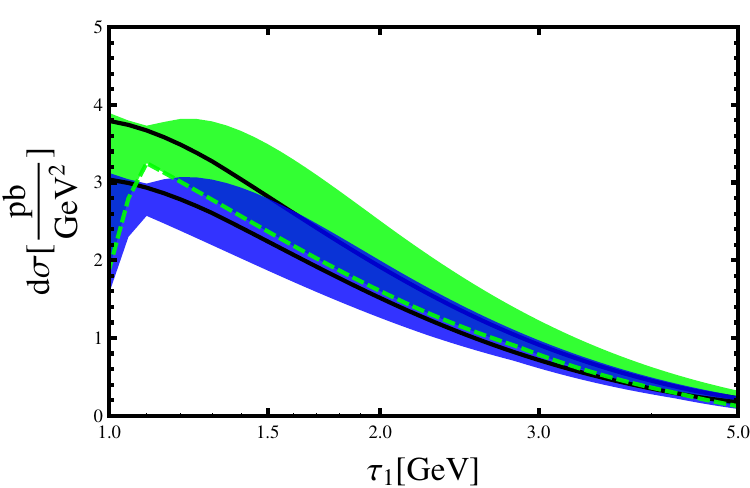}}
\subfigure [Au and Proton] { \label{fig:subfig1}\includegraphics[scale=0.5]{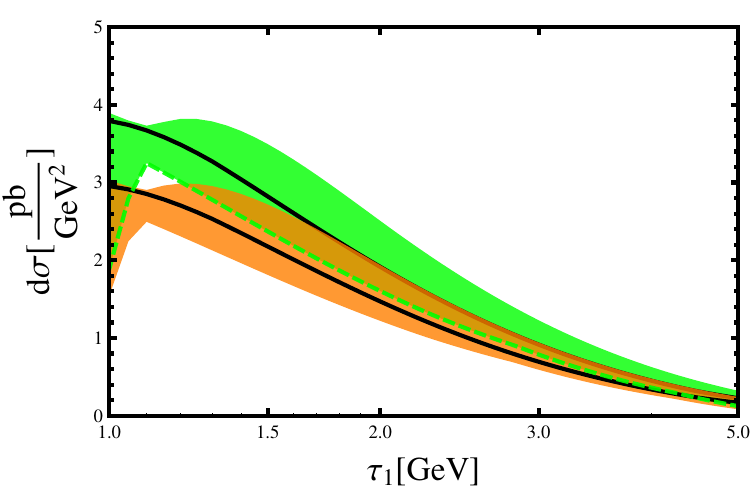}}
\subfigure [Ur and Proton] { \label{fig:subfig1}\includegraphics[scale=0.5]{UrpNNLLcomb.pdf}}
\caption{$\tau_1$-distributions with NNLL resummation for different nuclear targets for $Q_e=90$ GeV, 
 $P_{J_T}=20$ GeV, and $y=0$. In all figures, the green (upper) band corresponds to the NNLL resumed result for a proton target. The lower bands in different colors are the corresponding distributions for different nuclear targets.}
\label{tau1nuc}
\end{figure}

\begin{figure}
\subfigure [$\>$ C ] { \label{fig:subfig1}\includegraphics[scale=0.5]{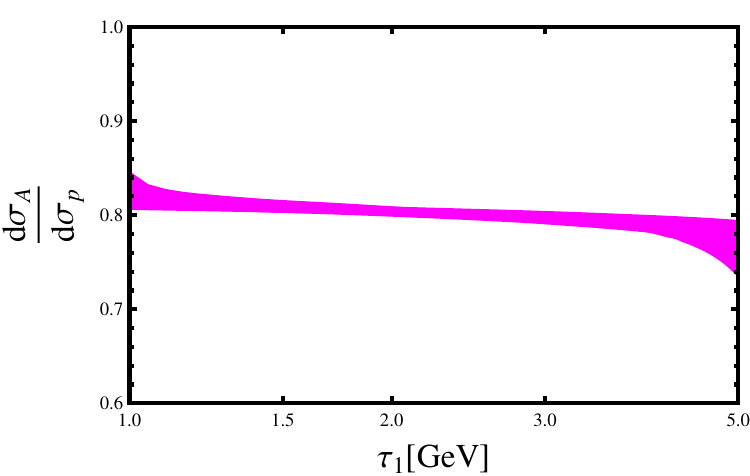}}
\subfigure [$\>$ Ca ] { \label{fig:subfig1}\includegraphics[scale=0.5]{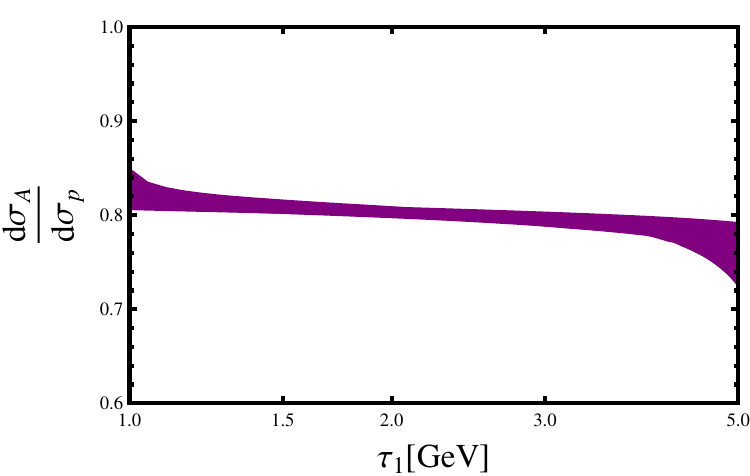}}
\subfigure [$\>$ Fe] { \label{fig:subfig1}\includegraphics[scale=0.5]{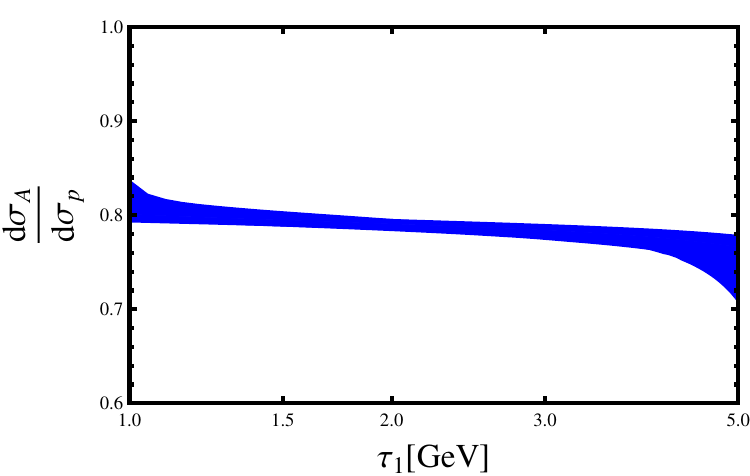}}
\subfigure [$\>$ Au ] { \label{fig:subfig1}\includegraphics[scale=0.5]{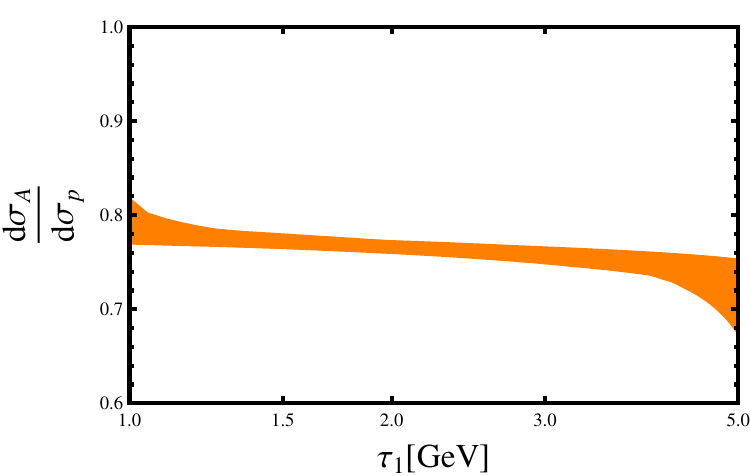}}
\subfigure [$\>$ Ur] { \label{fig:subfig1}\includegraphics[scale=0.5]{UrpNNLLratio.pdf}}
\subfigure [$\>$ C and Ur] { \label{fig:subfig1}\includegraphics[scale=0.5]{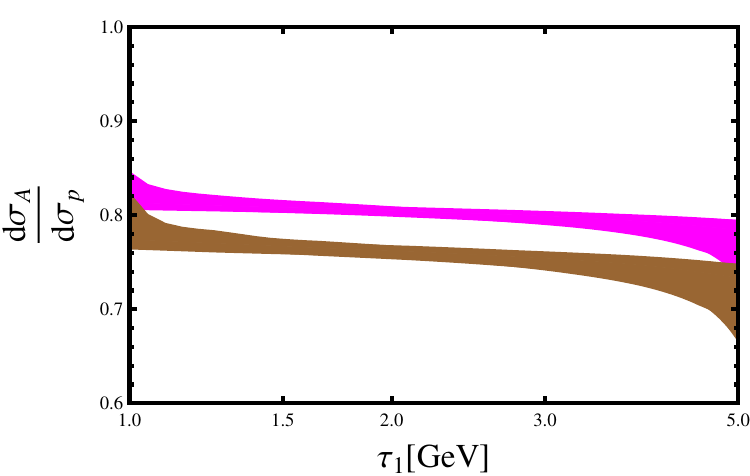}}
\caption{We show ratio $R_A=d\sigma_A/d\sigma_p$ in the $\tau_1$ distributions for various nuclear targets compared to the case of the proton target.  For easier visual comparison, we show the results for Carbon (C) and Uranium (Ur) together in subfigure $(\text{f})$. These results include resummation at the NNLL level of accuracy and are calculated at $Q_e=90$ GeV, $P_{J_T}=20$ GeV, and $y=0$, corresponding to the EMC region of the nuclear PDFs. }
\label{tau1rationuc}
\end{figure}

In Fig.~\ref{protontau1}, we show numerical results for the $\tau_1$ distribution for a proton target. The factorization formula of Eq.(\ref{factorization-EPS09}) was used to calculate this distribution for the kinematic configuration given by $Q_e=90$ GeV, $P_{J_T}=20$ GeV, and $y=0$, corresponding to typical EIC kinematics \cite{Boer:2011fh}. As discussed earlier, the $\tau_1$-distribution is affected by large Sudakov logarithms $\alpha_s^n\ln^{2n}(\tau_1/P_{J_T})$ in the region $\tau_1 \ll P_{J_T}$, so that the results of fixed order perturbation theory are no longer reliable and resummation is required. These Sudakov logarithms are associated with the veto on additional jets, enforced by the condition $\tau_1\ll P_{J_T}$ which restricts radiation between the hard jet and the nuclear beam direction to be soft ($E\sim \tau_1$), as shown in Fig. \ref{fig:process}. Fig.~\ref{protontau1} shows the result for the $\tau_1$-distribution after a resummation of the jet-veto logarithms. In particular, the red (lower) and green (upper) bands correspond to resummation at the NLL$^\prime$ and NNLL level of accuracy respectively. The NLL$^\prime$ resummation corresponds to NLL resummation combined with the product of the hard, beam, jet, and soft functions computed at NLO and using NLO PDFs. A summary of the counting of logs for resummation at different levels of accuracy can be found in Table 1 of Ref.\cite{Berger:2010xi}. The red (lower) and green (upper) bands in Fig.~\ref{protontau1}, are obtained from the envelope of the scale variations in Eq.(\ref{scalevar}). For reference, we show solid and dashed black curves corresponding to the scale choices  $(a)$ in Eq.(\ref{scalevar}) for $r=1$, for NNLL and NLL$^\prime$ resummation respectively. The red-dotted curve corresponds to the upper envelope of the NLL$^\prime$ (red) band, part of which is hidden by the NNLL (green) band.
\begin{figure}
\subfigure [$\>$ Proton ] { \label{fig:subfig1}\includegraphics[scale=0.5]{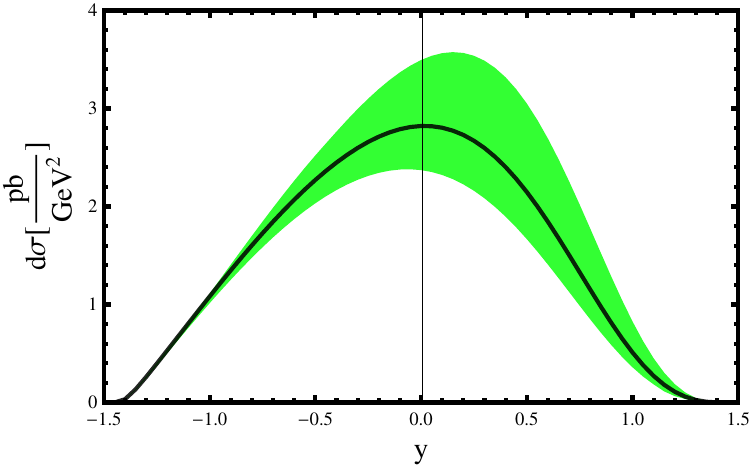}}
\subfigure [$\>$ Proton and C] { \label{fig:subfig1}\includegraphics[scale=0.5]{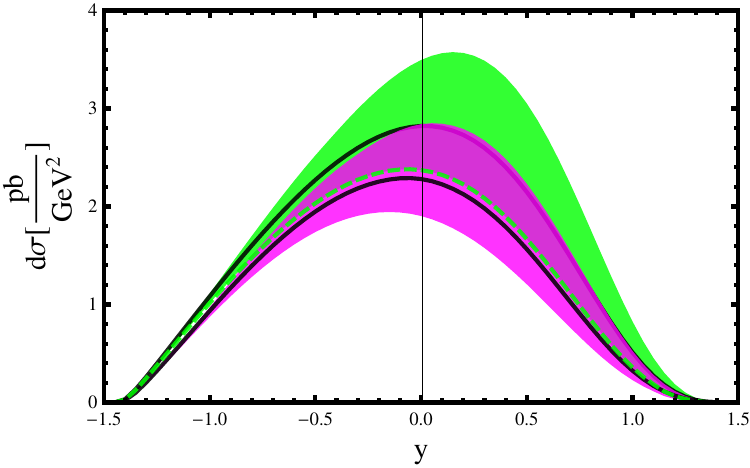}}
\subfigure [$\>$ Proton and Ca] { \label{fig:subfig1}\includegraphics[scale=0.5]{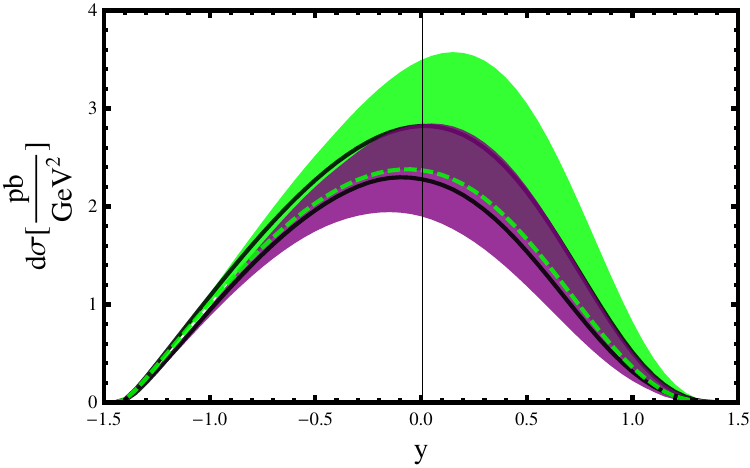}}
\subfigure [$\>$ Proton and Fe] { \label{fig:subfig1}\includegraphics[scale=0.5]{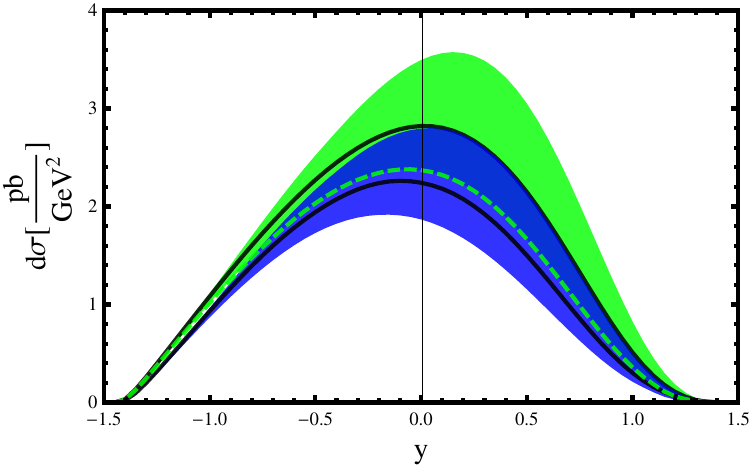}}
\subfigure [$\>$ Proton and Au] { \label{fig:subfig1}\includegraphics[scale=0.5]{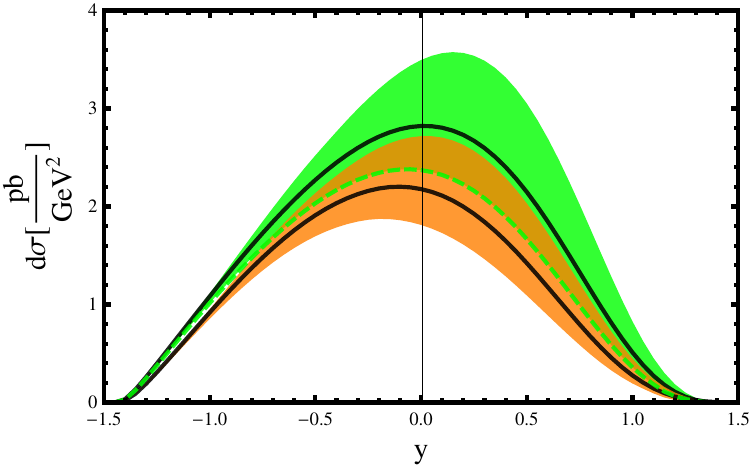}}
\subfigure [$\>$ Proton and Ur] { \label{fig:subfig1}\includegraphics[scale=0.5]{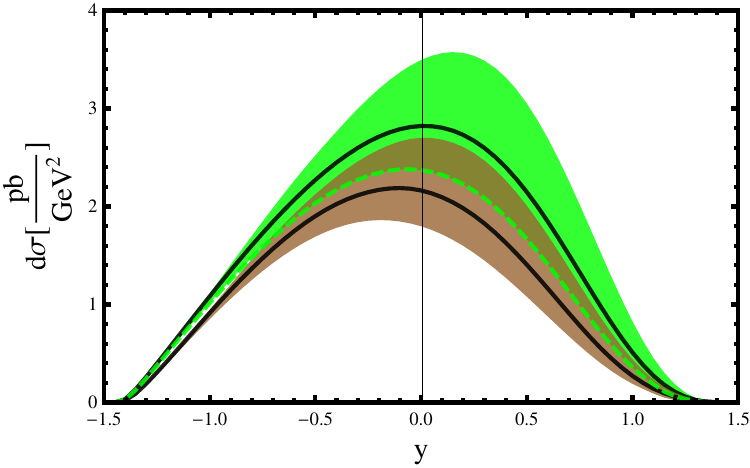}}
\caption{Rapidity ($y$) distributions with NNLL resummation for different nuclear targets for $Q_e=90$ GeV, 
 $P_{J_T}=20$ GeV, and $\tau_1=1.5$ GeV. In all figures, the green (upper) band corresponds to the NNLL resumed result for a proton target. The lower bands in different colors are the corresponding distributions for different nuclear targets.}
 \label{rapidity}
\end{figure}
\begin{figure}
\subfigure [$\>$ C ] { \label{fig:subfig1}\includegraphics[scale=0.5]{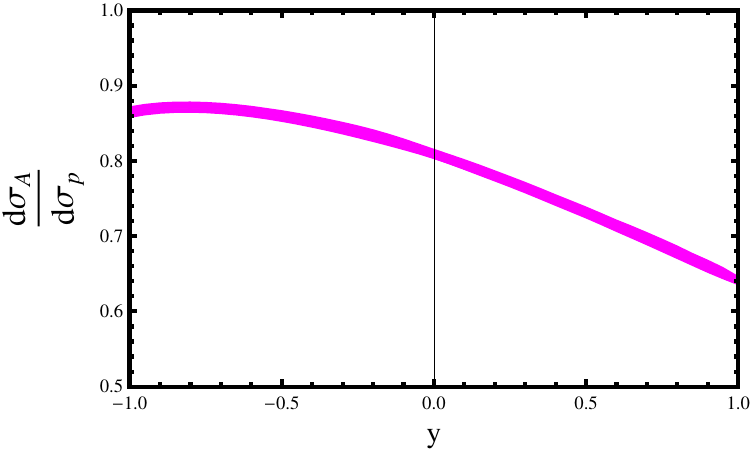}}
\subfigure [$\>$ Ca] { \label{fig:subfig1}\includegraphics[scale=0.5]{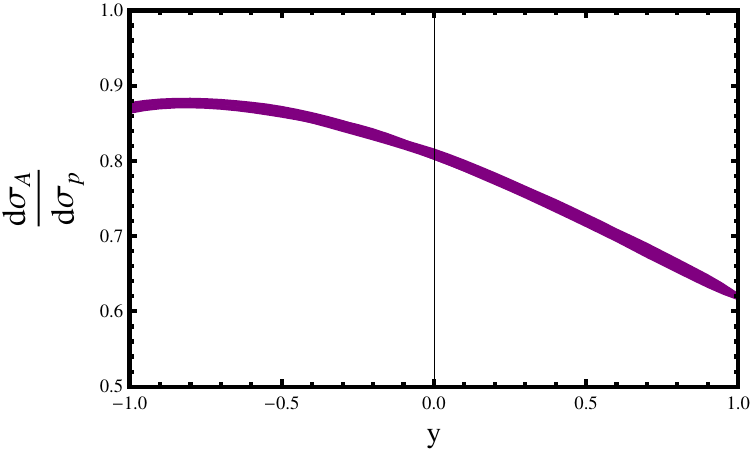}}
\subfigure [$\>$ Fe] { \label{fig:subfig1}\includegraphics[scale=0.5]{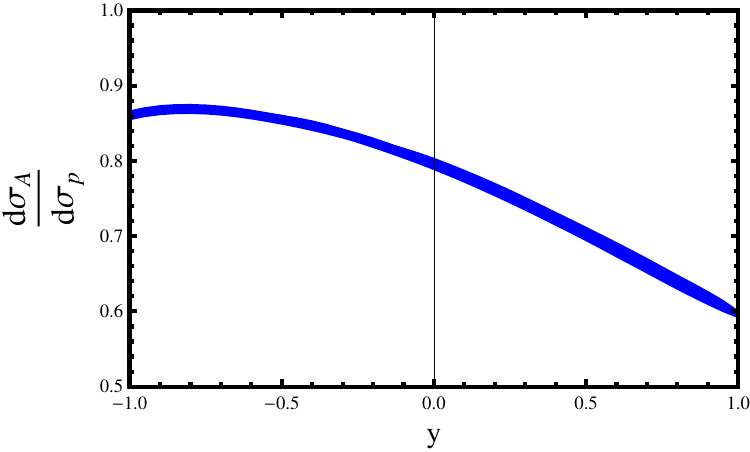}}
\subfigure [$\>$ Au] { \label{fig:subfig1}\includegraphics[scale=0.5]{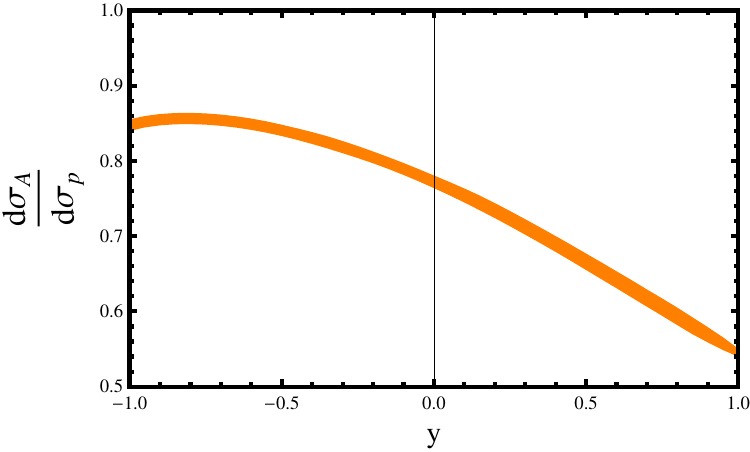}}
\subfigure [$\>$  Ur] { \label{fig:subfig1}\includegraphics[scale=0.5]{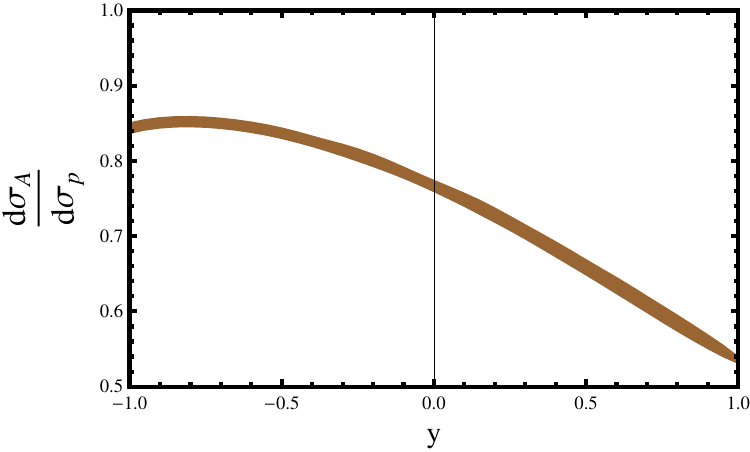}}
\subfigure [$\>$ C and Ur ] { \label{fig:subfig1}\includegraphics[scale=0.5]{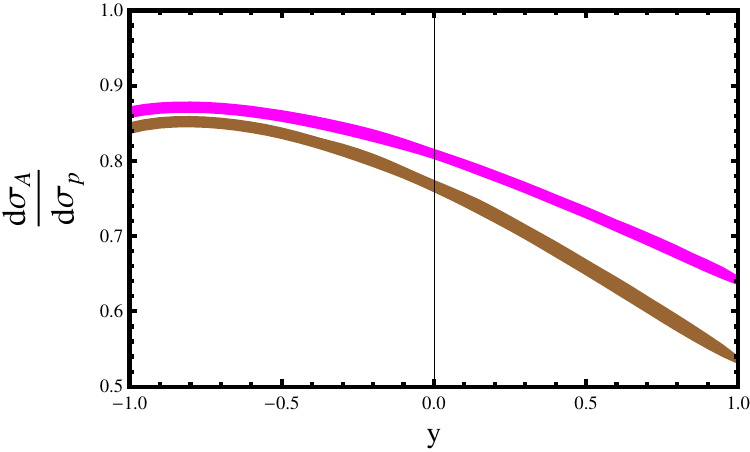}}
\caption{We show ratio $R_A=d\sigma_A/d\sigma_p$ in the rapidity $(y)$ distributions for various nuclear targets compared to the case of the proton target.  For easier visual comparison, we show the results for Carbon (C) and Uranium (Ur) together in subfigure $(\text{f})$. These results include resummation at the NNLL level of accuracy and are calculated at $Q_e=90$ GeV, $P_{J_T}=20$ GeV, and $\tau_1=1.5$ GeV, corresponding to the EMC region of the nuclear PDFs.}
\label{rapidity-ratio}
\end{figure}

Fig.~\ref{protontau1} shows the behavior of the cross-section as one implements  jet veto by restricting radiation at wide angles from the final-state jet and nuclear beam directions. As $\tau_1$ gets smaller, the final-state jet becomes narrower and wide-angle radiation becomes softer. The enhancement of the cross-section at small $\tau_1$ is a consequence of the cross-section being dominated by collinear emissions along the nuclear beam and final-state jet directions and soft emissions everywhere. In fact, the behavior of the cross-section in fixed order perturbation theory is singular in the limit $\tau_1 \to 0$. It is the resummation of the jet-veto Sudakov logarithms $\alpha_s ^n \ln^{2n} (\tau_1/P_{J_T})$ that tames the behavior of the cross-section at small $\tau_1$. 

Hard remissions between the nuclear beam and final state jet directions are allowed for larger $\tau_1 \sim P_{J_T}$. These emissions are perturbartively suppressed and are not accompanied by large Sudakov logarithms. This part of the spectrum can be described by fixed order perturbation theory. A matching calculation is required to smoothly connect the resummation region $\tau_1\ll P_{J_T}$ with the fixed-order perturbation theory region $\tau_1 \sim P_{J_T}$. We leave such a matching calculation for future work as the focus of this paper is on the resummation region. For this reason, Fig.~\ref{protontau1} is restricted to the region of small $\tau_1$. We have also not have shown the region $\tau_1<1$ GeV, since in this region the soft function ${\cal S}$ in Eq.(\ref{factorization-EPS09}), evaluated at the soft scale $\mu_S \sim \tau_1$, is affected by non-perturbative effects. We give numerical results for this non-perturbative soft region in section \ref{NPsoft-2}. 
\begin{figure}
\subfigure [$\>$ Proton ] { \label{fig:subfig1}\includegraphics[scale=0.5]{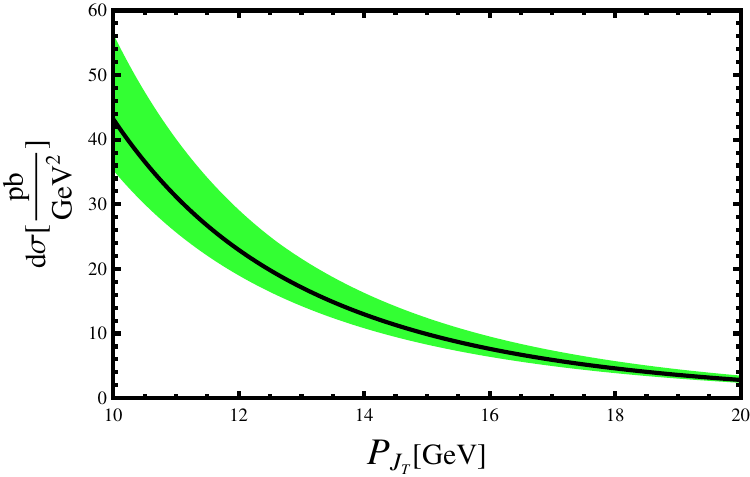}}
\subfigure [$\>$ Proton and C ] { \label{fig:subfig1}\includegraphics[scale=0.5]{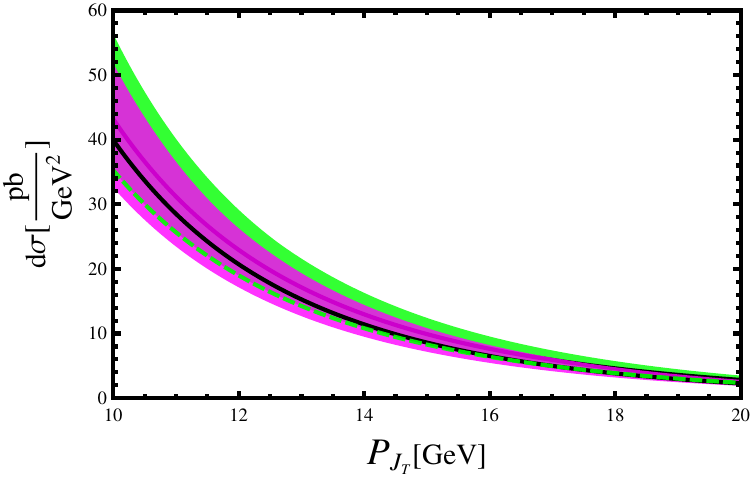}}
\subfigure [$\>$ Proton and Ca ] { \label{fig:subfig1}\includegraphics[scale=0.5]{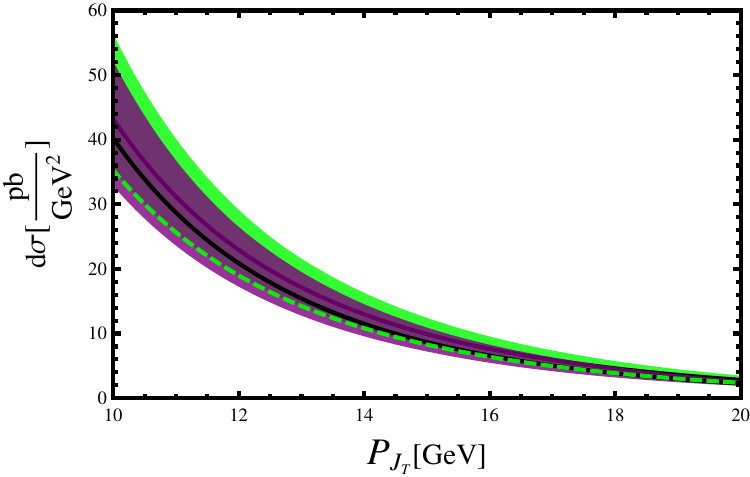}}
\subfigure [$\>$ Proton and Fe ] { \label{fig:subfig1}\includegraphics[scale=0.5]{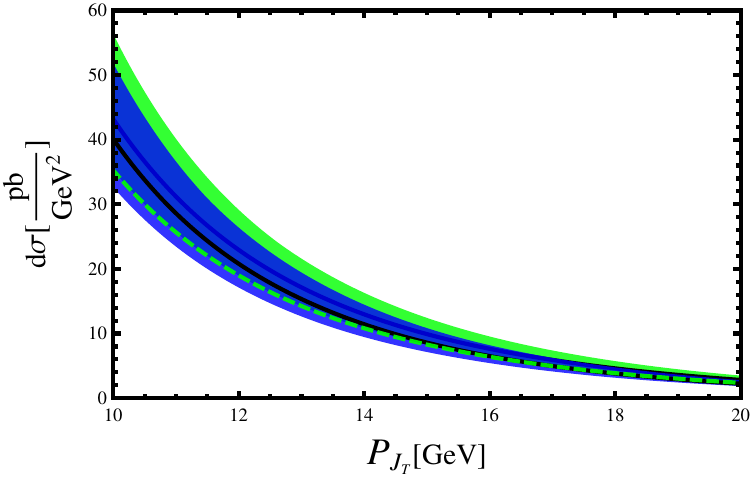}}
\subfigure [$\>$ Proton and Au ] { \label{fig:subfig1}\includegraphics[scale=0.5]{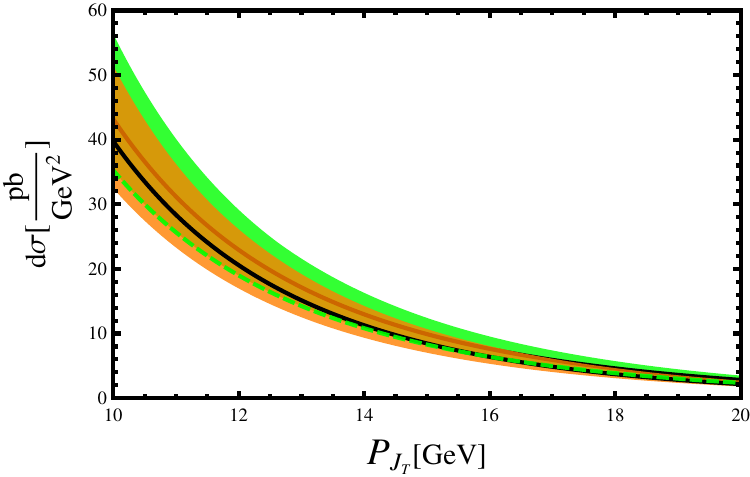}}
\subfigure [$\>$ Proton and Au ] { \label{fig:subfig1}\includegraphics[scale=0.5]{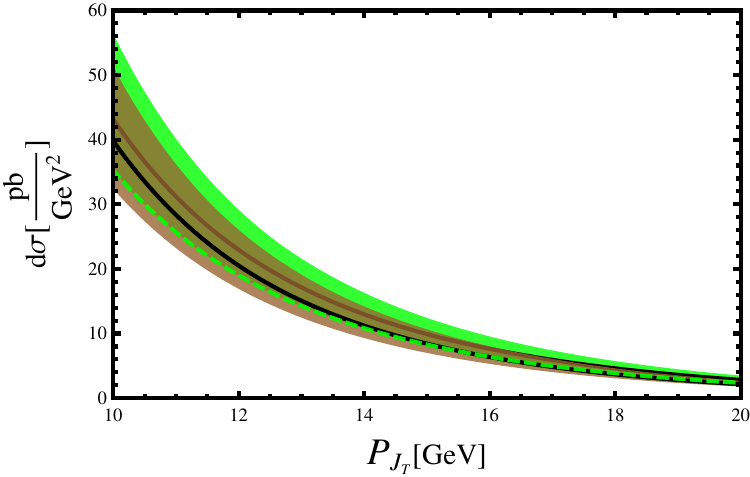}}
\caption{Jet transverse momentum ($P_{J_T}$) distributions with NNLL resummation for different nuclear targets for $Q_e=90$ GeV, 
 $y=0$ GeV, and $\tau_1=1.5$ GeV. In all figures, the green (upper) band corresponds to the NNLL resumed result for a proton target. The lower bands in different colors are the corresponding distributions for different nuclear targets.}
 \label{PJT}
\end{figure}
\begin{figure}
\subfigure [$\>$ C ] { \label{fig:subfig1}\includegraphics[scale=0.5]{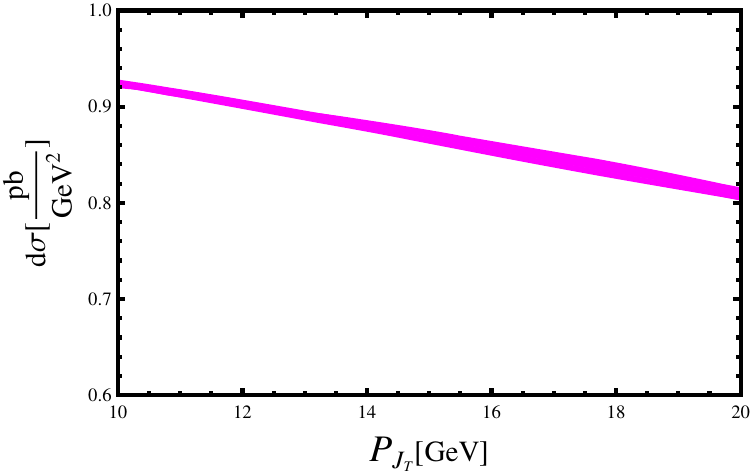}}
\subfigure [$\>$ Ca ] { \label{fig:subfig1}\includegraphics[scale=0.5]{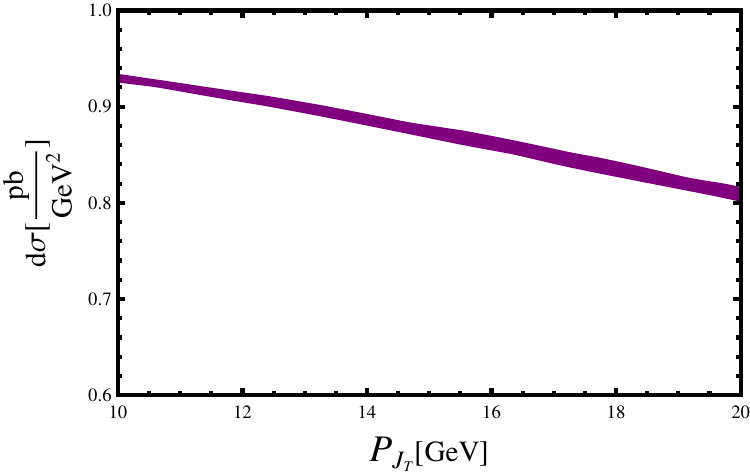}}
\subfigure [$\>$ Fe ] { \label{fig:subfig1}\includegraphics[scale=0.5]{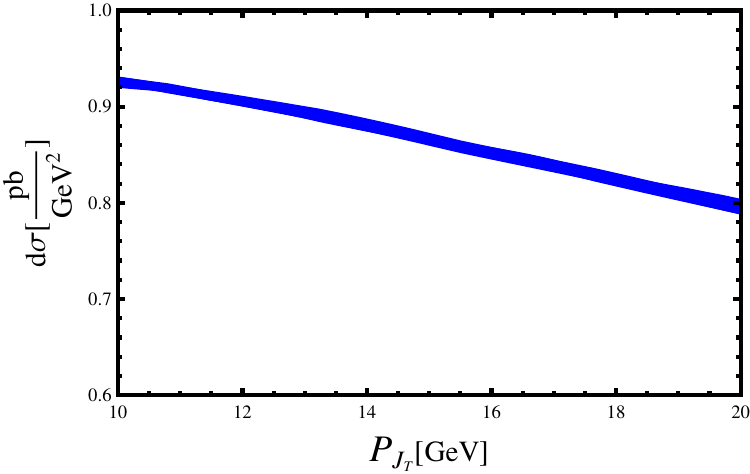}}
\subfigure [$\>$ Au ] { \label{fig:subfig1}\includegraphics[scale=0.5]{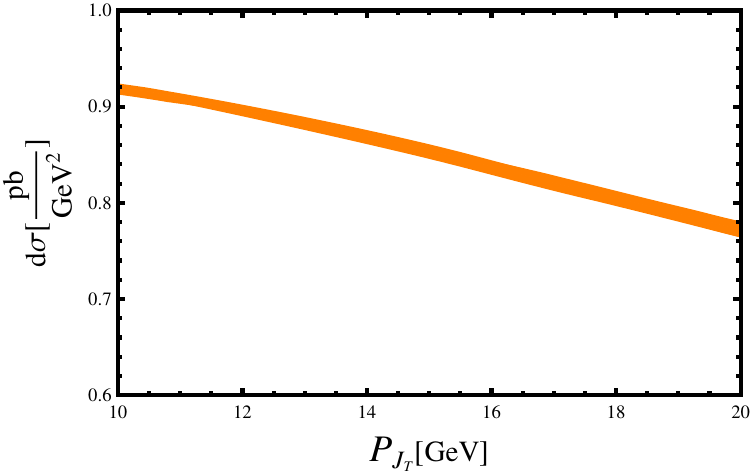}}
\subfigure [$\>$ Ur ] { \label{fig:subfig1}\includegraphics[scale=0.5]{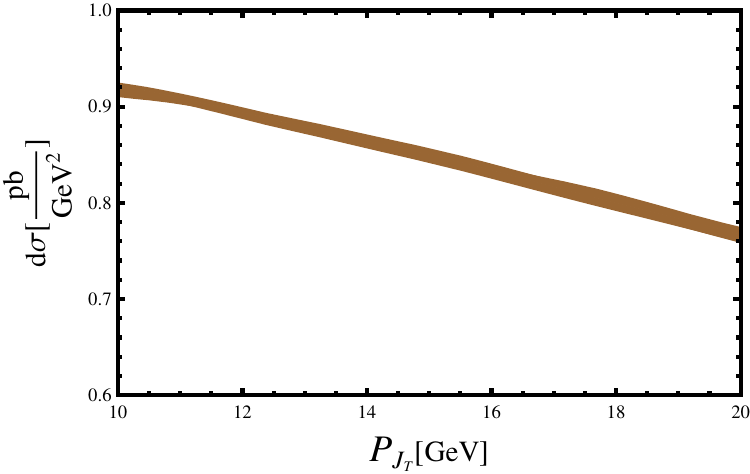}}
\subfigure [$\>$ C and Ur ] { \label{fig:subfig1}\includegraphics[scale=0.5]{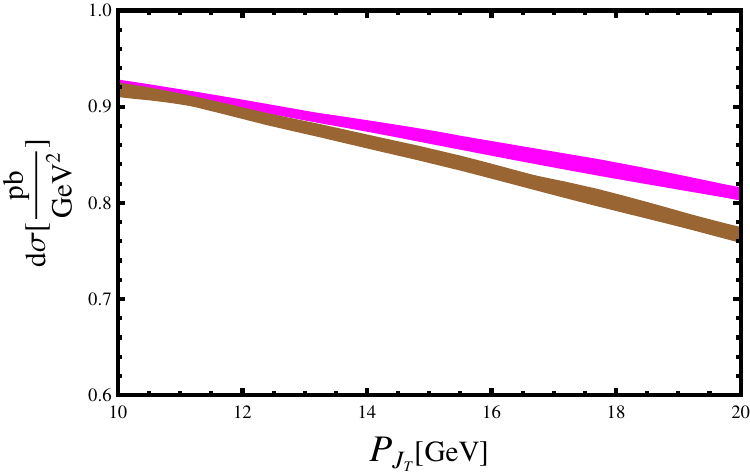}}
\caption{We show the ratio $R_A=d\sigma_A/d\sigma_p$ in the $P_{J_T}$ distributions for different nuclear targets relative to the proton for $Q_e=90$ GeV, $\tau_1=1.5$ GeV, and $y=0$.}
\label{PJT-ratio}
\end{figure}

In Fig.~\ref{tau1nuc}, we show the $\tau_1$ distributions with resummation at the NNLL level of accuracy for a variety of nuclear targets. In all plots, the green (upper) band corresponds to the $\tau_1$-distribution for a proton target and the lower bands in various colors correspond to distributions for heavier nuclear targets.  In Fig.~\ref{tau1rationuc}, we show the ratio $R_A$ of Eq.(\ref{ratio}) as a function of  $\tau_1$  for various nuclei at $Q_e=90$ GeV, $P_{J_T}=20$ GeV, and $y=0$.
The scale variation bands in Fig.~\ref{tau1rationuc} are obtained by computing the ratio $R_A$ using the same scale choices in $d\sigma_A$ and $d\sigma_p$ and then finding the envelope of the scale variations in Eq.(\ref{scalevar}). This procedure corresponds to the fact that the scales $\mu_H,\mu_B,\mu_J,\mu_S$, with typical scalings in Eq.(\ref{scales}), are determined by dynamics that are independent of the nuclear target.  The nuclear dependence only arises through the structure of the nuclear PDF which is evaluated at the beam scale $\mu_B$. As expected,  the scale variation uncertainty is dramatically reduced in the ratio as seen by comparing Figs.~\ref{tau1nuc} and \ref{tau1rationuc}. 

From Figs.~\ref{tau1nuc} and \ref{tau1rationuc}, we see that the cross-sections for heavier nuclei are generally suppressed relative to the proton. This can be understood by noting that for $Q_e=90$ GeV, $P_{J_T}=20$ GeV, and $y=0$, Eq.(\ref{xstar}) gives  $x_*\simeq 0.3$ for  the  lower limit of integration over Bjorken-$x$ in Eq.(\ref{factorization-EPS09}). From Figs.~\ref{Rfactors} and \ref{lumratio}, we see that this corresponds to probing the nuclear PDFs in the EMC region. In this region the parton density  in a proton bound inside a nucleus is suppressed compared to that of a free proton. As seen in Fig.~\ref{tau1rationuc}, the EMC effects are larger than the perturbative uncertainty quantified by the scale variation procedure. Thus, the ratio $R_A$ as a function of $\tau_1$ can be a sensitive probe of  such EMC effects.

In Fig.~\ref{rapidity} we show the rapidity distributions for various nuclear targets with NNLL resummation at $Q_e=90$ GeV, $P_{J_T}=20$ GeV, and $\tau_1 = 1.5$ GeV. Once again, in all figures the green (upper) band corresponds to the rapidity distribution for a proton target and the lower bands in various colors correspond to heavier nuclear targets. Here also we see the characteristic suppression for heavier nuclei compared to the proton target. This is shown more quantitatively in Fig.~\ref{rapidity-ratio}, where we show the ratio $R_A$ in Eq.(\ref{ratio}) as a function of rapidity for various  nuclei at $Q_e=90$ GeV, $P_{J_T}=20$ GeV, and $\tau_1=1.5$ GeV. The scale variation uncertainty is given by the width of the curves and once again we see a dramatic reduction of the perturbative uncertainty in the ratio $R_A$. The size of the suppression in the jet rapidity distributions for heavier nuclei, provides another measure of nuclear effects. As seen in Fig.~\ref{rapidity-ratio}, the deviation of $R_A$ from unity gets larger for increasing jet rapidity ($y$). This can be understood by noting that the value of $x_*$, as determined by Eq.(\ref{xstar}), increases with  the jet rapidity $y$. For the kinematics chosen, at $y=0$ we have $x_*\simeq 0.3$ and for larger values of $y$ we have correspondingly $x_*> 0.3$. From Figs.~\ref{Rfactors} and \ref{lumratio} we see that for increasing $y$, we are sensitive to the nuclear PDFs deeper into the EMC region. On the other hand, for more negative values of the jet rapidity $y$, we start becoming sensitive to the anti-shadowing region. As seen in Fig.~\ref{lumratio}, in the anti-shadowing region, the parton luminosity function of a bound proton is more similar to that of a free proton. Thus, as one goes to more negative values of the $y$, one is sensitive to both the anti-shadowing and the EMC regions so that the net effect is a smaller suppression.  The jet rapidity range $y\in [-1,1]$, covered in Figs.~\ref{rapidity} and \ref{rapidity-ratio}, corresponds to the range  $x_*\in [0.2,0.7]$. The overall effect  can be summarized by a decreasing $R_A$ for increasing $y$, as seen in Fig.~\ref{rapidity-ratio}. Note that this is in contrast to the $\tau_1$ distributions in Fig.~\ref{tau1rationuc} where $R_A$ is relatively flat as one varies $\tau_1$. This can be understood by noting from Eq.(\ref{xstar}), that the value of $x_*$ is independent of $\tau_1$, so that we are probing the same regions in the nuclear PDFs for different values of $\tau_1$. There is however a small indirect dependence on $\tau_1$ through the convolution structure in Eq.(\ref{factorization-EPS09}) which can affect the weighting of the different regions in Bjorken-$x$.

\begin{figure}
\subfigure [$\> Q_e=90\> $GeV ] { \label{fig:subfig1}\includegraphics[scale=0.5]{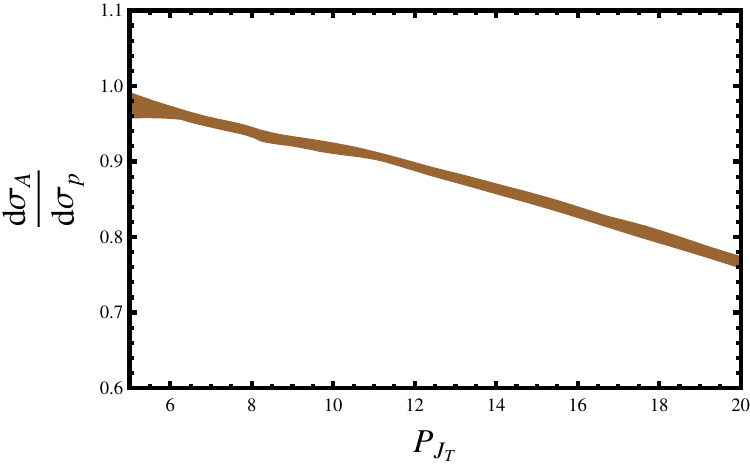}}
\subfigure [$\> Q_e=120 \>$GeV ] { \label{fig:subfig1}\includegraphics[scale=0.5]{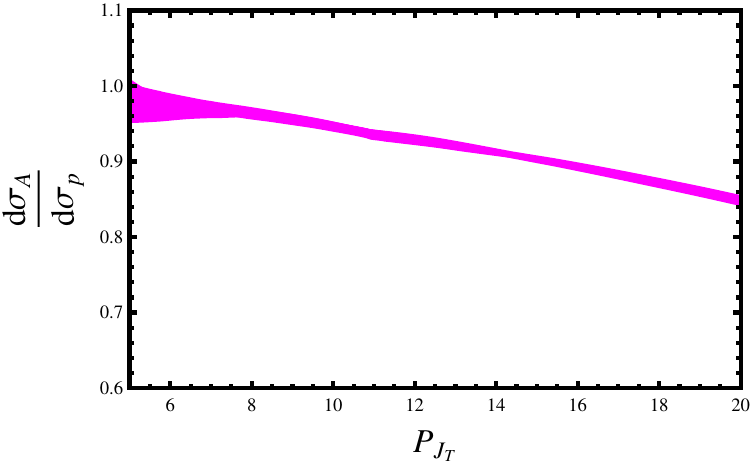}}
\subfigure [$\> Q_e=140 \>$GeV ] { \label{fig:subfig1}\includegraphics[scale=0.5]{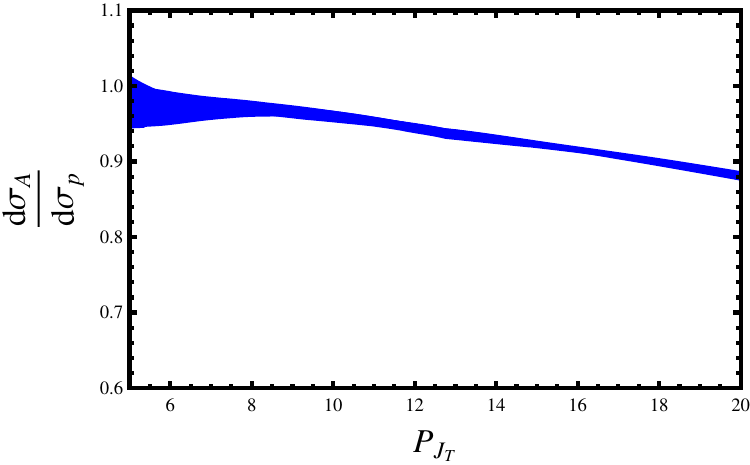}}
\subfigure [$\> Q_e=300\>$ GeV ] { \label{fig:subfig1}\includegraphics[scale=0.5]{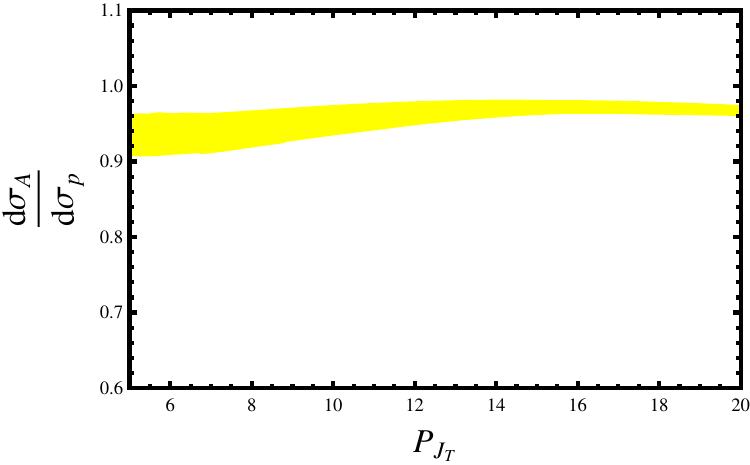}}
\subfigure [$\> Q_e=800\>$ GeV ] { \label{fig:subfig1}\includegraphics[scale=0.5]{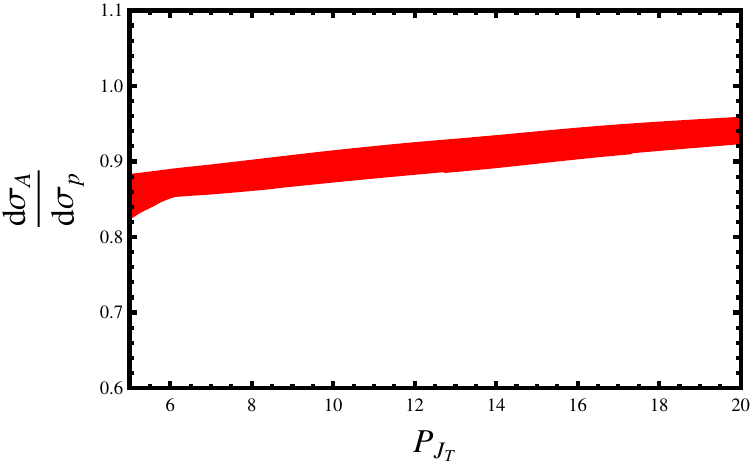}}
\subfigure [$\> Q_e=90,120,140, 300,800 \>$ GeV ] { \label{fig:subfig1}\includegraphics[scale=0.5]{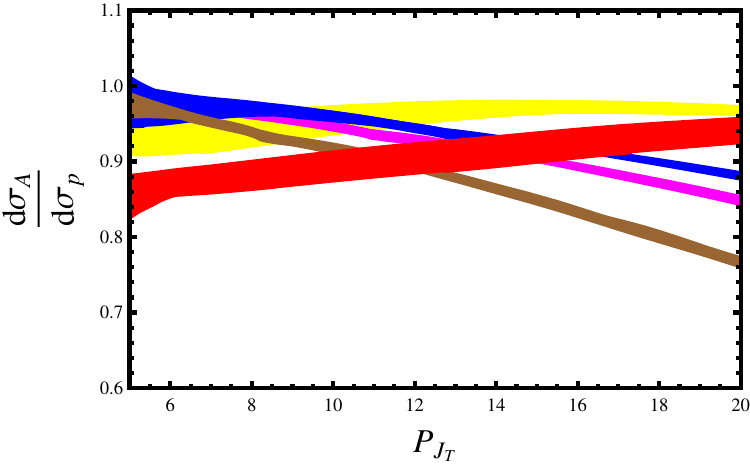}}
\caption{We show the ratio of $P_{J_T}$ distributions for Uranium relative to the proton for $Q_e=\{90, 120, 140, 300, 800\}$ GeV, $\tau_1=1.5$ GeV, and $y=0$. The different choices of $Q_e$ probe different ranges in Bjorken-$x$, as seen from  Eqs.(\ref{factorization-EPS09}) and (\ref{xstar}), yielding the different sizes and shapes for $R_A=d\sigma_A/d\sigma_p$.  }
\label{Qeratio}
\end{figure}

In Fig.~\ref{PJT} we show the $P_{J_T}$ distributions for various nuclei at $Q_e=90$ GeV, $y=0$, and $\tau_1=1.5$ GeV. Here also we see that the cross-section is suppressed for heavier nuclei compared to the proton due to the EMC effects that suppress the parton density in nucleons that are bound inside the nucleus.  In Fig.~\ref{PJT-ratio}, we show the ratio of the $P_{J_T}$ distributions of heavier nuclei to the that of the proton. The scale variation is again dramatically reduced in the ratio and is given by the width of the curves. We see that the relative difference in the cross-sections for heavier nuclei and the proton grows with increasing $P_{J_T}$. This is again a consequence of Eq.(\ref{xstar}) which shows that the value of $x_*$ grows with $P_{J_T}$. For $P_{J_T}$ in the range $[10\> \text{GeV}, 20 \>\text{GeV}]$,  $x_*$ takes on values in the range $\sim [0.1,0.3]$ respectively. From Fig.~\ref{lumratio} we see that for $P_{J_T}=10$ GeV we are closer to the anti-shadowing region where the parton luminosity for nucleons in heavier nuclei is similar to that of a free proton. For $P_{J_T}=20$ GeV, we are well into the EMC region where there is a significant suppression in the parton luminosity in heavier nuclei. As a result, we see the characteristic shape of $R_A$ as a function of $P_{J_T}$ which indicates an increased suppression for increasing $P_{J_T}$.

In the numerical results presented so far, the kinematic configurations chosen were sensitive to the anti-shadowing and EMC regions in Figs.~\ref{Rfactors} and \ref{lumratio}. One can also probe lower regions in Bjorken-$x$, such as the shadowing region, by choosing the appropriate kinematics. For illustration, in Fig.~\ref{Qeratio} we show the ratio $R_A$ as a function of $P_{J_T}$ for a Uranium target at $y=0$ and $\tau_1=1.5$ GeV for the five different values  $Q_e=90,120,140,300,800$ GeV. As seen from Eq.(\ref{xstar}), by increasing $Q_e$, one can probe lower values of $x_*$. For example, at $Q_e=300$ GeV and $P_{J_T}=5$ GeV we have $x_* \simeq 0.02$ which is in the shadowing region as seen in Figs.~\ref{Rfactors} and \ref{lumratio}. Thus, for this kinematic choice, the integration over Bjorken-$x$ in Eq.(\ref{factorization-EPS09}) covers the shadowing, anti-shadowing, and EMC regions. For $Q_e=800$ GeV, corresponding to LHeC kinematics, $x_* \simeq 0.006$ for $P_{J_T}=5$ GeV and $x_*\simeq 0.025$ for $P_{J_T}=20$ GeV so that  might start to probe small-$x$ saturation physics (see Ref.\cite{Albacete:2013tpa} for a recent review). In this case, large-$x$ physics can be isolated by going to much larger values of $P_{J_T}$.  Thus, the size and shape of the ratio $R_A$ as a function of $P_{J_T}$ and $Q_e$ can be a useful way to probe nuclear PDFs in different regions of Bjorken-$x$. Similar results can be obtained for distributions in the jet rapidity $y$ and $\tau_1$ as a function of $Q_e$.

The numerical results in Figs. \ref{tau1nuc} through \ref{Qeratio}, demonstrate that distributions in $\tau_1,P_{J_T},$ and $y$ for various nuclei and different values of $Q_e$, can be a powerful probe of nuclear PDFs, complementary to measurements of structure functions in inclusive deep inelastic scattering. Thus, a systematic program that measures distributions of various nuclei in the configuration space of $\{Q_e,\tau_1,P_{J_T},y\}$ can yield detailed information about nuclear structure.  

As discussed in section \ref{powercorrections}, these distributions will also be affected by power corrections. The scaling of these power corrections with the kinematic variables and their dependence on the nuclear targets was also discussed. In particular, the dominant nuclear-dependent power corrections have a kinematic scaling $\sim 1/(\tau_1 P_{J_T})$ rather than the typical scaling $\sim 1/Q^2$ (where $Q$ is the hard scale) in fully inclusive deep inelastic scattering. Using this information, deviations in the data from the leading twist predictions of Eq.(\ref{factorization-EPS09}) can be used as a probe of power corrections. In particular, the size of these deviations as a function of $\{A,Q_e,P_{J_T},y,\tau_1\}$ can provide detailed information on the behavior and size of the power corrections. Such a detailed study of power corrections is left as future work.

\subsection{Non-perturbative soft radiation effects}
\label{NPsoft-2}

\begin{figure}
\subfigure []  { \label{fig:subfig1}\includegraphics[scale=0.5]{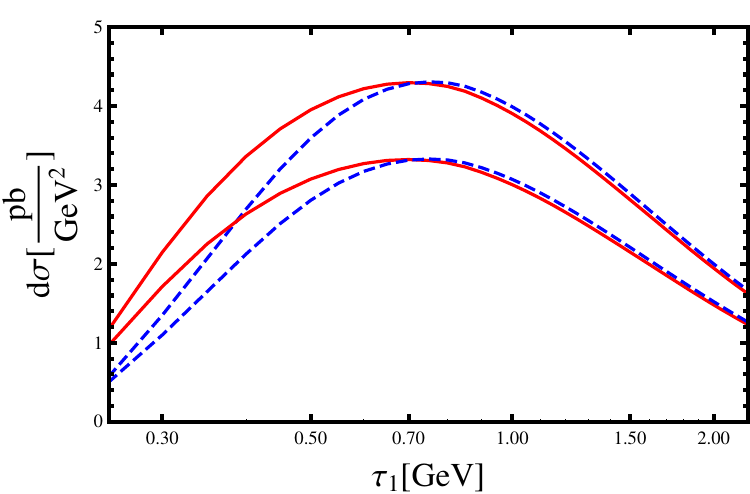}}
\subfigure[] { \label{fig:subfig1}\includegraphics[scale=0.5]{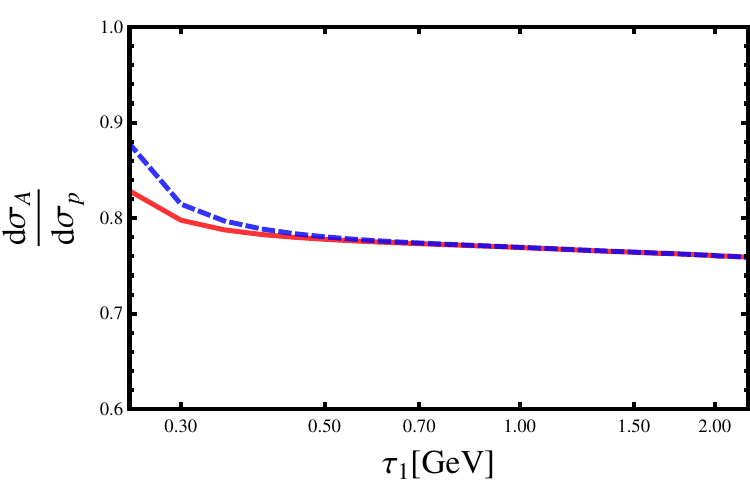}}
\caption{In sub-figure (a), we show the $\tau_1$-distributions for the proton and Uranium targets in the region $\tau_1 \sim \Lambda_{QCD}$.  The solid red (dashed blue) curves correspond to the soft function model I (II) in Eqs.(\ref{Fmodnum}) and (\ref{models}). The top (bottom) two curves are for the proton (Uranium) target. In sub-figure (b) we show the ratio $R_A=d\sigma_A/d\sigma_p$ as a function of $\tau_1$ for the Uranium target using the  soft function model I (II) as denoted by the solid red (dashed blue) curves. This plot shows that the model dependence of the soft function (seen in sub-figure (a)) largely cancels out in the ration $R_A$ since the solid red (dashed blue) curves, corresponding to models I (II) respectively, largely overlap. The plots are for the kinematic configuration $Q_e=90$ GeV, $P_{J_T}=20$ GeV, and $y=0$.}
\label{NP}
\end{figure}

In the numerical results presented so far, we have restricted to the region $\tau_1 > 1$ GeV so that the soft function ${\cal S}$ in Eq.(\ref{factorization-EPS09}) remains perturbatively calculable.  The soft function ${\cal S}$, which describes the dynamics of soft radiation with energy $E\sim \tau_1$, naturally lives at the scale $\mu_S \sim \tau_1$. Thus,  in the region $\tau_1\sim \Lambda_{QCD}$ the soft function becomes non-perturbative. As discussed in section \ref{npsoft}, for phenomenological purposes we implement a model for the soft function, as shown in Eqs.(\ref{softmod}) and (\ref{smodnorm}), as a convolution between the perturbative soft function ${\cal S}_{\text{part}.}$ and a model function function $S_{\text{mod}.}$. As explained in section \ref{npsoft}, such a parameterization  has the property that for $\tau_1\gg \Lambda_{QCD}$ the soft function model reduces to the perturbative result ${\cal S}_{\text{part}.}$ as desired.   From Eqs.(\ref{softmod}) through (\ref{fac-pos-4}), the soft function model can be parameterized by the function $F_{\text{mod}.}$  which is related to ${\cal S}_{\text{mod}.}$ as in Eq.(\ref{fmoduy}). For the purposes of generating numerical results, we employ the parameterization 
\bea
\label{Fmodnum}
F_{\text{mod}.}(u) &=& \frac{N(a,b,\Lambda)}{\Lambda} \Big( \frac{u}{\Lambda} \Big )^{a-1}\text{Exp}\Bigg [-\frac{(u-b)^2}{\Lambda^2} \Bigg ],
\eea
where the values of the parameters $a,b,\Lambda$ determine the model. The normalization $N(a,b,\Lambda)$ is chosen to satisfy the condition
\bea
\int_0^\infty du\> F_{\text{mod}.}(u) &=& 1,
\eea
which is equivalent to the normalization condition in Eq.(\ref{smodnorm}). The parameters $a,b,\Lambda$ are chosen so that $F_{\text{mod.}}$ peaks in the region $u\sim \Lambda_{QCD}$, which ensures that the soft function reduces to the perturbtive result for $\tau_1\gg \Lambda_{QCD}$ up to power corrections in $\Lambda_{QCD}/\tau_1$, as explained in the discussion around Eq.(\ref{softmod-ope}). 

For the soft scale $\mu_S$ appearing in ${\cal S}_{\text{part}.}$ in the soft function model of Eq.(\ref{softmod}), we make the choice
\bea
\mu_S = \tau_1 \sqrt{1 + \left (\frac{\tau_1^{\text{min}}}{\tau_1}\right )^2},
\eea 
with $\tau_1^{\text{min}}=1$ GeV. This choice has the property that in the limit $\tau_1\to 0$ the soft scale in ${\cal S}_{\text{part.}}$ remains perturbative $\mu_S \to 1$ GeV. For $\tau_1\gg \tau_1^{\text{min}}$, the soft scale reduces to $\mu_S \sim \tau_1$.

In Fig.~\ref{NP} (a), we show the $\tau_1$-distributions for a proton and Uranium target in the region that includes $\tau_1\sim \Lambda_{QCD}$ at $Q_e=90$ GeV, $P_{J_T}=20$ GeV, and $y=0$. Two curves are shown for the proton (top two curves) and Uranium (bottom two curves) targets. The two curves for each target correspond to using the two sets of model parameters
\bea
\label{models}
\text{Model I}:&& a=2.0, b=-0.2, \Lambda =0.2 \>\text{GeV}, \nn \\
\text{Model II}: && a=1.2, b=-0.1, \Lambda =0.3 \>\text{GeV}, 
\eea
where Model I and Model II correspond to the solid-red and blue-dashed curves respectively in Fig.~\ref{NP} (a). We see that for small values of $\tau_1\sim \Lambda_{QCD}$, there is a significant difference in the distributions. However, in the  region $\tau_1> 1$ GeV, the curves for models I and II converge to the perturbative result as expected. 

Since the soft function is universal and independent of the nuclear target, the model parameters $a,b,\Lambda$ can be extracted from measurements of the $\tau_1$ distributions in the region $\tau_1 < 1 $ GeV using a proton target. Similarly, one can also measure distributions in $y$ and $P_{J_T}$ in the region $\tau_1 \sim \Lambda_{QCD}$ in order to extract the soft function parameters. Once the parameters are extracted from data on the proton target, the soft function model can be used as a known input for the heavier nuclei.

In Fig.~\ref{NP} (b), we show the ratio $R_A$ for the Uranium target as a function of $\tau_1$. Again, the solid-red and blue-dashed curves correspond to using models I and II in Eq.(\ref{models}) respectively. We see that in the ratio $R_A$, the model dependence is greatly reduced as seen by large overlap of the two curves corresponding to the two different model soft functions. Thus, in addition to the reduction in the perturbative uncertainties, there is also a reduction in the uncertainty associated with the non-perturbative soft function, when considering the ratio $R_A$.

\section{Conclusions}
\label{conclusions}
In this paper, we studied electron-nucleus collisions with one final state jet $e^- + N_A\to J +X$, as a probe of nuclear structure and dynamics. We used a factorization framework to calculate the cross-section differential in 1-jettiness ($\tau_1$) and the transverse momentum ($P_{J_T}$) and rapidity ($y$) of the jet. The 1-jettiness variable $\tau_1$ is a global event shape that quantitatively   characterizes the degree to which the final state looks 1-jet-like and provides enhanced sensitivity to radiation at wide angles from the jet and nuclear beam directions. A veto on additional jets is imposed by restricting to the region $\tau_1 \ll P_{J_T}$, allowing only soft radiation ($E\sim \tau_1$) between the beam and jet directions. This phase space restriction induces Sudakov jet-veto logarithms $\sim \alpha_s^n\ln ^{2n} (\tau_1/P_{J_T})$ that can spoil the convergence of perturbation theory and requires resummation. Implementing the jet-veto and performing the resummation using the 1-jettiness global event shape, allows for better theoretical control compared to methods that depend on the details of a jet algorithm. This  allows one to perform analytic calculations at higher orders in perturbation theory and resummation, allowing for improved accuracy.

Distributions in $\tau_1$ provide a quantitative measure of the hadronic activity or the pattern of radiation between the beam and jet directions. By studying this distribution for a range of nuclear targets and at different kinematics, one can probe various aspects of nuclear physics. At leading twist, the factorization formula takes the schematic form in Eqs.(\ref{schem-1}) and (\ref{schem-2}) and directly probes the nuclear PDFs. This can be understood by noting that in the region $\Lambda_{QCD}\ll \tau_1 \ll P_{J_T}$, the leading-twist cross-section is given entirely in terms of perturbatively calculable universal functions (independent of nuclear target) and the nuclear PDFs. Thus, comparing distributions in $\{\tau_1,P_{J_T},y\}$ for a range of nuclear targets and center of mass energies allows for a systematic study of the nuclear PDFs. 

Power corrections beyond leading twist will probe dynamical nuclear effects such as higher twist correlations and nuclear modification effects such as jet quenching and energy loss mechanisms of fast-moving partons through cold nuclear matter. We gave a discussion of the various sources of power corrections and the ratios of energy scales that determine their sizes based on the power counting. The effective field theory framework allows one to systematically derive operator definitions of the power corrections to the leading twist factorization formula. We leave such a detailed study of power corrections for future work. From a phenomenological point of view, the size and shape of the various power corrections can be investigated by looking at the size of deviations between data and the leading twist prediction for a range of nuclear targets and kinematics. One typically expects that the nuclear-medium-induced power corrections will have a much larger effect for heavier nuclei and one might correspondingly expect larger deviations from the leading twist predictions for heavier nuclei. 

As a first step, in this paper we give numerical results at leading twist with resummation at the next-to-next-leading (NNLL) logarithmic order for distributions in $\{\tau_1,P_{J_T}, y\}$ for the nuclear targets: Carbon, Calcium, Iron, Gold, and Uranium. We also give results for the ratio of these distributions between heavy nuclei and the proton. We find that there is a dramatic reduction in the scale variation uncertainty, as expected, when considering such ratios of distributions.  In the region $\tau_1\sim \Lambda_{QCD}$, the soft function in the leading twist formula becomes non-perturbative since the energy of the soft radiation has the scaling $E \sim \tau_1$. In this case, we employed a model for the soft function such that it has the correct renormalization group properties and reduces to the perturbative result for $\tau_1\gg \Lambda_{QCD}$. Furthermore, this soft function is independent of the nuclear target. This universality can be exploited to extract the non-perturbative soft function from data on the proton target and then used for the case of heavier nuclei. We also showed that in the ratio of distributions between a heavy nucleus and the proton, the dependence on the parameters of the non-perturbative soft function model largely cancel.

Our leading twist numerical results indicate that distributions in $\{\tau_1,P_{J_T},y\}$ are quite sensitive to differences in the nuclear PDFs. By choosing appropriate kinematics one can probe various regions in Bjorken-$x$ of the nuclear PDFs. This allows one to conduct  studies of nuclear phenomena such as shadowing, anti-shadowing, and the EMC effect. 

We conclude by noting that this is just a first step in using  event shapes for exclusive jet production as a probe of nuclear dynamics. There are many further directions to pursue, including  constructing new observables that are variants of the one we studied in order to probe nuclear dynamics in different ways. For example, one can be separately differential in the contributions to 1-jettiness from the beam and jet regions, employ a standard jet shape analysis while still retaining information about wide angle soft radiation, construct analogous observables in different reference frames, study exclusive multi-jet production, and perform detailed studies of the various nuclear-medium-induced power corrections. One might also consider extending such methods  to studies of p-A and A-A collisions. We look forward to such further developments which can be  part of the broad program of physics envisioned by the EIC and LHeC proposals for a future electron-ion collider.

\acknowledgements

We thank Alberto Accardi and Frank Petriello for useful discussions and comments. This work was supported in part by  the U.S. Department of Energy under contract numbers ~ DE-AC02-05CH11231 (ZK), DE-AC02-98CH10886 (JQ), DE-AC02-06CH11357 (XL) and the grants  DE-FG02-95ER40896 (XL) and DE-FG02-08ER4153 (XL), and the U.S. National Science Foundation under grant NSF-PHY-0705682 (SM).  

\appendix

\section{Field-theoretic definitions}
\label{QFT-def}

The electromagnetic quark current at the jet production vertex for flavor $q$ is given by
\bea
J^q_\mu (0) & =& \bar{\psi}_q\gamma_\mu \psi_q (0),
\eea
and is matched onto the an operator in the SCET as
\bea
J^q_\mu (0)  & =& \int d\omega_A \int d\omega_J \>C(\omega_A \omega_J,\mu )\>\bar{\chi}_{q,\omega_J}Y_{n_J}^\dagger\gamma_\mu \>Y_{n_A}\chi_{q,\omega_A} (0),
\eea
where $C(\omega_A \omega_J,\mu )$ is the Wilson coefficient that contains the physics of the hard scale and is related to the hard function $H$ that appears in Ea.(\ref{factor-1}) as
\bea
H(\omega_A \omega_J,\mu) &=& | C(\omega_A \omega_J,\mu)|^2.
\eea
The variables $\omega_{A,J}$ are label momenta that denote the large light-cone momentum components of the collinear quark fields along the nuclear beam and jet directions respectively. The fields $\chi$ denote collinear quark fields $\xi$ dressed by collinear Wilson lines $W$ that sum up collinear emissions
\bea
\chi_{n} (x) &=& W^\dagger_{n} \chi_{n} (x) , \qquad W_{n} = \sum_{\text{perms.}}\> \text{Exp}\> \Big [-\frac{g }{\bn\cdot {\cal P}}\bn \cdot A_{n,q}^c (x)\Big ],
\eea
and the $Y$ denote soft Wilson lines that sum up eikonal soft emissions
\bea
Y_{n_A} (x) = \text{P} \> \text{Exp}\> \Big [ i g \int_{-\infty}^x ds\> n_A \cdot A_s (s \>n_A^\mu )\Big ], \qquad Y_{n_J}(x)= \overline{\text{P}} \> \text{Exp}\> \Big [ -i g \int^{\infty}_x ds\> n_J \cdot A_s (s\> n_J^\mu )\Big ]. \nn \\
\eea
For detailed explanations of the notations used above, we refer the reader to the original SCET papers in Refs.~\cite{Bauer:2000ew,Bauer:2000yr,Bauer:2001ct,Bauer:2001yt,Bauer:2002nz,Beneke:2002ph}.

The quark jet function $J^q$ in Eq.(\ref{factor-1}) is defined as
\bea
J^q(s_J=\omega_J r^+,\mu) &=& \frac{1}{4\pi N_c \omega_J} \>\text{Im}\> \Big [i\int d^4x \>e^{i r\cdot x} \langle 0 | T\{\bar{\chi}_{n,\omega_J}(0)\bnslash_J\chi_{n_J} (x)\}|0\rangle\Big ] \nn \\
\eea
and the beam function is defined as
\bea
\label{bf}
B^q(x,t,\mu) &=& \frac{1}{2x \bn_A\cdot p_A} \int \frac{db^-}{4\pi}\> e^{-i\frac{t b^-}{2x \bn_{A}\cdot p_{A}}} \sum_{\text{pols.}} \langle p_A|\bar{\chi}_{n_A}\delta (x\bn_A\cdot p_A - \bar{{\cal P}}^\dagger) (b^-) \frac{\bnslash_A}{2} \chi_{n_A} (0) | p_A\rangle. \nn \\
\eea
Finally, the soft function in Eq.(\ref{factor-1}) is defined in terms of the generalized hemisphere soft function ${\cal S}(k_a,k_J,\mu)$ through Eq.(\ref{soft-1}). The definition of ${\cal S}(k_a,k_J,\mu)$ is given by
\bea
\label{QFT-definition}
{\cal S}(k_a,k_J,\mu) &=& \frac{1}{N_c}\sum_{X_s} \>\text{Tr}\>\langle 0 | \bar{T} [Y_{n_A}^\dagger Y_{n_J} ] (0) \delta(k_a - \frac{q_A \cdot K_{X_s}^{(a)}}{Q_a})\delta(k_J - \frac{q_J\cdot K_{X_s}^{(J)}}{Q_J}) |X_s \rangle \nn \\
&\times& \langle X_s | T  [Y_{n_J}^\dagger Y_{n_A} ] (0) | 0 \rangle
\eea
where we have defined
\bea
\label{psoft}
K_{X_s}^{(J)} &=& \sum_{k\in X_s} p_k \>\theta (\frac{2q_A\cdot p_k}{Q_a} - \frac{2q_J\cdot p_k}{Q_J} ), \qquad K_{X_s}^{(a)} = \sum_{k\in X_s} p_k \>\theta (\frac{2q_J\cdot p_k}{Q_J} - \frac{2q_A\cdot p_k}{Q_a} ). \nn \\
\eea

\section{Fixed order NLO expressions}
\label{nlo}

In momentum space and at tree-level, the hard, jet, beam, and soft functions are given by
\bea
H^{(0)}(\xi^2, \mu_H) &=& 1, \nn \\
J^{q(0)}(s_J, \mu_J) &=& \delta(s_J), \nn \\
{\cal I}^{qi(0)}\left(\frac{x}{z}, t_a, \mu_J\right)&=& \delta_{qi}\delta(1-\frac{x}{z})\delta(t_a), \nn \\
{\cal S}_{\text{part}.}^{(0)} (k_a,k_J,\mu_S) &=& \delta(k_a) \delta(k_J).
\eea
The NLO expressions for the hard \cite{Manohar:2003vb,Bauer:2003di} and  jet \cite{Mannel:2000aj,Bosch:2004th} functions are given by 
\bea
H^{(1)}(\xi^2,\mu) &=&  \frac{\alpha_s C_F}{4\pi} \Big [ -2 \ln^2 \frac{\xi^2}{\mu^2} + 6 \ln \frac{\xi^2}{\mu^2} - 16 + \frac{\pi^2}{3}\Big ],\nn \\
J^{q(1)}(s,\mu) &=&  \frac{\alpha_s C_F}{4\pi}  \Bigg \{  \delta(s) \Big (  7-\pi^2 \Big ) -  \frac{3}{\mu^2} \Big [ \frac{\mu^2 \theta(s)}{s}\Big ]_+ + \frac{4}{\mu^2}\Big [\frac{\mu^2 \theta(s) \ln (s/\mu^2)}{s} \Big ]_+\Bigg \},\nn \\
\eea
and the beam function coefficients \cite{Stewart:2010qs} and the soft function \cite{Jouttenus:2011wh} are given by
\bea
{\cal I}_{n;qq}^{(1)}(x,t,\mu) &=& \frac{\alpha_s C_F}{2\pi}\Bigg \{\delta(t) \Bigg [-\frac{\pi^2}{6}\delta(1-x) -\frac{1+x^2}{1-x}\ln x +(1-x) \Bigg ] \nn \\
&+&\delta(t) \Big [\frac{\ln (1-x)}{1-x} \Big ]_+(1+x^2) + \frac{2}{\mu^2}\Big [ \frac{\ln (t/\mu^2)}{t/\mu^2}\Big ]_+\delta(1-x)\nn \\
&+&\frac{1}{\mu^2}\Big [ \frac{\mu^2}{t}\Big ]_+\frac{1+x^2}{(1-x)_+} \Bigg \},\nn \\
{\cal I}_{n;qg}^{(1)}(x,t,\mu) &=& \frac{\alpha_s T_F}{2\pi} \Bigg \{ \frac{1}{\mu^2} \Big [ \frac{\mu^2}{t} \Big ]_+ (1-2x +2x^2 ) \nn \\
&+& \delta(t)\Big [ (1-2x +2x^2 ) (\ln \frac{1-x}{x} -1)+ 1\Big ]  \Bigg \}, 
\eea

\bea
{\cal S}^{(1)}_{\text{part}.}(k_a,k_J,\mu) &=&  -\frac{\alpha_s C_F}{4\pi } \Bigg \{  \frac{8\>\delta(k_j)}{\tilde{\mu}}\Big [ \frac{\theta(k_a) \tilde{\mu} \ln k_a/\tilde{\mu}}{k_a}\Big ]_+ +  \frac{8\>\delta(k_a)}{\tilde{\mu}}\Big [ \frac{\theta(k_j) \tilde{\mu} \ln k_j/\tilde{\mu}}{k_j}\Big ]_+ \nn \\
&-&\frac{\pi^2}{3}\delta (k_a)  \delta(k_j)  \Bigg \}, 
\eea
where we have defined
\bea
\tilde{\mu} &=& \mu\sqrt{\hat{s}_{aJ}}, \qquad
\hat{s}_{aJ} = \frac{2 q_A\cdot q_J}{Q_a Q_J}.
\eea

In generating numerical results we  worked with the position-space version of the factorization formula as given in Eq.(\ref{fac-pos-4}). The  jet, beam, and soft functions in momentum space are given in terms of their position-space analogs as
\bea
J(s_J,\mu;\mu_J) &=& \int \frac{dy_J}{2\pi} \> e^{iy_Js_J} J(y_J,\mu;\mu_J),\nn \\
B^q_A(x,t,\mu) &=& \int \frac{dy}{2\pi}\> e^{ity} B^q_A(x,y,\mu),\nn \\
S(k_a,k_J,\mu;\mu_S) &=& \int \frac{dy_{k_a} dy_{k_J}}{4\pi^2} \> e^{iy_{k_a}k_a+iy_{k_J}k_J} S(y_{k_a},y_{k_J},\mu;\mu_S).
\eea
At  tree-level the position-space  jet, beam, and soft functions are given by
\bea
J^{q(0)}(y_J, \mu_J) &=& 1, \nn \\
{\cal I}^{qi(0)}\left(\frac{x}{z}, y_a, \mu_J\right)&=& \delta_{qi}\delta(1-\frac{x}{z}), \nn \\
{\cal S}_{\text{part}.}^{(0)} (y_a,y_J,\mu_S) &=& 1.
\eea
At NLO, the corresponding expressions are 
\bea
J^{q(1)}(y_J, \mu)
&=& \frac{\alpha_s C_F}{4\pi}\Big [  7-\frac{2\pi^2}{3}  +3 \ln(iy_J\mu^2e^{\gamma_E})+ 2\ln^2(i y_J \mu^2e^{\gamma_E}) \Big ], \nn \\
{\cal I}_{n;qq}^{(1)}(x,y_{t_a},\mu) 
&=& \frac{\alpha_s C_F}{2\pi}\Big [ -\frac{1+x^2}{1-x}\ln x +(1-x)+ \Big [\frac{\ln (1-x)}{1-x} \Big ]_+(1+x^2) \nn \\
&-&\ln(iy_{t_a}\mu e^{\gamma_E})\frac{1+x^2}{(1-x)_+}+ \ln^2(i y_{t_a} \mu^2e^{\gamma_E}) \delta(1-x) \Big ],\nn \\
{\cal I}_{n;qg}^{(1)}(x,y_{t_a},\mu) 
&=& \frac{\alpha_s T_F}{2\pi} \Bigg \{ -\ln(iy_{t_a}\mu e^{\gamma_E}) (1-2x +2x^2 ) \nn \\
&+& (1-2x +2x^2 ) (\ln \frac{1-x}{x} -1)+ 1  \Bigg \}, \nn \\
{\cal S}^{(1)}_{\text{part}.}(y_{k_a},y_{k_J},\mu) &=& - \frac{\alpha_s C_F}{4\pi} \Big [  4 \ln^2(iy_{k_a}\tilde{\mu} e^{\gamma_E})  +4\ln^2(iy_{k_J}\tilde{\mu} e^{\gamma_E}) +\pi^2  \Big ]. \nn \\
\eea
In arriving at these results in position space we made use of the identities
\bea
\int_0^\infty dz \> e^{-i z y}\> \Big [ \frac{\theta(z)}{z} \Big ]_+ &=& -\ln (i y e^{\gamma_E}), \nn \\
\int_0^\infty dz \> e^{-i z y}\> \Big [ \frac{\theta(z)\ln z}{z} \Big ]_+ &=& \frac{1}{2}\ln^2 (i y e^{\gamma_E}) + \frac{\pi^2}{12}. \nn \\
\eea

\section{Renormalization group evolution}
\label{resum}
In this section we collect useful formulae that were used in determining the RG evolution of the various quantities in the factorization formula given in Eqs.(\ref{factor-1}) and (\ref{beam}). In particular, we collect formulae for the RG evolution of the hard ($H$), beam ($B$), jet ($J$) and  soft (${\cal S}$) functions.
\subsection{Hard function}
The anomalous dimension $\gamma_H$ of the hard function is defined by
\bea
\label{hardanom}
\mu \frac{d}{d\mu} H(Q^2,\mu) &=& \gamma_H \>H(Q^2,\mu),\nn \\
\eea
and can be written as
\bea
\gamma_H &=& \gamma_c + \gamma_c^*,
\eea
where $\gamma_c$ is the anomalous dimension of the Wilson coefficient $C(Q^2,\mu)$ which satisfies $H(Q^2,\mu) =  | C(Q^2,\mu) |^2$. 
The general form of the anomalous dimension $\gamma_c$ is
\bea
\gamma_c &=& \sum_{(i,j)} \frac{T_i\cdot T_j}{2}\gamma_{\text{cusp}}(\alpha_s) \ln \frac{\mu^2}{-s_{ij}} + \sum_i \gamma^i(\alpha_s),
\eea
where $s_{ij}=2\sigma_{ij}p_i\cdot p_j+ i0$ and $\sigma_{ij}=+1$ if the momenta $p_i$ and $p_j$ are both incoming or outgoing and $\sigma_{ij}=-1$ otherwise.
 $\gamma_{\text{cusp}}$ is related to the cusp anomalous dimension in the the fundamental and adjoint representations $\Gamma_{\text{cusp}}^F(\alpha_s)$ and $\Gamma_{\text{cusp}}^A(\alpha_s)$ respectively as
\bea
\frac{\Gamma_{\text{cusp}}^F(\alpha_s)}{C_F} &=& \frac{\Gamma_{\text{cusp}}^A(\alpha_s)}{C_A} = \gamma_{\text{cusp}}(\alpha_s).
\eea
The cusp and non-cusp anomalous dimensions and the beta function have expansions in $\alpha_s$ given by
\bea
\gamma_{\text{cusp}} [\alpha_s] &=& \sum_{n=0}^\infty \Big (\frac{\alpha_s}{4\pi}\Big )^{n+1}\gamma_n^{\text{cusp}},\>\> 
 \gamma^i[\alpha_s] = \sum_{n=0}^\infty \Big (\frac{\alpha_s}{4\pi}\Big )^{n+1}\gamma_n^i,\>\>
 \beta[\alpha_s] = -2\alpha_s \sum_{n=0}^\infty \Big (\frac{\alpha_s}{4\pi}\Big )^{n+1}\beta_n. \nn \\
\eea
For NNLL resummation we need $\gamma_{\text{cusp}}$ \cite{Korchemsky:1987wg, Moch:2004pa}, $\gamma^i$ \cite{Moch:2005id}, and $\beta$\cite{Tarasov:1980au,Larin:1993tp} to 3-loops, 2-loops, and 3-loops respectively along with NLO PDFs. The 1-loop, 2-loop, and 3-loop cusp anomalous dimension coefficients are given by 
\bea
\gamma_0^{\text{cusp}} &=& 4 , \nn \\
\gamma_1^{\text{cusp}}&=& 4  \Big [ \Big ( \frac{67}{9}-\frac{\pi^2}{3}\Big ) C_A-\frac{20}{9}T_Fn_f\Big ], \nn \\\
\gamma_2^{\text{cusp}} &=& 4  \Big [ C_A^2 \Big (\frac{245}{6} -\frac{134\pi^2}{27}+ \frac{11\pi^4}{45} + \frac{22}{3}\zeta_3 \Big )+C_A T_F n_f\Big (-\frac{418}{27}+ \frac{40\pi^2}{27}-\frac{56}{3}\zeta_3 \Big )\nn \\
&+& C_FT_Fn_f\Big ( -\frac{55}{3} + 16 \zeta_3\Big ) - \frac{16}{27}T_F^2n_f^2\Big ], \nn \\
\eea
and the beta function coefficients up to 3-loops are given by
\bea
\beta_0 &=& \frac{11}{3} C_A - \frac{4}{3}T_F n_f, \nn \\
\beta_1 &=& \frac{34}{3}C_A^2 -\frac{20}{3}C_AT_fn_f-4C_FT_Fn_f, \nn \\
\beta_2 &=& \frac{2857}{54}C_A^3 + T_f n_f (2 C_F^2 -\frac{205}{9}C_F C_A-\frac{1415}{27}C_A^2) + T_f^2n_f^2(\frac{44}{9}C_F + \frac{158}{27}C_A). \nn \\
\eea
We define two useful quantities $S(\mu_f,\mu_i)$ and $A(\mu_f,\mu_i)$ are needed for the evolution of the hard, jet, beam, and soft functions as
\bea
 S(\mu_f,\mu_i) &=& -\int_{\alpha_s(\mu_i)}^{\alpha_s(\mu_f)} \frac{d\alpha}{\beta[\alpha]}\gamma_{\text{cusp}}[\alpha] \int_{\alpha_s(\mu_i)}^\alpha \frac{d\alpha '}{\beta[\alpha ']}, \nn \\
A(\mu_f,\mu_i) &=& -\int_{\alpha_s(\mu_i)}^{\alpha_s(\mu_f)}\frac{d\alpha}{\beta[\alpha]}\gamma_{\text{cusp}}[\alpha],\nn \\
\eea
The expansion of these quantities in $\alpha_s$ up to terms needed for NNLL resummation are given by
\bea
S(\mu_f,\mu_i) &=& \frac{\gamma_0^{\text{cusp}}}{4\beta_0^2}\Bigg \{ \frac{4\pi}{\alpha_s(\mu_i)} \Big ( 1-\frac{1}{r}-\ln r\Big ) + \Big ( \frac{\gamma_1^{\text{cusp}}}{\gamma_0^{\text{cusp}}}-\frac{\beta_1}{\beta_0}\Big )(1-r+\ln r )+\frac{\beta_1}{2\beta_0}\ln^2r\nn \\
&+& \frac{\alpha_s(\mu_i)}{4\pi}\Bigg [\Big ( \frac{\beta_1\gamma_1}{\beta_0\gamma_0^{\text{cusp}}} - \frac{\beta_2}{\beta_0} \Big )(1-r+r\ln r) +\Big ( \frac{\beta_1^2}{\beta_0^2}-\frac{\beta_2}{\beta_0}\Big )(1-r)\ln r \nn \\
&-&\Big(\frac{\beta_1^2}{\beta_0^2}-\frac{\beta_2}{\beta_0}-\frac{\beta_1 \gamma_1^{\text{cusp}}}{\beta_0\gamma_0^{\text{cusp}}}+\frac{\gamma_2^{\text{cusp}}}{\gamma_0^{\text{cusp}}}\Big)\frac{(1-r)^2}{2}\Bigg ]\Bigg \} \nn \\
\eea
and
\bea
\label{Aevo}
A(\mu_f,\mu_i)&=& \frac{\gamma_0^{\text{cusp}}}{2\beta_0}\left\{
\log r + \,
\frac{\alpha_s(\mu_i)}{4\pi}\left(\frac{\gamma_1^{\text{cusp}}}{\gamma_0^{\text{cusp}}}-\frac{\beta_1}{\beta_0} \right)\,
(r-1) \,
\right.\nn \\
&&\left.\,
+ \frac{\alpha_s^2(\mu_i)}{16\pi^2}\,
\left[\frac{\gamma_2^{\text{cusp}}}{\gamma_0^{\text{cusp}}}-\frac{\beta_2}{\beta_0}
-\frac{\beta_1}{\beta_0}\,
\left(\frac{\gamma_1^{\text{cusp}}}{\gamma_0^{\text{cusp}}}-\frac{\beta_1}{\beta_0} \right)
\right]\frac{r^2-1}{2}
\right\}.
\eea
The solution to the RG equation in Eq.(\ref{hardanom}) gives the evolution factor
\bea
H(Q^2,\mu,\mu_H) &=& U_H(Q^2,\mu,\mu_H)H(Q^2,\mu_H), \nn \\
U_H(Q^2,\mu,\mu_H) &=& \exp \Big[  4 C_FS(\mu,\mu_H)-2A_H (\mu,\mu_H)  \Big ]\Big ( \frac{\mu_H^2}{Q^2} \Big )^{2C_FA(\mu,\mu_H)}, \nn \\
\eea
where
\bea
A_H(\mu_f,\mu_i) &=& -\int_{\alpha_s(\mu_i)}^{\alpha_s(\mu_f)}\frac{d\alpha}{\beta[\alpha]}\gamma^{q}_H[\alpha].\nn \\
\eea
and the 1-loop and 2-loop non-cusp anomalous dimensions for quark fields  are given by
\bea
\gamma_{H_0}^q &=& -6C_F , \nn \\
\gamma_{H_1}^q &=& C_F^2 (-3+4\pi^2-48\zeta_3) +  C_FC_A (-\frac{961}{27}-\frac{11\pi^2}{3}+52 \zeta_3) +  C_FT_F n_f (\frac{260}{27}+\frac{4\pi^2}{3}),\nn \\
\eea
The expansion of $A_{H}(\mu_f,\mu_i)$ is given by replacing $\gamma^{\text{cusp}}_{0,1}\to \gamma^q_{H_0,H_1}$ in Eq.(\ref{Aevo}).

\subsection{Beam, jet, and soft functions}
The RG equations for the beam, jet, and soft functions are given by the convolution equations
\bea
\label{RGbjs}
\mu \frac{d}{d\mu} B^q_A(x,t,\mu) &=& \int dt' \> \gamma_B(t-t',\mu) \>B^q_A(x,t',\mu),\nn \\
\mu \frac{d}{d\mu} J(s,\mu) &=& \int ds' \> \gamma_J(s-s',\mu) J(s',\mu),\nn \\
\mu \frac{d}{d\mu}{\cal S}(k_a,k_J,\mu) &=& \int dk_a' \int dk_J'\>\gamma_S(k_a-k_a',k_J-k_J',\mu){\cal S}(k_a',k_J',\mu),
\eea
where the anomalous dimension for the soft function $\gamma_S$ takes the separable form
\bea
\gamma_S(k_a,k_J,\mu) &=& \delta(k_a) \gamma_S(k_J,\mu) + \delta(k_J) \gamma_S(k_a,\mu).
\eea
The anomalous dimensions for the jet, beam, and soft functions have the general form
\bea
\gamma_J(s,\mu) &=& -2 C_F\gamma_{\text{cusp}} (\alpha_s) \>\frac{1}{\mu^2}\Big ( \frac{\mu^2\theta(s)}{s}\Big )_+ + \gamma^i(\alpha_s)\> \delta(s), \nn \\
\gamma_B(t,\mu) &=& -2C_F \gamma_{\text{cusp}}(\alpha_s)\frac{1}{\mu^2} \Big [ \frac{\mu^2 \theta(t)}{t}\Big ]_+ + \gamma_B^q (\alpha_s) \delta(t), \nn \\
\gamma_S(k,\mu) &=& 2C_F \gamma_{\text{cusp}}(\alpha_s)\>\frac{1}{\tilde{\mu}} \Big ( \frac{\tilde{\mu}}{k}\Big )_+ + \gamma^s(\alpha_s)\delta(k),
\eea
where we have defined the scale $\tilde{\mu} \equiv \mu\sqrt{\hat{s}_{aJ}}$ in the soft function anomalous dimension. 

It is often simpler to work in the Fourier transformed space of the beam, jet, and soft functions. For example, the factorization formula in Eq.(\ref{fac-pos-4}) is expressed in terms of the Fourier transformed quantities. The beam, jet, and soft functions and their position space analogs are related by
\bea
B^q_A(x,t,\mu) &=& \int \frac{dy}{2\pi}\> e^{ity} B^q_A(x,y,\mu),\nn \\
J(s_J,\mu;\mu_J) &=& \int \frac{dy_J}{2\pi} \> e^{iy_Js_J} J(y_J,\mu;\mu_J),\nn \\
S(k_a,k_J,\mu;\mu_S) &=& \int \frac{dy_{k_a} dy_{k_J}}{4\pi^2} \> e^{iy_{k_a}k_a+iy_{k_J}k_J} S(y_{k_a},y_{k_J},\mu;\mu_S),
\eea
Going into position space, the  RG equations take the simpler form
\bea
\mu \frac{d}{d\mu} J(y,\mu) &=& \gamma_J(y,\mu) J(y,\mu), \nn \\
\mu \frac{d}{d\mu}B^q_A(x,y,\mu)&=& \gamma_B(y,\mu) B^q_A(x,y,\mu), \nn \\
\mu\frac{d}{d\mu}{\cal S}(y_a,y_J) &=& \Big [ \gamma_S(y_a,\mu) + \gamma_S(y_J,\mu)\Big ]{\cal S}(y_a,y_J), 
\eea
where the position space anomalous dimension is defined as
\bea
\gamma_B(y,\mu) &=& \int dt e^{-ity} \gamma_B(t,\mu), \nn \\
\gamma_J(y,\mu) &=& \int ds \> e^{-iys} \gamma_J(s,\mu),\nn \\
\gamma_S(y,\mu) &=& \int dk \> e^{-iky}\> \gamma_S(k,\mu). \nn \\
\eea
These position space anomalous dimensions take the general form
\bea
\gamma_B(y,\mu) &=& 2C_F \gamma_{\text{cusp}}(\alpha_s) \ln(i y \mu^2 e^{\gamma_E}) + \gamma^q(\alpha_s), \nn \\
\gamma_J(y,\mu) &=& 2C_F \gamma_{\text{cusp}}(\alpha_s) \ln(i y \mu^2 e^{\gamma_E}) + \gamma^q(\alpha_s), \nn \\
\gamma_S(y,\mu)&=&-2C_F \gamma_{\text{cusp}}(\alpha_s) \ln(i y \mu e^{\gamma_E}) + \gamma^s(\alpha_s).  
\eea
The evolution equations in position space are given in terms of the evolution factors $U_i$ as
\bea
B_A^q(x,y,\mu;\mu_B) &=& U_B(y,\mu,\mu_B) B_A^q(x,y,\mu_B), \nn \\
J(y,\mu;\mu_J) &=& U_J(y,\mu,\mu_B)J(y,\mu_J),\nn \\
{\cal S}(y_a,y_J,\mu;\mu_S) &=& U_S(y_a,y_J,\mu,\mu_S){\cal S}(y_a,y_J,\mu;\mu_S),
\eea
are are given by
\bea
U_B(y_{t_a},\mu_f,\mu_i) &=& \text{exp}\Big [ -4 C_F S(\mu_f,\mu_i) - A_B(\mu_f,\mu_i)\Big ] \Big ( i y_{t_a} \mu_i^2 e^{\gamma_E}\Big )^{-2C_F A(\mu_f,\mu_i)},\nn \\
U_J(y,\mu_f,\mu_i) &=& \text{exp}\Big [ -4 C_F S(\mu_f,\mu_i) - A_J(\mu_f,\mu_i)\Big ] \Big ( i y \mu_i^2 e^{\gamma_E}\Big )^{-2C_F A(\mu_f,\mu_i)},\nn \\
U_S(y_a,y_J,\mu,\mu_S) &=& \Big [ y_a y_J(i\mu_S e^{\gamma_E}\sqrt{\hat{s}_{aJ}})^2\Big ]^{2C_FA(\mu,\mu_S)}\text{exp}\Big [ 4C_FS(\mu,\mu_S)-A_S(\mu,\mu_S)\Big], \nn \\
\eea
where we have defined the quantities
\bea
A_B(\mu_f,\mu_i)&=& -\int_{\alpha_s(\mu_i)}^{\alpha_s(\mu_f)}\frac{d\alpha}{\beta[\alpha]}\gamma^{q}_B[\alpha],\nn \\
A_J(\mu_f,\mu_i)&=&= -\int_{\alpha_s(\mu_i)}^{\alpha_s(\mu_f)}\frac{d\alpha}{\beta[\alpha]}\gamma^{q}_J[\alpha],\nn \\
A_S(\mu,\mu_S) &=& -\int_{\alpha_s(\mu_i)}^{\alpha_s(\mu_f)}\frac{d\alpha}{\beta[\alpha]}\gamma_{S}[\alpha],
\eea
and 
\bea
\gamma^q_B &=& \gamma^q_J, \qquad \gamma_{S} = - \gamma^q_J -\gamma^q_B-\gamma^q_H.
\eea
The $\alpha_s$ expansion of $\gamma^q_J$ is given by
\bea
 \gamma^q_J[\alpha_s] = \sum_{n=0}^\infty \Big (\frac{\alpha_s}{4\pi}\Big )^{n+1}\gamma_{J_n}^q,
\eea
and the terms needed for NNLL resummation are
\bea
\gamma^{q}_{J_0} &=& 6 C_F, \nn \\
\gamma^{q}_{J_1} &=& C_F \Big [(\frac{146}{9}-80\zeta_3)C_A +(3-4\pi^2+ 48\zeta_3) C_F +(\frac{121}{9} +\frac{2\pi^2}{3})\beta_0 \Big ].
\eea


\bibliographystyle{h-physrev3.bst}
\bibliography{disjettiness}

\begin{thebibliography}{10}

\bibitem{Arsene:2004fa}
BRAHMS Collaboration, I.~Arsene {\em et~al.},
\newblock Nucl.Phys. {\bf A757}, 1 (2005), nucl-ex/0410020.

\bibitem{Back:2004je}
B.~Back {\em et~al.},
\newblock Nucl.Phys. {\bf A757}, 28 (2005), nucl-ex/0410022.

\bibitem{Adams:2005dq}
STAR Collaboration, J.~Adams {\em et~al.},
\newblock Nucl.Phys. {\bf A757}, 102 (2005), nucl-ex/0501009.

\bibitem{Adcox:2004mh}
PHENIX Collaboration, K.~Adcox {\em et~al.},
\newblock Nucl.Phys. {\bf A757}, 184 (2005), nucl-ex/0410003.

\bibitem{Muller:2012zq}
B.~Muller, J.~Schukraft, and B.~Wyslouch,
\newblock Ann.Rev.Nucl.Part.Sci. {\bf 62}, 361 (2012), 1202.3233.

\bibitem{Aamodt:2010jd}
ALICE Collaboration, K.~Aamodt {\em et~al.},
\newblock Phys.Lett. {\bf B696}, 30 (2011), 1012.1004.

\bibitem{CMS:2012aa}
CMS Collaboration, S.~Chatrchyan {\em et~al.},
\newblock Eur.Phys.J. {\bf C72}, 1945 (2012), 1202.2554.

\bibitem{Milov:2011jk}
A.~Milov,
\newblock J.Phys. {\bf G38}, 124113 (2011), 1107.0460.

\bibitem{Gyulassy:1993hr}
M.~Gyulassy and X.-n. Wang,
\newblock Nucl.Phys. {\bf B420}, 583 (1994), nucl-th/9306003.

\bibitem{Baier:1996sk}
R.~Baier, Y.~L. Dokshitzer, A.~H. Mueller, S.~Peigne, and D.~Schiff,
\newblock Nucl.Phys. {\bf B484}, 265 (1997), hep-ph/9608322.

\bibitem{Zakharov:1997uu}
B.~Zakharov,
\newblock JETP Lett. {\bf 65}, 615 (1997), hep-ph/9704255.

\bibitem{Wiedemann:2000za}
U.~A. Wiedemann,
\newblock Nucl.Phys. {\bf B588}, 303 (2000), hep-ph/0005129.

\bibitem{Gyulassy:2000er}
M.~Gyulassy, P.~Levai, and I.~Vitev,
\newblock Nucl.Phys. {\bf B594}, 371 (2001), nucl-th/0006010.

\bibitem{Wang:2001ifa}
X.-N. Wang and X.-f. Guo,
\newblock Nucl.Phys. {\bf A696}, 788 (2001), hep-ph/0102230.

\bibitem{Arnold:2002ja}
P.~B. Arnold, G.~D. Moore, and L.~G. Yaffe,
\newblock JHEP {\bf 0206}, 030 (2002), hep-ph/0204343.

\bibitem{Ovanesyan:2011xy}
G.~Ovanesyan and I.~Vitev,
\newblock JHEP {\bf 1106}, 080 (2011), 1103.1074.

\bibitem{Vitev:2008rz}
I.~Vitev, S.~Wicks, and B.-W. Zhang,
\newblock JHEP {\bf 0811}, 093 (2008), 0810.2807.

\bibitem{Vitev:2009rd}
I.~Vitev and B.-W. Zhang,
\newblock Phys.Rev.Lett. {\bf 104}, 132001 (2010), 0910.1090.

\bibitem{D'Eramo:2010xk}
F.~D'Eramo, H.~Liu, and K.~Rajagopal,
\newblock Int.J.Mod.Phys. {\bf E20}, 1610 (2011), 1010.0890.

\bibitem{Ploskon:2009zd}
STAR Collaboration, M.~Ploskon,
\newblock Nucl.Phys. {\bf A830}, 255C (2009), 0908.1799.

\bibitem{Kapitan:2011xy}
STAR Collaboration, J.~Kapitan,
\newblock (2011), 1111.1892.

\bibitem{:2012is}
ATLAS Collaboration, G.~Aad {\em et~al.},
\newblock Phys.Lett. {\bf B719}, 220 (2013), 1208.1967.

\bibitem{Tarafdar:2012ef}
PHENIX Collaboration, S.~Tarafdar,
\newblock (2012), 1208.0456.

\bibitem{Boer:2011fh}
D.~Boer {\em et~al.},
\newblock (2011), 1108.1713.

\bibitem{AbelleiraFernandez:2012cc}
LHeC Study Group, J.~Abelleira~Fernandez {\em et~al.},
\newblock J.Phys. {\bf G39}, 075001 (2012), 1206.2913.

\bibitem{AbelleiraFernandez:2012ni}
J.~Abelleira~Fernandez {\em et~al.},
\newblock (2012), 1211.4831.

\bibitem{Antonelli:1999kx}
V.~Antonelli, M.~Dasgupta, and G.~P. Salam,
\newblock JHEP {\bf 0002}, 001 (2000), hep-ph/9912488.

\bibitem{Dasgupta:2001sh}
M.~Dasgupta and G.~Salam,
\newblock Phys.Lett. {\bf B512}, 323 (2001), hep-ph/0104277.

\bibitem{Dasgupta:2001eq}
M.~Dasgupta and G.~Salam,
\newblock Eur.Phys.J. {\bf C24}, 213 (2002), hep-ph/0110213.

\bibitem{Dasgupta:2002bw}
M.~Dasgupta and G.~P. Salam,
\newblock JHEP {\bf 0203}, 017 (2002), hep-ph/0203009.

\bibitem{Catani:1996vz}
S.~Catani and M.~Seymour,
\newblock Nucl.Phys. {\bf B485}, 291 (1997), hep-ph/9605323.

\bibitem{Graudenz:1997gv}
D.~Graudenz,
\newblock (1997), hep-ph/9710244.

\bibitem{Adloff:1997gq}
H1 Collaboration, C.~Adloff {\em et~al.},
\newblock Phys.Lett. {\bf B406}, 256 (1997), hep-ex/9706002.

\bibitem{Aktas:2005tz}
H1 Collaboration, A.~Aktas {\em et~al.},
\newblock Eur.Phys.J. {\bf C46}, 343 (2006), hep-ex/0512014.

\bibitem{Adloff:1999gn}
H1 Collaboration, C.~Adloff {\em et~al.},
\newblock Eur.Phys.J. {\bf C14}, 255 (2000), hep-ex/9912052.

\bibitem{Breitweg:1997ug}
ZEUS Collaboration, J.~Breitweg {\em et~al.},
\newblock Phys.Lett. {\bf B421}, 368 (1998), hep-ex/9710027.

\bibitem{Chekanov:2002xk}
ZEUS Collaboration, S.~Chekanov {\em et~al.},
\newblock Eur.Phys.J. {\bf C27}, 531 (2003), hep-ex/0211040.

\bibitem{Chekanov:2006hv}
ZEUS Collaboration, S.~Chekanov {\em et~al.},
\newblock Nucl.Phys. {\bf B767}, 1 (2007), hep-ex/0604032.

\bibitem{Stewart:2010tn}
I.~W. Stewart, F.~J. Tackmann, and W.~J. Waalewijn,
\newblock Phys.Rev.Lett. {\bf 105}, 092002 (2010), 1004.2489.

\bibitem{Collins:1989gx}
J.~C. Collins, D.~E. Soper, and G.~F. Sterman,
\newblock Adv.Ser.Direct.High Energy Phys. {\bf 5}, 1 (1988), hep-ph/0409313.

\bibitem{Kang:2011jw}
Z.-B. Kang, A.~Metz, J.-W. Qiu, and J.~Zhou,
\newblock Phys.Rev. {\bf D84}, 034046 (2011), 1106.3514.

\bibitem{Kang:2012zr}
Z.-B. Kang, S.~Mantry, and J.-W. Qiu,
\newblock Phys.Rev. {\bf D86}, 114011 (2012), 1204.5469.

\bibitem{Guo:2000nz}
X.-f. Guo and X.-N. Wang,
\newblock Phys.Rev.Lett. {\bf 85}, 3591 (2000), hep-ph/0005044.

\bibitem{Wang:2002ri}
E.~Wang and X.-N. Wang,
\newblock Phys.Rev.Lett. {\bf 89}, 162301 (2002), hep-ph/0202105.

\bibitem{Stewart:2009yx}
I.~W. Stewart, F.~J. Tackmann, and W.~J. Waalewijn,
\newblock Phys.Rev. {\bf D81}, 094035 (2010), 0910.0467.

\bibitem{Stewart:2010pd}
I.~W. Stewart, F.~J. Tackmann, and W.~J. Waalewijn,
\newblock Phys.Rev.Lett. {\bf 106}, 032001 (2011), 1005.4060.

\bibitem{Berger:2010xi}
C.~F. Berger, C.~Marcantonini, I.~W. Stewart, F.~J. Tackmann, and W.~J.
  Waalewijn,
\newblock JHEP {\bf 1104}, 092 (2011), 1012.4480.

\bibitem{Liu:2012zg}
X.~Liu, S.~Mantry, and F.~Petriello,
\newblock Phys.Rev. {\bf D86}, 074004 (2012), 1205.4465.

\bibitem{Jouttenus:2013hs}
T.~T. Jouttenus, I.~W. Stewart, F.~J. Tackmann, and W.~J. Waalewijn,
\newblock (2013), 1302.0846.

\bibitem{Jouttenus:2011wh}
T.~T. Jouttenus, I.~W. Stewart, F.~J. Tackmann, and W.~J. Waalewijn,
\newblock Phys.Rev. {\bf D83}, 114030 (2011), 1102.4344.

\bibitem{Thaler:2011gf}
J.~Thaler and K.~Van~Tilburg,
\newblock JHEP {\bf 1202}, 093 (2012), 1108.2701.

\bibitem{Bauer:2000ew}
C.~W. Bauer, S.~Fleming, and M.~E. Luke,
\newblock Phys.Rev. {\bf D63}, 014006 (2000), hep-ph/0005275.

\bibitem{Bauer:2000yr}
C.~W. Bauer, S.~Fleming, D.~Pirjol, and I.~W. Stewart,
\newblock Phys.Rev. {\bf D63}, 114020 (2001), hep-ph/0011336.

\bibitem{Bauer:2001ct}
C.~W. Bauer and I.~W. Stewart,
\newblock Phys.Lett. {\bf B516}, 134 (2001), hep-ph/0107001.

\bibitem{Bauer:2001yt}
C.~W. Bauer, D.~Pirjol, and I.~W. Stewart,
\newblock Phys.Rev. {\bf D65}, 054022 (2002), hep-ph/0109045.

\bibitem{Bauer:2002nz}
C.~W. Bauer, S.~Fleming, D.~Pirjol, I.~Z. Rothstein, and I.~W. Stewart,
\newblock Phys.Rev. {\bf D66}, 014017 (2002), hep-ph/0202088.

\bibitem{Beneke:2002ph}
M.~Beneke, A.~Chapovsky, M.~Diehl, and T.~Feldmann,
\newblock Nucl.Phys. {\bf B643}, 431 (2002), hep-ph/0206152.

\bibitem{Kang:2013nha}
D.~Kang, C.~Lee, and I.~W. Stewart,
\newblock (2013), 1303.6952.

\bibitem{Fleming:2006cd}
S.~Fleming, A.~K. Leibovich, and T.~Mehen,
\newblock Phys.Rev. {\bf D74}, 114004 (2006), hep-ph/0607121.

\bibitem{Luo:1994np}
M.~Luo, J.-w. Qiu, and G.~F. Sterman,
\newblock Phys.Rev. {\bf D50}, 1951 (1994).

\bibitem{Kang:2011bp}
Z.-B. Kang, I.~Vitev, and H.~Xing,
\newblock Phys.Rev. {\bf D85}, 054024 (2012), 1112.6021.

\bibitem{Dusling:2009ni}
K.~Dusling, F.~Gelis, T.~Lappi, and R.~Venugopalan,
\newblock Nucl.Phys. {\bf A836}, 159 (2010), 0911.2720.

\bibitem{Stewart:2010qs}
I.~W. Stewart, F.~J. Tackmann, and W.~J. Waalewijn,
\newblock JHEP {\bf 1009}, 005 (2010), 1002.2213.

\bibitem{Mantry:2009qz}
S.~Mantry and F.~Petriello,
\newblock Phys.Rev. {\bf D81}, 093007 (2010), 0911.4135.

\bibitem{Ligeti:2008ac}
Z.~Ligeti, I.~W. Stewart, and F.~J. Tackmann,
\newblock Phys.Rev. {\bf D78}, 114014 (2008), 0807.1926.

\bibitem{Hoang:2007vb}
A.~H. Hoang and I.~W. Stewart,
\newblock Phys.Lett. {\bf B660}, 483 (2008), 0709.3519.

\bibitem{Gelis:2010nm}
F.~Gelis, E.~Iancu, J.~Jalilian-Marian, and R.~Venugopalan,
\newblock Ann.Rev.Nucl.Part.Sci. {\bf 60}, 463 (2010), 1002.0333.

\bibitem{Eskola:2009uj}
K.~Eskola, H.~Paukkunen, and C.~Salgado,
\newblock JHEP {\bf 0904}, 065 (2009), 0902.4154.

\bibitem{deFlorian:2003qf}
D.~de~Florian and R.~Sassot,
\newblock Phys.Rev. {\bf D69}, 074028 (2004), hep-ph/0311227.

\bibitem{deFlorian:2011fp}
D.~de~Florian, R.~Sassot, P.~Zurita, and M.~Stratmann,
\newblock Phys.Rev. {\bf D85}, 074028 (2012), 1112.6324.

\bibitem{Hirai:2007sx}
M.~Hirai, S.~Kumano, and T.-H. Nagai,
\newblock Phys.Rev. {\bf C76}, 065207 (2007), 0709.3038.

\bibitem{Kovarik:2010uv}
K.~Kovarik {\em et~al.},
\newblock Phys.Rev.Lett. {\bf 106}, 122301 (2011), 1012.0286.

\bibitem{Owens:2012bv}
J.~Owens, A.~Accardi, and W.~Melnitchouk,
\newblock (2012), 1212.1702.

\bibitem{Stewart:2011cf}
I.~W. Stewart and F.~J. Tackmann,
\newblock Phys.Rev. {\bf D85}, 034011 (2012), 1107.2117.

\bibitem{Albacete:2013tpa}
J.~L. Albacete, A.~Dumitru, and C.~Marquet,
\newblock (2013), 1302.6433.

\bibitem{Manohar:2003vb}
A.~V. Manohar,
\newblock Phys.Rev. {\bf D68}, 114019 (2003), hep-ph/0309176.

\bibitem{Bauer:2003di}
C.~W. Bauer, C.~Lee, A.~V. Manohar, and M.~B. Wise,
\newblock Phys.Rev. {\bf D70}, 034014 (2004), hep-ph/0309278.

\bibitem{Mannel:2000aj}
T.~Mannel and S.~Recksiegel,
\newblock Phys.Rev. {\bf D63}, 094011 (2001), hep-ph/0009268.

\bibitem{Bosch:2004th}
S.~Bosch, B.~Lange, M.~Neubert, and G.~Paz,
\newblock Nucl.Phys. {\bf B699}, 335 (2004), hep-ph/0402094.

\bibitem{Korchemsky:1987wg}
G.~Korchemsky and A.~Radyushkin,
\newblock Nucl.Phys. {\bf B283}, 342 (1987).

\bibitem{Moch:2004pa}
S.~Moch, J.~Vermaseren, and A.~Vogt,
\newblock Nucl.Phys. {\bf B688}, 101 (2004), hep-ph/0403192.

\bibitem{Moch:2005id}
S.~Moch, J.~Vermaseren, and A.~Vogt,
\newblock JHEP {\bf 0508}, 049 (2005), hep-ph/0507039.

\bibitem{Tarasov:1980au}
O.~Tarasov, A.~Vladimirov, and A.~Y. Zharkov,
\newblock Phys.Lett. {\bf B93}, 429 (1980).

\bibitem{Larin:1993tp}
S.~Larin and J.~Vermaseren,
\newblock Phys.Lett. {\bf B303}, 334 (1993), hep-ph/9302208.

\end{thebibliography}

\end{document}